\documentclass[11pt, a4paper, twoside]{article}

\usepackage[utf8]{inputenc}
\usepackage[T1]{fontenc}
\usepackage{lmodern}
\usepackage{microtype} 
\usepackage{setspace}
\AddToHook{env/table/begin}{\singlespacing}
\AddToHook{env/figure/begin}{\singlespacing}

\usepackage[
    margin=1in,
    headheight=14pt,
    includeheadfoot
]{geometry}

\usepackage{amsmath, amssymb, amsthm, mathtools}
\IfFileExists{bbm.sty}{\usepackage{bbm}}{} 

\usepackage{booktabs}      
\usepackage{array}         
\usepackage{dcolumn}       
\usepackage{threeparttable}
\usepackage{multirow}
\usepackage{caption}
\usepackage{subcaption}
\usepackage{graphicx}
\graphicspath{{Updated_Figures_Oct9/}{Figures/}{./}}
\usepackage{float}
\usepackage{xcolor}
\usepackage{tikz}
\usepackage{pgfplots}
\pgfplotsset{compat=1.18}

\definecolor{oxfordblue}{RGB}{0, 33, 71}
\definecolor{crimsonred}{RGB}{153, 0, 0}
\definecolor{slategray}{RGB}{112, 128, 144}

\usepackage[round, authoryear]{natbib}
\usepackage{fancyhdr}
\usepackage[
    colorlinks=true,
    linkcolor=oxfordblue,
    citecolor=oxfordblue,
    urlcolor=crimsonred,
    bookmarks=true
]{hyperref}
\usepackage[nameinlink, noabbrev]{cleveref}

\title{%
    \vspace{-1.5cm}%
    \textbf{What is Work For?}\\[0.35em]
    {\Large \textbf{Labor, Politics, and Meaning in the Age of AI}}%
    \thanks{I thank Ben Ansell, Joshua Martin, David Rueda, Alex Imas, Alexander Kuo, Ranjit Lall, Maria del Rio-Chanona, Tris Papakonstantinou, Stephanie Chan, Nina Ong, Claire Conneely, Jess Hill, and Charlie Buckley for their helpful feedback and support with this paper. Original survey data ($N = 8,234$ across the United Kingdom, United States, Denmark, and India) was collected in partnership with More in Common and supported by Google DeepMind. The views expressed in this paper do not necessarily reflect those of the University of Oxford or Google DeepMind.}%
}

\author{%
    \textbf{Julian Jacobs} \\
    \small University of Oxford and Google DeepMind\\ \\
}

\date{%
    \today
}

\begin{document}

\maketitle

\begin{abstract}
\noindent
This paper investigates the role of work in providing meaning, purpose, community, and other non-monetary benefits. Using an original cross-national survey of 8,234 working-age adults across the UK, US, Denmark, and India, I study the factors that motivate people in their work, with an eye toward the future AI economic transformation. I find that $81\%$  of respondents report being satisfied with their job, and a majority across all four countries are strongly motivated by the non-monetary components of their employment. Using an original ``Work Values'' Index, I show that one third of working respondents are motivated intrinsically by their job---valuing their role for its own sake. Nearly one quarter are driven by social and community-related aspects of their employment, $17\%$  by social status, and just over one quarter by the transactional financial benefits of their job. Although a majority of people would leave their role if offered their current salary as a guaranteed income, one fifth of workers would stay even if offered up to $120\%$  of their salary. Intrinsically motivated workers were disproportionately likely to refuse a basic income at any substitution level. Groups attached to their job for non-monetary reasons were also more likely to pursue volunteering or work-like activities if they received guaranteed income substitution but could no longer perform their role. Crucially, these non-monetary work motivations are evenly distributed across the income distribution and between white- and blue-collar workers. I argue that these findings have implications for managing the AI transition, understanding attitudes to AI, and making sense of policy preferences. Guaranteeing a basic income may indeed be necessary for managing any future AI labor shock, yet it could ultimately fail to compensate workers for any loss of the non-monetary benefits they derive from work.

\vspace{0.3cm}
\end{abstract}

\thispagestyle{empty}
\clearpage
\setcounter{page}{1}

\section{Introduction}
\label{sec:intro}

How much do people value the work that they do every day? In standard neoclassical economic theory, labor and leisure are trade-offs, with employment treated purely as a source of \textit{disutility}---something people would prefer to do less of if given an option (see Adam Smith's ``toil and trouble,'' \citealp{Smith1998}; \citealp{Lazear2018}). Yet a diversity of literature in economic sociology and behavioral philosophy suggests that labor provides non-monetary benefits, which include a source of personal identity, structure, communal belonging, and social status \citep{Durkheim1893, Marshall1920, Spencer2015, Valentino2021, KorinekJuelfs2022}. Indeed, subjective perceptions of social belonging are often entangled with self-perceptions of value, prestige, and importance within the workplace \citep{BlauDuncan1967, RidgewayWalker1995, Lamont2000, Valentino2021}. 

This paper explores the value that people place on their employment, with a particular focus on the AI economic transition. It is well established that technological automation can impose short- and medium-run challenges for some workers, who may experience hardship due to under-employment, stagnating or lower real wages, and redundant skill-sets \citep{Autor2015, AcemogluRestrepo2017, AcemogluRestrepo2020}. Yet when a technology dispenses with some forms of work, it also may create a loss of the \textit{non-monetary} value that a person's employment offered---for instance, social opportunities or a feeling of purpose. Recent digital automation played a role in rendering some middle-wage occupations obsolete, potentially shifting some displaced workers into roles that were more precarious and less linked to their prior skill-sets \citep{Autor2008, AcemogluAutor2011, Oesch2013, Kalleberg2009}. Social scientists have recently theorized that this labor dislocation can---in addition to imposing economic hardship---erode workers' sense of \textit{intrinsic purpose} and usher in feelings of \textit{status anxiety} or ``ontological insecurity,'' whereby a person feels aggrieved at the loss of meaning and social standing their job afforded them \citep{Giddens1991, CassarMeier2018, Kurer2020, MettenBayerlein2023}. Political scientists, in particular, have used versions of this theory to explain digitalization's role in kindling populist spirit across developed economies \citep{Gingrich2019, KurerGallego2019, Kurer2020, Jacobs2024}.

The role of the non-pecuniary\footnote{Throughout this paper, I use the terms ``non-monetary'' and ``non-pecuniary'' interchangeably to refer to the rewards of employment beyond wages and material benefits.} value of work in shaping the AI economic transition has, to date, been an under-explored topic of empirical research. Indeed, if a person attaches significant weight to their work, outside of financial considerations, income transfers alone (e.g.\ through basic incomes) may ultimately fail to replace what a person loses if AI automates their job. Reckoning with this question is particularly important today, as advanced AI systems diffuse throughout economic activities, creating the potential for entirely new areas of labor substitution \citep{JacobsImas2026}.

To better understand the non-monetary value that people attach to their work and what it tells us about how societies may navigate the AI economic transition, I developed an original cross-national representative survey ($N = 8,234$), run by More in Common, across four economies: the United Kingdom, the United States, Denmark, and India. This work captures several measures of how much people care about their work for non-pecuniary reasons, workers' perceived exposure to automation, and general sentiments about AI's likely effects on society. In particular, this research makes five core contributions to existing literature. 

First, across all occupations, $81\%$ of respondents reported being satisfied with their jobs, and I show that a majority of workers are attached to their employment for non-monetary reasons. Specifically, I have developed a four-category \textit{Work Values Profile} classification, which shows whether a worker is primarily motivated by transactional, intrinsic, community, or status-related considerations. Just over one-quarter ($26\%$) of respondents have a \textit{Transactional} relationship with their employment, valuing it for pay and benefits. By contrast, the data shows that one-third of workers in our sample carry an \textit{Intrinsic} motivation, whereby they report obtaining a sense of purpose, learning, or a feeling of societal contribution from their work. Meanwhile, nearly one-quarter ($23\%$) of respondents have a \textit{Community} orientation, reporting that they value their employment because it brings them a sense of social belonging or an opportunity for socializing. Furthermore, $17\%$ of respondents are \textit{Status-Driven}, saying they value their employment because it offers them recognition and prestige. An especially important observation---with implications for how policy responses to technological change ought to be designed---is that these Work Values Profiles cut very evenly across occupation categories and income levels. In other words, a person's likelihood of valuing their work for non-monetary reasons is \textit{not} obviously moderated by how much a person makes or what they do. 

Second, I show that most people would choose to voluntarily leave their job if they were offered a full guaranteed income replacement. Yet a sizable fraction---$20\%$ of workers---would choose to stay in their job even if offered up to $120\%$ as guaranteed income. This share stays roughly consistent across the income distribution, at $21\%$ for low-income, $18\%$ for middle-income, and $22\%$ for high-income workers. The share is slightly higher for ``white-collar'' workers ($21\%$) than for ``blue-collar'' workers ($17\%$). Work Values Profiles, meanwhile, are indeed associated with a person's likelihood of demanding higher levels of wage replacement to voluntarily leave work. Specifically, in a model that compares profiles while controlling for income, occupation, demographic observables, and country fixed effects, we see that \textit{Intrinsically} motivated workers demand a 10-percentage-point higher raw median guaranteed income ($80\%$ versus $70\%$ on the discrete 10-percentage-point steps of the survey auction) to leave work when compared with \textit{Transactional} workers. In continuous regression terms, this amounts to a statistically significant covariate-adjusted mean premium of around two percentage points higher guaranteed wage replacement for \textit{Intrinsic} workers relative to \textit{Transactional} ones ($p < 0.01$)---rising to five to six percentage points ($p < 0.01$) when accounting for the nearly one-in-four \textit{Intrinsic} workers who refuse to quit even at the $120\%$ ceiling. \textit{Community}-oriented workers demand a smaller premium than \textit{Intrinsic} workers, though still a statistically significant one, over \textit{Transactional} workers among job-secure workers and in the $>120\%$ refusal cohort ($p < 0.05$), possibly owing to the more readily available presence of non-work social opportunities. Curiously, workers with \textit{Status}-oriented profiles require the lowest level of wage replacement to quit working, with a raw median of $60\%$ (10 percentage points below \textit{Transactional} workers in the auction exercise) and a statistically significant two percentage points lower covariate-adjusted mean than transactional workers in our regression model ($p < 0.01$). 

Third, I show that how people weigh the non-monetary value of work, as measured by Work Values Profiles, is associated with divergences in how they would choose to spend their time if they were no longer employed but received income substitution. In particular, a majority ($55\%$) of \textit{Status}-motivated workers say they would continue to pursue formal \textit{work} activities, such as starting a business or seeking another job. This can partly explain why people who are status-motivated gave up work at lower levels of wage substitution: doing so allows them to pursue other status-seeking opportunities, perhaps more unencumbered by their current role. On the other hand, most ($57\%$) workers motivated by the \textit{intrinsic} value of their work would fill their time with volunteering or caregiving, potentially reflecting continued mission-attachment. Similarly, $51\%$ of \textit{Community}-oriented workers would turn to caregiving or volunteering, alongside $46\%$ who would continue to pursue \textit{work-like} activities. By complete contrast, $79\%$ of \textit{Transactional} workers would choose to functionally retire and spend their time with leisure or travel. Each of these divergences is highly statistically significant ($p < 0.01$) in multivariate regression models controlling for income, occupation, demographic observables, and country fixed effects.

Fourth, I show that perceptions of AI's economic impacts may play a role in moderating people's work attachment, including the wage that they would require to stop working. Specifically, if someone believes that AI is more likely to automate their job, their demanded wage replacement shrinks across all Work Values Profiles, at a statistically significant two to four percentage points per standard deviation of perceived threat ($p < 0.01$). These results are robust to a battery of controls on observables; however, I note that the model cannot infer the directionality of this relationship (i.e.\ more pessimistic people might demand lower wages, while also believing AI will automate their job). Nonetheless, this research finds a relationship between perceptions of AI's threat and demanded replacement wages. I note, in particular, that across Work Values Profiles, the slope of the downward shift is significantly steeper among \textit{Intrinsic} and \textit{Community} oriented workers than among \textit{Transactional} workers. Future research, with a causal identification approach, can help elucidate this relationship further. 

Finally, I directly engage with contemporary political economy literature, exploring a theoretical puzzle about why technological automation has historically been associated with rising support for anti-establishment right-wing parties even when welfare transfers are available. Across all four countries, and with particular depth in a supplementary United Kingdom module capturing institutional alienation and redistributive ideology, I find evidence that political reactions to AI may operate through both material and non-monetary channels. In baseline evaluations, \textit{Transactional} workers view AI's broader political impact most negatively, sit furthest to the economic left in support of redistribution, and express the strongest preference for dismantling existing political institutions. At the same time, when workers anticipate that AI will genuinely harm job quality and employment in their industry, non-monetary workers---those motivated by \textit{intrinsic} purpose, \textit{workplace community}, and \textit{social status}---exhibit more intense reactions and the steepest increases in political pessimism and perceptions that the political system is ``rigged.'' 

This paper is organized across eight sections. Section \ref{sec:lit_review} reviews the literature on technological change, the non-monetary value of work, and political backlash to economic shocks, while Section \ref{sec:theory} formalizes a theoretical framework for understanding the non-pecuniary value of labor, including with an eye toward understanding socio-political behavior. Section \ref{sec:data} describes the survey design used for this research project, the Work Values Profiles, variable definitions, and initial descriptive findings. Section \ref{sec:results_reservation} discusses empirical findings about how workers value the non-pecuniary aspects of their employment. Section \ref{sec:results_political} analyzes the interaction between AI threats, work values profiles, age cohorts, and occupational tiers in shaping political behavior. Section \ref{sec:discussion} discusses these findings in relation to my core hypotheses as well as several limitations of this paper's analysis, and Section \ref{sec:conclusion} concludes with a brief discussion of this paper's implications for researchers, policymakers, and workers. 

Altogether, this paper argues that work continues to provide a significant source of value to people's lives outside of pure economic considerations. The presence of this non-monetary dimension can be seen across income groups and occupational categories, suggesting far more complex sets of social, psychological, and individual reasons for valuing work. If AI does indeed usher in a significant shock to the labor market that displaces many workers, this research paper suggests that supplementing incomes (e.g.\ through basic income programs) may be a \textit{necessary} but \textit{insufficient} mechanism to support workers. An effective societal response may need to reckon with the simultaneous shocks to both income \textit{and} the deeply-embedded social, cultural, political, and normative value that work provides.

\section{Literature}
\label{sec:lit_review}

Technological change is instrumental in helping social scientists understand economic history, and modern capitalism. The evidence is clear that productivity-boosting innovation has been vital to sustained economic growth, as well as in shaping global and domestic distributions of power \citep{Autor2015}. Over the past 40 years, major technology shocks have amalgamated with financialization \citep{Epstein2005, Krippner2005, AnsellAhlquist2017, Wiedemann2021}, trade exposure \citep{AutorDornHanson2013, PierceSchott2016}, and institutional liberalization \citep{Blyth2013, WesternRosenfeld2011} to shift the global balance of economic power. 

Given the magnitude of technology's economic effects, it is perhaps unsurprising that it has also consistently ushered in anxieties over the potential impacts of new innovations on labor automation \citep{Mokyr2015}. During the early Industrial Revolution, Charlotte \citet[p.~50]{Bronte1849} observed how newly introduced textile machinery in northern England ``threw thousands out of work, and left them without legitimate means of sustaining life.'' David Ricardo likewise argued ``the substitution of machinery for human labour is often very injurious to the interests of the class of labourers'' \citep{Kurz1984}, while Karl Marx and Friedrich Engels warned that rapid mechanization risked deepening working-class alienation and immiseration \citep{EngelsMcLellan2009}. By 1927, U.S. Secretary of Labor James Davis asked whether extraordinary industrial mechanization merely ``dispenses with [workers]'' \citep{NYT1927}. Jobs have, of course, persisted. And so mainstream social science often dismissed fears about automation as the ``Luddite fallacy,'' arguing that productivity-boosting innovation ultimately creates more better-paying employment than it destroys over the long run. 

Yet an extensive body of modern literature shows that technological change---including modern digitalization and digital automation---has produced sharp divisions between labor market winners and losers \citep{Alvaredo2018, Atkinson2015, Milanovic2016, Piketty2014}. These effects appear to have emerged through at least two mechanisms. First, macroeconomic evidence suggests that prior moments of technological progress have compressed labor's aggregate share of national income, while capital's share of income has grown \citep{Allen2009, Elsby2016, AcemogluRestrepo2017}. Meanwhile, technology has previously shifted demand for skills in the labor market. Skill-biased technological change (SBTC) defines the phenomenon where new innovations shift which skillsets are rewarded by markets. Previously, workers whose cognitive skills were complemented by digital technologies saw rising real wages, while workers in routine middle-skill occupations often faced wage stagnation and declining employment shares \citep{Autor2008, AcemogluAutor2011, GoosManningSalomons2014}.\footnote{Traditionally, the routine nature of work was a significant factor in predicting how SBTC would impact demand for work, with more routine work experiencing generally higher automation risk. Yet there is preliminary evidence that this measure may be less helpful in understanding AI labor impacts, since AI automation exposure is present across \textit{both} routine and non-routine cognitive labor \citep{IscenkoEtAl2026}.}

The corresponding political implications of these labor market effects have been a central focus of comparative political economy. Scholars have previously argued that the rise of right-wing populism and anti-establishment voting across developed democracies represents an ``economic backlash'' driven by the experience of falling behind for many workers \citep{EvansTilley2017, Im2019, Gingrich2019, KurerGallego2019, BicchiKuoGallego2024}. Yet the exact mechanics of this backlash present a theoretical puzzle. Standard theoretical frameworks assume that voters respond primarily to material economic incentives. If technological displacement inflicts material damage on workers' livelihoods, affected individuals will likely demand greater government spending and support left-wing pro-redistribution parties, according to these theoretical assumptions \citep{Margalit2011, SwankBetz2003}. Why, then, does automation exposure seem to kindle authoritarian nostalgia, out-group hostility, and voting for right-wing populist parties that often do not support wider welfare transfers \citep{GidronHall2017, Kurer2020, Jacobs2024}? Perhaps even more importantly, when states provide welfare benefits and income transfers to compensate workers who fall behind, why does this fail to ward against populist drift among technology-displaced workers \citep{Gingrich2019, Kurer2020}? 

This puzzle is central to this paper's theoretical contribution. Specifically, I look into the role of the non-pecuniary value of work in explaining the mismatch between pro-redistribution material incentives on the one hand and real-world right-wing populist electoral behavior on the other. Prior literature shows that work is unsurprisingly a multi-dimensional social concept, offering much more to workers than a simple exchange of leisure hours for wages. Within sociology and psychology, for instance, employment is often regarded as a source of personal identity, temporal structure, purpose, and social status \citep{Durkheim1893, Marshall1920, Spencer2015, KorinekJuelfs2022}. Yet a recurring question within this literature concerns how each of these dimensions operates. In practice, social status or purpose, for instance, might be features of a comparatively small portion of the labor market, or else they might be present across broad swathes of workers. 

Moreover, there is a question about what these non-monetary components mean in practice, given their inherently subjective and relativistic nature \citep{SinghManoux2005, Miyakawa2012, RidgewayWalker1995, BlauDuncan1967}. One articulation of this concept is that modern labor markets comprise a \textit{heterarchy} of prestige, whereby multiple competing and subjective hierarchies of status, prestige, and morality define individuals' self-perception of occupational worth \citep{Valentino2021}. For instance, industrial craft workers, skilled tradespeople, and administrative personnel often possess deep occupational pride in their ``blue-collar'' work or ``producerist ethic,'' which prizes tangible skills, self-reliance, and hard work \citep{Bonikowski2016, Lamont2000}. These hierarchies are often ``sticky'' across generations \citep{Hodge1964, Bourdieu1984}. 

With this framework in mind, we can better understand why rapid technological change---such as that which might soon be ushered in by AI---can be disruptive, beyond pure economic considerations. Indeed, productivity-boosting technological change that alters demand for skills can shift established heterarchies. When automation dispenses with occupations, many displaced workers lose both their income and the non-pecuniary value of their prior job \citep{Oesch2013, GoosManningSalomons2014}. In the political economy literature, one prominent explanation concerns \textit{status anxiety}: a form of ontological insecurity whereby an individual might perceive an erosion of their place in the social order \citep{Giddens1991, Kinnvall2004, MettenBayerlein2023, Kurer2020}. What is also fundamental, however, is the loss of \textit{intrinsic meaning} and craft or professional purpose derived from the act of working itself. This is a reward that work offers that, unlike external status or material consumption, cannot be carried into a different role or replaced by passive cash transfers \citep{RyanDeci2000, CassarMeier2018}. 

According to this theory, material income and social status can shift inversely and decouple \citep{GidronHall2017, Kurer2020, Vlandas2019}. A worker whose job is changed by AI might maintain their employment and receive compensatory government welfare transfers while simultaneously experiencing significant feelings of eroded status, and the loss of the non-pecuniary value of their work. Social psychological literature suggests that the resulting threats to social standing can manifest in higher rates of authoritarianism, nostalgia for a perceived golden age, and heightened out-group hostility \citep{Lipset1959, TajfelTurner1979, Riek2006, Kupper2010}. And so, when technological shocks threaten work, status anxiety can exert a rightward pressure, toward parties with anti-establishment and nostalgic rhetoric. This force appears to compete with---and can triumph over---the simultaneous leftward pull ushered in by material economic insecurity. 

\section{Theory}
\label{sec:theory}

I now formalize this theory, distinguishing between the broader concept of the \textit{non-monetary (non-pecuniary) value of work} ($N_W$) and three conceptually distinct non-monetary sub-dimensions of employment. These include \textit{intrinsic meaning} ($I_W$), \textit{communal workplace belonging} ($C_W$), and \textit{social status} ($S_W$). Building on prior literature, \textit{intrinsic utility} ($I_W$) captures rewards derived from the act of working itself---including personal purpose and fulfillment and societal contribution \citep{Durkheim1893, RyanDeci2000, CassarMeier2018}. When work provides \textit{intrinsic} value, it does so for its own sake because the worker finds their role inherently meaningful. \textit{Communal workplace utility} ($C_W$), meanwhile, captures the value of work in the form of social integration, relational belonging, and daily camaraderie with coworkers and clients \citep{Durkheim1893, LorenziniGiugni2012}. Finally, a measure of \textit{status utility} ($S_W$) captures a person's perception that their occupation provides them with ``social status,'' prestige, and public recognition from groups in society \citep{Weber1978, TajfelTurner1979, GidronHall2017, Valentino2021}. \citet{Shayo2009} suggests that individuals weight both material consumption and the relative status of the groups with which they identify when making employment decisions. For instance, a worker may shift toward alternative identities or entrepreneurial pursuits when their immediate economic role fails to satisfy subjective perceptions of high social standing. Across all three non-monetary dimensions, commitment to a current job depends on the value a worker gets from their occupation ($I_W, C_W, S_W$) compared to what they would experience if they were not working in that role ($I_{NW}, C_{NW}, S_{NW}$). Since intrinsic utility ($I_W$) is generated directly through active participation in one's occupation, an intrinsically attached worker cannot simply replicate it via passive cash transfers ($I_W > I_{NW}$). Workplace community ($C_W$) is similarly evaluated relative to the community a worker would have outside of employment ($C_{NW}$)---while leaving a job sacrifices daily camaraderie with coworkers, free time and guaranteed income allow a worker to invest more in family, friendships, and local community life, partially substituting for workplace ties ($C_W \ge C_{NW}$). Similarly, for a status-driven worker who prizes external recognition ($\beta_S \gg 0$), commitment to a current job depends on whether that role delivers a superior subjective perception of prestige relative to outside alternatives ($S_{NW}$) or whether guaranteed income can instead serve as a financial floor to pursue higher-status work or entrepreneurship ($S_{NW} \ge S_W$). We can therefore model an individual's labor market participation and welfare preferences by capturing the utility of employment ($U_W$), decomposed by material earnings ($M$) and the umbrella non-monetary value of labor ($N_W \equiv \beta_I I_W + \beta_C C_W + \beta_S S_W$) across intrinsic ($I_W$), communal ($C_W$), and subjective social status ($S_W$) benefits:
\begin{equation}
    U_W = \alpha M + N_W = \alpha M + \beta_I I_W + \beta_C C_W + \beta_S S_W
    \label{eq:utility_work}
\end{equation}
where $\alpha > 0$ represents the marginal utility of material income, $\beta_I \ge 0$ represents the marginal utility of intrinsic purpose, $\beta_C \ge 0$ represents the marginal utility of communal workplace belonging, and $\beta_S \ge 0$ represents the marginal utility of one's subjective perception of their employment social standing. 

Now consider what happens when a person is outside the workforce ($U_{NW}$), whether due to automation displacement or voluntary exit onto a state-provided income replacement program (e.g. unemployment insurance). When out of work, a person might receive a monetary government transfer $T \ge 0$ and experiences non-work non-monetary utility $N_{NW} \equiv \beta_I I_{NW} + \beta_C C_{NW} + \beta_S S_{NW}$ across non-work intrinsic engagement $I_{NW}$, non-work community $C_{NW}$, and non-work societal standing $S_{NW}$:
\begin{equation}
    U_{NW} = \alpha T + N_{NW} = \alpha T + \beta_I I_{NW} + \beta_C C_{NW} + \beta_S S_{NW}
    \label{eq:utility_nonwork}
\end{equation}
Here, a person will be indifferent between continuing to work and not working when $U_W = U_{NW}$:
\begin{equation}
    \alpha M + \beta_I I_W + \beta_C C_W + \beta_S S_W = \alpha T^* + \beta_I I_{NW} + \beta_C C_{NW} + \beta_S S_{NW}
\end{equation}
Solving for the reservation transfer $T^*$ necessary to compensate a worker sufficiently to get them to voluntarily stop working yields:\footnote{Workers also can receive non-monetary benefits or costs from employment beyond $I_W$, $C_W$, and $S_W$. This includes daily temporal structure, habit, or workplace amenities. This is absorbed by the individual disturbance term $\varepsilon_i$. Section~\ref{sec:results_reservation} shows that the profile premiums hold when controlling directly for overall job satisfaction.}
\begin{equation}
    T^* = M + \underbrace{\frac{N_W - N_{NW}}{\alpha}}_{\text{Non-Monetary Premium}} = M + \underbrace{\frac{\beta_I}{\alpha} \left( I_W - I_{NW} \right)}_{\text{Intrinsic Premium}} + \underbrace{\frac{\beta_C}{\alpha} \left( C_W - C_{NW} \right)}_{\text{Community Premium}} + \underbrace{\frac{\beta_S}{\alpha} \left( S_W - S_{NW} \right)}_{\text{Status Premium}}
    \label{eq:reservation_transfer}
\end{equation}

For a worker who views their occupation as a \textit{disutility} or a \textit{transactional} exchange with their employer for money ($\beta_I / \alpha \to 0$, $\beta_C / \alpha \to 0$, and $\beta_S / \alpha \to 0$), the required transfer can simply be written as $T^* \approx M$. In this transactional case, losing a job would not entail a social, psychological, or status-related cost. And so, if the state simply provides an income replacement $T$ equal to forgone wages $M$ (both current and future-expected), the individual is fully compensated, within the model. Things become more complicated when we imagine a person who attaches a significant amount of non-monetary value to their occupation ($N_W > N_{NW}$), where each of the three non-transactional dimensions---intrinsic, community, and status---can individually generate a premium relative to their non-work alternative. This leaves us with a broader non-monetary premium, $\frac{N_W - N_{NW}}{\alpha}$. 

For instance, to voluntarily leave their occupation, an intrinsically motivated worker ($\beta_I \gg 0$) might demand a transfer $T^*$ that exceeds their current salary $M$ by an \textit{intrinsic premium}, $\frac{\beta_I}{\alpha}(I_W - I_{NW})$. A community-oriented worker ($\beta_C \gg 0$) could similarly demand a \textit{community premium}, $\frac{\beta_C}{\alpha}(C_W - C_{NW})$, to the extent that workplace belonging exceeds the community they would have outside of work ($C_{NW}$). By contrast, for a \textit{status-driven} worker ($\beta_S \gg 0$), the \textit{status premium}, $\frac{\beta_S}{\alpha}(S_W - S_{NW})$, is positive only if their work confers a heightened subjective self-perception of social status or prestige, relative to non-work alternatives. Theoretically, when work provides irreplaceable meaning, the total value of this non-monetary premium could be virtually infinite; however, later in this paper we will attempt to measure each component precisely across those samples in this paper's survey. 

And so, let's suppose a labor shock eliminates a worker's job and that a government responds by offering a standard safety net transfer, where $T \le M$. The worker then experiences an uncompensated deficit:
\begin{equation}
    \Delta U_{\text{deficit}} = U_W - U_{NW} = \alpha(M - T) + \underbrace{\beta_I(I_W - I_{NW}) + \beta_C(C_W - C_{NW}) + \beta_S(S_W - S_{NW})}_{N_W - N_{NW}}
\end{equation}
More importantly, even if full wage insurance ($T = M$) reduces the material deficit $\alpha(M-T)$, a worker with high non-pecuniary attachment experiences an uncompensated non-monetary deficit $N_W - N_{NW} = \beta_I(I_W - I_{NW}) + \beta_C(C_W - C_{NW}) + \beta_S(S_W - S_{NW})$. And so even full income substitution is insufficient to compensate for lost employment.

\begin{figure}[tbp]
    \centering
    \singlespacing
    \resizebox{\textwidth}{!}{%
    \begin{tikzpicture}[
        >=stealth,
        header/.style={
            font=\footnotesize\bfseries,
            text=oxfordblue, align=flush center
        },
        shockbox/.style={
            draw=oxfordblue, line width=0.9pt, rounded corners=4pt,
            fill=white, align=flush center, inner sep=7pt,
            text width=3.15cm, font=\scriptsize
        },
        mechbox/.style={
            draw=oxfordblue, line width=0.85pt, rounded corners=4pt,
            fill=white, align=flush center, inner sep=7pt, minimum height=2.95cm,
            text width=4.35cm, font=\scriptsize
        },
        mechbox_nm/.style={
            draw=crimsonred!85!black, line width=0.85pt, rounded corners=4pt,
            fill=white, align=flush center, inner sep=7pt, minimum height=2.95cm,
            text width=4.35cm, font=\scriptsize
        },
        outbox_trans/.style={
            draw=oxfordblue, line width=0.9pt, rounded corners=4pt,
            fill=oxfordblue!5, align=left, inner sep=7pt, minimum height=2.95cm,
            text width=5.25cm, font=\scriptsize
        },
        outbox_nm/.style={
            draw=crimsonred, line width=0.9pt, rounded corners=4pt,
            fill=crimsonred!6, align=left, inner sep=7pt, minimum height=2.95cm,
            text width=5.25cm, font=\scriptsize
        },
        arrow_blue/.style={->, line width=0.9pt, oxfordblue},
        arrow_red/.style={->, line width=0.9pt, crimsonred},
        edgelbl/.style={
            font=\scriptsize\itshape,
            text=slategray, align=center
        },
        edgelbl_red/.style={
            font=\scriptsize\itshape,
            text=crimsonred, align=center
        }
    ]

    \node[header] at (0.0, 3.85) {1. Technological Shock};
    \node[header] at (5.25, 3.85) {2. Work Values Profiles};
    \node[header] at (12.85, 3.85) {3. Worker Preferences};
    \draw[oxfordblue!35, line width=0.5pt] (-1.82, 3.50) -- (15.72, 3.50);

    \node[shockbox] (shock) at (0.0, 0.0) {
        \textbf{Perceived AI \\ Replacement Threat}\\[4pt]
        Perceived risk that AI will replace one's job threatens material income ($M$) and the non-monetary value of work ($N_W$)
    };

    \node[mechbox] (trans) at (5.25, 1.80) {
        \textbf{Transactional Work Values}\\
        \textit{(Pay \& Benefits)}\\[4pt]
        Work is valued primarily as an exchange for pay ($U_W \approx \alpha M$). Welfare transfers can replace lost earnings.
    };

    \node[mechbox_nm] (nonmon) at (5.25, -1.80) {
        \textbf{Non-Monetary Work Values}\\
        \textit{(Intrinsic $I_W$, Community $C_W$, Status $S_W$)}\\[4pt]
        Work provides purpose, belonging, or social status beyond pay ($N_W > N_{NW}$). Transfers cannot replace lost employment.
    };

    \node[outbox_trans] (out_trans) at (12.85, 1.80) {
        {\centering \textbf{Transactional Outcomes}\par}\vspace{3pt}
        $\bullet$ \textbf{Reservation Wage:} Lower wage replacement demanded to quit; shift to leisure under basic income.\\[3pt]
        $\bullet$ \textbf{Political Attitudes:} Less of a negative effect of AI job threat, relative to workers who also stand to lose significant non-pecuniary value.
    };

    \node[outbox_nm] (out_nm) at (12.85, -1.80) {
        {\centering \textbf{Non-Monetary Outcomes}\par}\vspace{3pt}
        $\bullet$ \textbf{Reservation Wage:} Higher wage premium to quit ($T^* > M$), due to the non-pecuniary value of work for this group.\\[3pt]
        $\bullet$ \textbf{Political Attitudes:} Lost standing and feeling left behind $\rightarrow$ perceptions of a rigged system and demand for greater redistribution.
    };

    \draw[arrow_blue] (shock.east) -- ++(0.55, 0) |- (trans.west);
    \draw[arrow_red]  (shock.east) -- ++(0.55, 0) |- (nonmon.west);

    \draw[arrow_blue] (trans.east) -- node[edgelbl, above=3pt] {Transfers} node[edgelbl, below=3pt] {compensate} (out_trans.west);
    \draw[arrow_red]  (nonmon.east) -- node[edgelbl_red, above=3pt] {Transfers} node[edgelbl_red, below=3pt] {insufficient} (out_nm.west);

    \end{tikzpicture}%
    }
    \caption{\textbf{Theorized Relationship Between Non-Pecuniary Work Values and Attitudes Toward Work and AI.} For transactional workers who value employment mainly for pay ($T^* \approx M$), welfare transfers can compensate for lost earnings. For workers who attach non-monetary value to their occupation ($I_W, C_W, S_W$), passive transfers cannot replace the lost meaning, community, or standing of employment ($N_W > N_{NW}$).}
    \label{fig:theory_diagram}
\end{figure}
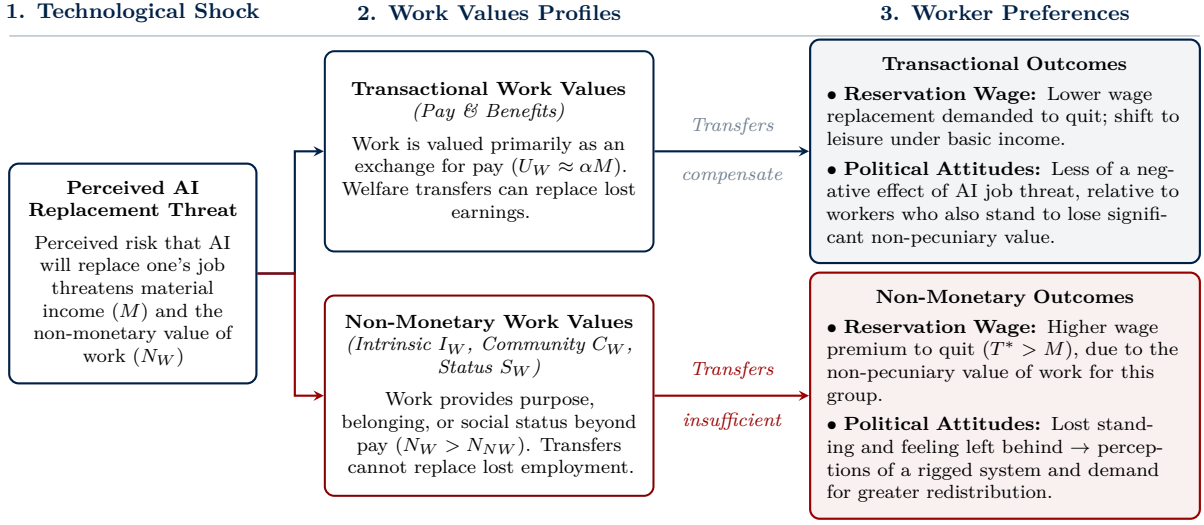

This theory directly engages with existing political economy literature, which seeks to explain how technological shocks have affected political attitudes, preferences for redistribution, and support for populist parties. As discussed in Section~\ref{sec:lit_review}, prior literature indicates that workers react negatively to automating technology shocks, in part through support for anti-establishment populist parties \citep{Im2019, KurerGallego2019, Kurer2020, BicchiKuoGallego2024, Jacobs2024}. What is less clear, based on this prior literature, is why this support does not appear to be effectively moderated by welfare transfers \citep{Gingrich2019, Kurer2020}. As we can see in Figure~\ref{fig:theory_diagram}, the non-monetary value of work provides one possible explanation for this, building off prior literature on status and occupational worth \citep{GidronHall2017, KurerGallego2019, Kurer2020, Valentino2021}. Specifically, for workers who attach a meaningful non-monetary value to the work they do, transfers would be insufficient to fully compensate them for lost employment. And so the well-understood mechanisms by which economic malaise can usher in political backlash---feelings of being left behind, lost standing, worsened quality of life---can take root. 

It is with this theory that I seek to test three primary empirical hypotheses across our four-country survey dataset. First, I hypothesize (\textit{Hypothesis 1}) that in direct comparisons with transactionally motivated people, workers who place a significant importance on the non-monetary value of labor (including intrinsic, community, or status-related components) will require higher wages to voluntarily exit employment. Second (\textit{Hypothesis 2}), workers who value their employment particularly due to status, intrinsic, or communal motivations will pursue work-like or prosocial activities (such as seeking new employment, entrepreneurship, volunteering, or caregiving, rather than pure leisure) in scenarios where they receive income transfers but are no longer able to work in their current role. And third (\textit{Hypothesis 3}), workers who place a significant value on the non-monetary component of work are most likely to express concern about AI-related impacts on their employment, and among those who perceive a high threat of displacement, they will react more strongly in their labor market and political attitudes than transactional workers.

\section{Data, Design, and the Non-Monetary Value of Work}
\label{sec:data}

I now proceed by discussing how this paper will test its hypotheses about the role of the non-pecuniary value of work in shaping labor market attachment and socio-political attitudes. To study these questions, I conducted an original cross-national survey of 8,234 working-age adults across four high- and emerging-income democracies. These countries are Denmark ($N=2,032$), India ($N=2,072$), the United Kingdom ($N=2,052$), and the United States ($N=2,078$), respectively. This survey was fielded in June and July of 2026, in partnership with More in Common (a polling organization), which used national quota sampling on age, gender, educational attainment, region, and other observables to generate a representative sample of each country's adult working population.\footnote{All subsequent regression models estimate unweighted OLS or logit specifications that include these demographic variables and country fixed effects in the conditioning set.}

The four countries were selected because they represent divergent economic contexts and market structures. We might accordingly expect divergences in the ways in which people value their work---and respond to AI---to be partially conditioned by these national contexts. The United Kingdom, for starters, is a mature hybrid post-industrial service economy, in which the public is generally familiar with AI systems---two thirds of British adults, for example, say they have used AI chatbots. Yet it is also a country marked by high levels of economic pessimism, which is apparent in the British public's general sentiments about AI's impacts. Public attitudes toward AI's impact on jobs sit at $-36$, while just 14\% of Britons expect AI's economic impacts to benefit ordinary citizens.

\begin{figure}[!htbp]
    \centering
    \includegraphics[width=0.86\linewidth]{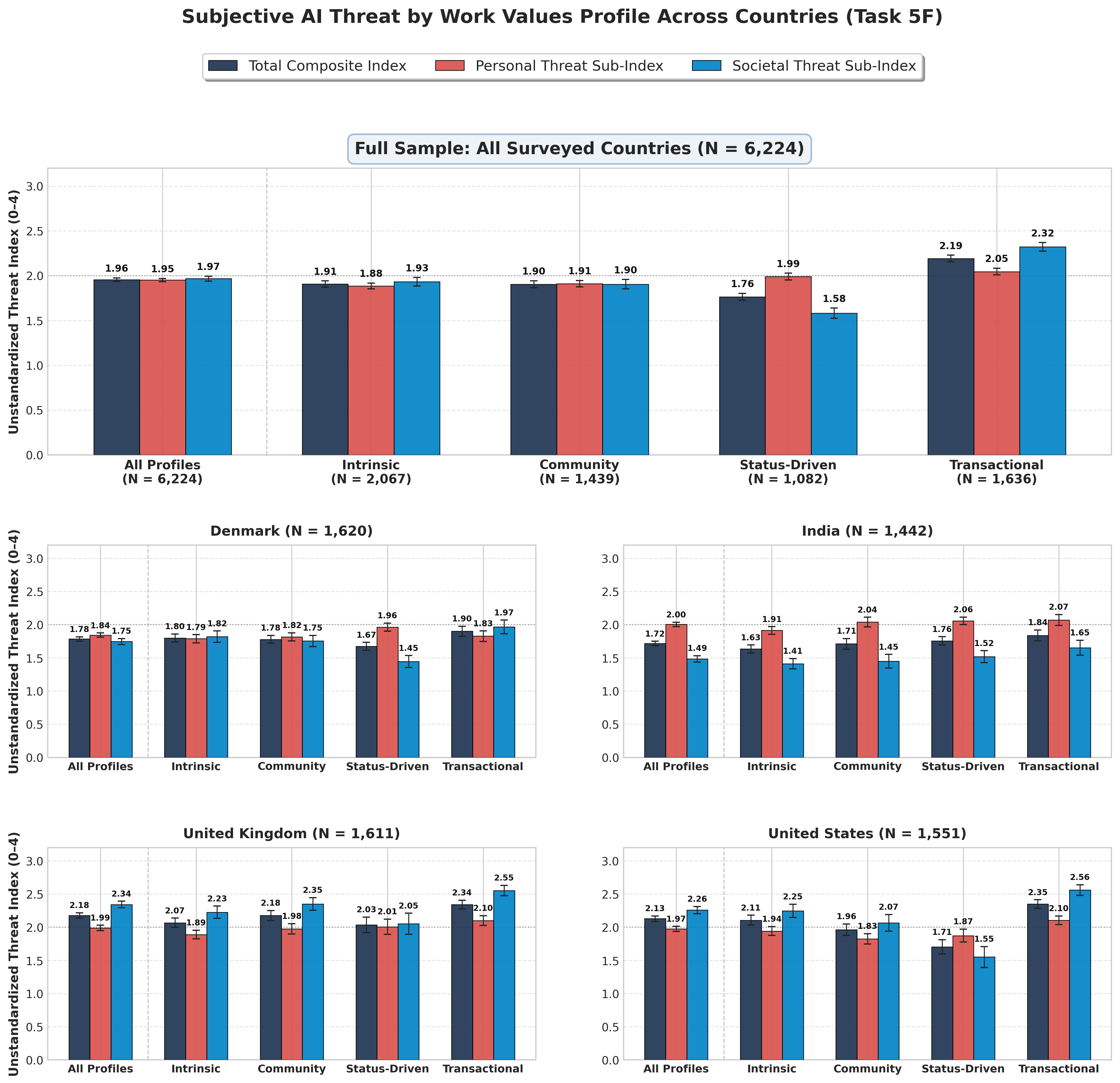}
    \caption{\textbf{Subjective AI Threat by Work Values Profile Across Surveyed Countries}}
    \label{fig:ai_threat_by_work_values}
    \vspace{0.10cm}
    \begin{minipage}{\linewidth}
        \footnotesize 
        \textit{Notes:} Group means and 95\% confidence intervals ($1.96 \times \text{SE}$ whiskers) for unstandardized AI threat indices rescaled to a common [0, 4] metric where 0 represents minimum perceived threat, 2 represents the scale midpoint, and 4 is the maximum perceived threat. Within each panel, the leftmost group (`All Profiles') shows the average across all classified respondents ($N=6,224$), followed by the four Work Values Profiles. The Total Index averages all 9 evaluative items. The Personal Threat Sub-Index averages the four workplace and industry items. The Societal Threat Sub-Index averages the five macro-level societal items. Top hero panel displays the pooled four-country classified sample ($N=6,224$); lower panels display individual country samples.
    \end{minipage}
\end{figure}

The United States, meanwhile, is a highly dynamic liberal economy, with flexible labor markets and high average wages, but significant levels of inequality and a long-standing socio-cultural skepticism of state interference with free enterprise. As we will see later, Americans are disproportionately likely to have a transactional relationship with their employment, relative to other countries. And they are also more likely to emphasize individual self-reliance over public insurance or safety nets. Denmark is a coordinated social democracy, and home to its famous ``flexicurity'' system, which amalgamates an ease of hiring and firing with generous support for active labor market policies and safety nets, which are meant to provide a cushion for workers who experience job loss. Danish workers reported the highest job satisfaction rate among countries in this study, with 82\% indicating they are satisfied or very satisfied. They also hold the lowest fear of personal AI job replacement ($+24$ net positive expectations regarding AI's macroeconomic impact). Finally, India is a developing economy with expanding digital services and strong global demand for its software, IT, and digital outsourcing services. An important note is that, due to internet coverage and literacy constraints, online survey sampling in India captures a disproportionately educated, urban, and ``white-collar'' segment of the country's working population. Several findings from the surveyed Indian sample should therefore be approached with caution, including, for instance, that the country exhibited the highest attachment to work, with net $+90$ calling it important to their personal identity versus $+62$ in the United States and $+52$ in the United Kingdom and Denmark. This optimism was, however, mirrored by a higher level of concern about AI-enabled automation, with 27\% of Indian respondents naming it as the single greatest threat to their careers over the next decade. Such marked results may indeed be a feature of India's outsource-heavy digital economy; yet, it also is almost certainly a symptom of over-representation of certain populations inherent in the online sample.\footnote{Throughout this paper's analysis, all models include country fixed effects alongside education, income, and occupation controls. Country-by-country breakdowns (Table~\ref{tab:work_profiles}, Figure~\ref{fig:ai_threat_by_work_values}, Figure~\ref{fig:reservation_wage_curve}, and Figure~\ref{fig:app_important_to_identity}) as well as Western-only sensitivity models show that all core behavioral and attitudinal patterns hold independently within the three Western democracies (Denmark, the United Kingdom, and the United States; $N=6,162$ total, $N=4,782$ classified).}

Table~\ref{tab:demographics} shows the core demographic summary statistics across all four survey samples, including age, gender, higher educational attainment, part-time employment status, occupational category, household income terciles, and settlement urbanness. Across the three Western samples, demographic characteristics are closely aligned in average age (40 to 44 years), university attainment (43\% to 48\%), and white-collar employment shares (65\% to 71\%), whereas the online Indian sample is notably younger (mean age 34.3 years) and predominantly comprises university-educated (77\%) and white-collar (84\%) workers.

\begin{table}[!htbp]
    \centering
    \begin{threeparttable}
        \caption{\textbf{Demographics of Surveyed Population}}
        \label{tab:demographics}
        \footnotesize
        \renewcommand{\arraystretch}{1.08}
        \setlength{\tabcolsep}{0pt}
        \begin{tabular*}{\textwidth}{@{\extracolsep{\fill}}lcccc@{}}
            \toprule
            \textbf{Demographic Variable} & 
            \textbf{Denmark} & 
            \textbf{India} & 
            \textbf{UK} & 
            \textbf{US} \\
            \midrule
            Age (years) [Mean (SD)] & 39.88 (13.16) & 34.26 (11.93) & 42.54 (14.06) & 43.84 (14.93) \\
            Female [\%] & 45.0\% & 37.0\% & 49.0\% & 46.0\% \\
            Higher Education (Bachelor's+) [\%] & 43.0\% & 77.0\%* & 44.0\% & 48.0\% \\
            Part-Time Employment [\%] & 16.0\% & 17.0\% & 23.0\% & 20.0\% \\
            White-Collar Occupational Tier [\%] & 65.0\% & 84.0\%* & 70.0\% & 71.0\% \\
            High Household Income Tercile [\%] & 30.0\% & 30.0\% & 30.0\% & 29.0\% \\
            Middle Household Income Tercile [\%] & 42.0\% & 34.0\% & 39.0\% & 37.0\% \\
            Low Household Income Tercile [\%] & 28.0\% & 36.0\% & 31.0\% & 34.0\% \\
            Urbanness (1--8 Scale) [Mean (SD)] & 6.13 (2.12) & 5.69 (2.33) & 7.17 (1.48) & 6.01 (2.18) \\
            \midrule
            \textbf{Total Surveyed Sample ($N$)} & \textbf{2,032} & \textbf{2,072} & \textbf{2,052} & \textbf{2,078} \\
            \bottomrule
        \end{tabular*}
        \begin{tablenotes}[flushleft]
            \footnotesize
            \item \textit{Notes:} Continuous variables report the mean with standard deviation in parentheses. The binary indicators report the sample percentage. I note again that India's online survey sample captures disproportionately the urban, educated, and white-collar subset of the country's workforce (for example, 77\% are university graduates and 84\% are white-collar workers). Urbanness is measured on a 1--8 scale ($1=\text{Water/Uninhabited}$, $2=\text{Very Low Density Rural}$, $\dots$, $8=\text{Urban Centre}$) derived from the European Commission GHS-SMOD Degree of Urbanisation (DEGURBA) and matched to postal codes.
        \end{tablenotes}
    \end{threeparttable}
\end{table}

A central part of this paper is understanding why some workers value their employment beyond financial considerations, and how this valuation shapes their attitudes toward the AI transition. To better understand this---and empirically operationalize the concepts laid out in Section~\ref{sec:theory} of this paper---the survey asked respondents to complete a nine-item attitudinal set of questions, capturing their work priorities. For each dimension we looked at, respondents rated its importance to their working life on a four-point Likert scale, ranging from ``very important'' to ``not important at all.''\footnote{Incomplete or non-substantive responses (``Don't know'') were coded as missing, and items were reverse-coded onto a 1--4 scale so that $4 = \text{Very important}$ and $1 = \text{Not important at all}$.}

I then consolidated the eight core work commitment items into four indices that directly align with the theoretical concepts laid out in prior literature and earlier in this paper.\footnote{This consolidates eight of the nine items, excluding ``having time and space to be myself,'' which captures work--life boundaries and time away from work rather than attachment to work itself.} The resulting work-values indices form a core method through which we attempt to test this paper's underlying theory. The \textit{Transactional Index} ($M$) measures the average importance assigned to (1) ``salary / financial compensation'' and (2) ``benefits (e.g., healthcare, pension, paid leave).'' The \textit{Intrinsic Index} ($I_W$) captures the average importance assigned to work that is fulfilling for its own sake. That is, work that generates (1) ``a sense of purpose or personal fulfillment,'' (2) ``opportunities to learn and develop new skills,'' and (3) ``contributing to society or making a difference.'' The \textit{Community Index} ($C_W$) captures the average importance assigned to horizontal workplace sociability: (1) ``a sense of belonging or community at work'' and (2) ``opportunities to meet people and socialize.'' Finally, the \textit{Status-Driven Index} ($S_W$) captures the importance assigned to subjective self-perceptions of (1) ``social status or recognition from others.''

With these four domain indices in mind, I then assigned each active worker to one of four mutually exclusive \textit{Work Values Profiles}. This was done using a two-step classification rule. To do this and capture each worker's priorities (while resolving ties without Gaussian ceiling distortion), I took two analytical steps.\footnote{Standardizing multi-item and single-item Likert indices via Gaussian $z$-scores distorts comparisons. This is because the single-item Status measure has greater bottom-tail variance ($\text{SD}=0.97$), which inflates its ceiling $z$-score and would force all  respondents who rate Status as ``Very important'' into the Status-Driven profile.} In Step 1, each respondent is evaluated on their 1--4 Likert domain averages. Respondents whose highest-priority domain score did not exceed the importance threshold ($\le 2.0$ on the reverse-coded 1--4 Likert scale, i.e., $\ge 3.0$ on the original scale, $N=111$) or who rated all four domains the same with zero differentiation (4-way straight-liners, $N=771$) are assigned alongside non-working adults ($N=1,128$) to the unclassified ``Other'' category ($N=2,010$, or 24\% of the total sample). This group is excluded from profile analyses. Meanwhile, if any single domain achieves the highest importance score ($N=4,224$), the respondent is assigned directly to that profile. In Step 2, when two or three domains tie at any respondent's maximum score $m$ ($N=2,000$), the tie is broken in favor of the index for which achieving a score of at least $m$ is rarest in the active workforce ($\arg\min_{d} \Pr(\bar{X}_d \ge m)$).\footnote{This tie-breaking rule can be justified on several grounds. Cases where all three items in the \textit{Intrinsic} battery are rated at the ceiling ($4.0=\text{Very important}$) occur for 21.4\% of workers, compared to 22.3\% for the two-item \textit{Community} battery, 27.6\% for single-item \textit{Status}, and 27.9\% for two-item \textit{Transactional}. Since we resolve ties in favor of the empirically rarer characteristic, we capture the dimension on which a respondent stands out most relative to other respondents. Since tied respondents also rated pay and benefits at their maximum, assigning ceiling ties to the non-monetary profile makes our regression contrasts against \textit{Transactional} profiles conservative. Robustness checks using models without tie-breaking produce consistent results.} As a consequence, all comparative analysis is conducted among the remaining $N=6,224$ respondents.

Table~\ref{tab:work_profiles} shows the distribution of the resultant Work Values Profiles across the surveyed population. Among the $N=6,224$ classified working-age adults across all four countries, one third hold an Intrinsic orientation ($N=2,067$, or 33 percent), 26\% hold a Transactional orientation ($N=1,636$), 23\% hold a Community orientation ($N=1,439$), and 17\% hold a Status-Driven orientation ($N=1,082$). As alluded to briefly earlier, there are interesting cross-national divergences that appear in line with intuition and comparative politics frameworks. Once again, the United States---home to a liberal market economy and weaker public safety net---has the largest pool of transactionally motivated workers in this survey (38\% of classified workers, vs.\ 34\% Intrinsic, 18\% Community, and 11\% Status-Driven). In the United Kingdom, Transactional and Intrinsic profiles are nearly even at the top of the distribution (about one third each). Meanwhile, 24\% of Britons are motivated by their workplace community and social life, while just 10\% are status-driven. By contrast, in Denmark---where social security and institutions are most generous---the share of transactionally oriented workers drops to 19 percent, with intrinsically motivated workers at 30 percent, community-motivated workers at 28 percent, and status-driven workers at 22 percent. The Indian sample was, meanwhile, similarly driven by intrinsically motivated (35 percent), status-driven (27 percent), and community-motivated (22 percent) orientations, with transactional profiles sitting at just 15 percent.

\begin{table}[!htbp]
    \centering
    \begin{threeparttable}
        \caption{\textbf{Distribution of Work Values Profiles by Country}}
        \label{tab:work_profiles}
        \footnotesize
        \renewcommand{\arraystretch}{1.08}
        \setlength{\tabcolsep}{0pt}
        \begin{tabular*}{\textwidth}{@{\extracolsep{\fill}}lccccc@{}}
            \toprule
            \textbf{Country} & 
            \textbf{Transactional} & 
            \textbf{Intrinsic} & 
            \textbf{Community} & 
            \textbf{Status-Driven} & 
            \textbf{All Classified ($N$)} \\
            \midrule
            Denmark & 311 (19.20\%) & 489 (30.19\%) & 457 (28.21\%) & 363 (22.41\%) & 1,620 (100.0\%) \\
            India   & 222 (15.40\%) & 511 (35.44\%) & 321 (22.26\%) & 388 (26.91\%) & 1,442 (100.0\%) \\
            UK      & 517 (32.09\%) & 547 (33.95\%) & 379 (23.53\%) & 168 (10.43\%) & 1,611 (100.0\%) \\
            US      & 586 (37.78\%) & 520 (33.53\%) & 282 (18.18\%) & 163 (10.51\%) & 1,551 (100.0\%) \\
            \midrule
            \textbf{Pooled Classified} & \textbf{1,636 (26.29\%)} & \textbf{2,067 (33.21\%)} & \textbf{1,439 (23.12\%)} & \textbf{1,082 (17.38\%)} & \textbf{6,224 (100.0\%)} \\
            \bottomrule
        \end{tabular*}
        \begin{tablenotes}[flushleft]
            \footnotesize
            \item \textit{Notes:} Mutually exclusive profiles constructed from the 1--4 Likert job importance battery among qualified respondents ($N=6,224$). (1) \textbf{Transactional Profile} ($M$) is defined as placing the highest priority on pay and benefits. (2) \textbf{Intrinsic Profile} ($I_W$) places highest priority on task-inherent purpose, skill learning, and societal contribution. (3) \textbf{Community Profile} ($C_W$) places the highest priority on workplace belonging and socializing. (4) \textbf{Status-Driven Profile} ($S_W$) most emphasizes social status and recognition. Non-working respondents not administered the battery ($N=1,128$) and working respondents with four-way straight-lined scores, all-``Don't know'' responses, or no index meeting the $>2.0$ ($<3.0$ raw) importance threshold ($N=882$) are excluded. 
        \end{tablenotes}
    \end{threeparttable}
\end{table}

A question for this methodology is whether these Work Values profiles capture distinct orientations toward work rather than simply proxying for economic class or political beliefs. To test this, I use a multivariate linear probability model that predicts profile membership across demographic and political covariates among classified respondents ($N=6,082$ and $N=5,391$; reported in Appendix Table~\ref{tab:app_profile_validation} ). Controlling for country-level fixed effects, high household income does not significantly predict the likelihood of any work values profile. In line with prior literature, a person's occupation and partisan voting preferences, similarly, have virtually no association with these profiles.

\begin{figure}[!htbp]
    \centering
    \includegraphics[width=0.95\linewidth]{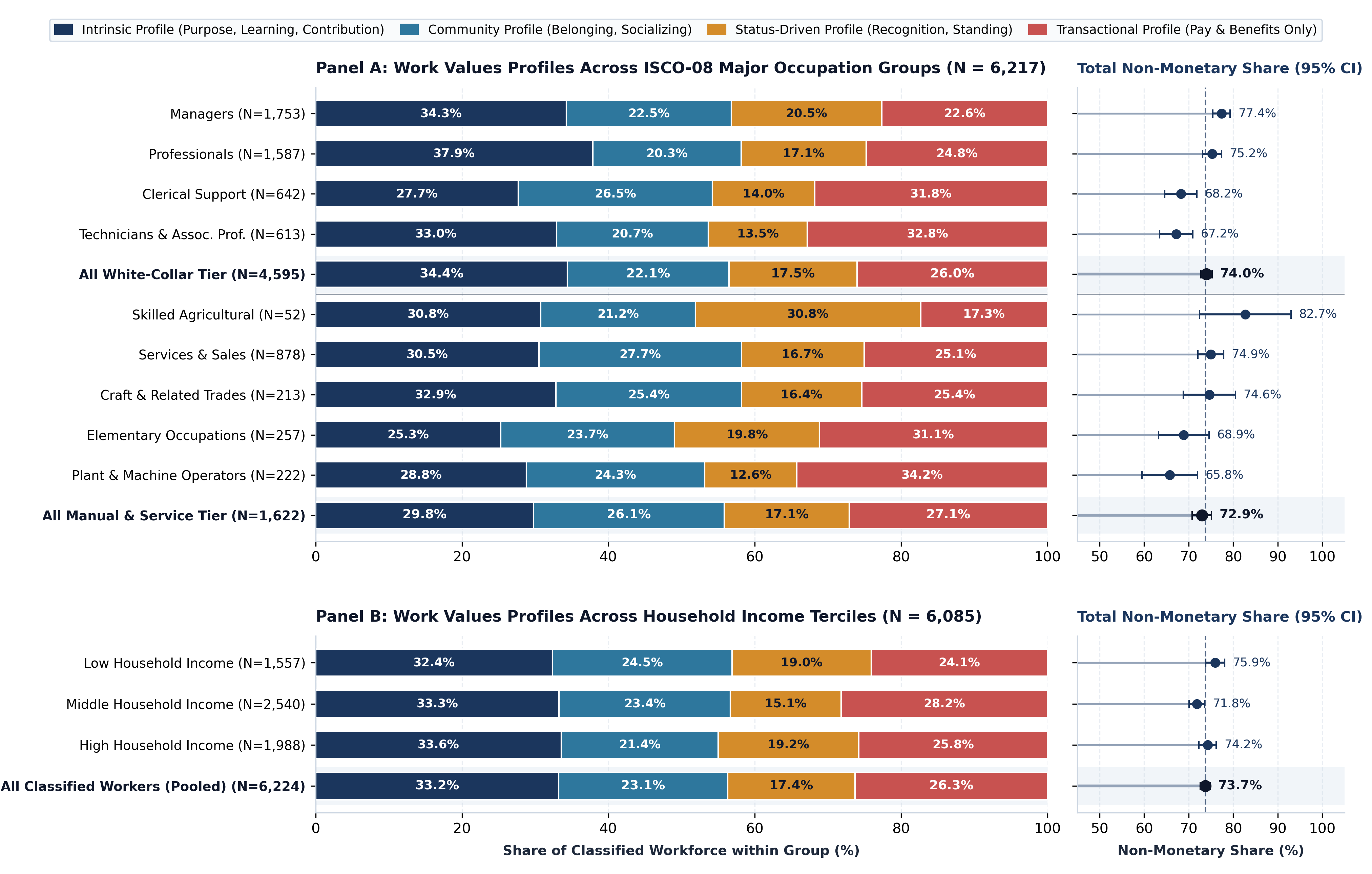}
    \caption{\textbf{Distribution of Work Values Profiles Across ISCO-08 Major Occupation Groups and Household Income Terciles}}
    \label{fig:work_values_by_occupation_income}
    \vspace{0.10cm}
    \begin{minipage}{\linewidth}
        \footnotesize
        \textit{Notes:} Left panels display the within-group percentage breakdown across the four Work Values Profiles (\textit{Intrinsic}, \textit{Community}, \textit{Status-Driven}, and \textit{Transactional}), over ISCO-08 major occupation groups (Panel A, aggregating White-Collar ISCO 1--4 [$N = 4,595$] vs.\ Manual \& Service ISCO 5--9 [$N = 1,622$] tiers; total $N = 6,217$) and household income terciles (Panel B; $N = 6,085$). Right panels plot the total non-pecuniary share across Work Values Profiles (with 95\% confidence intervals); the vertical dashed line denotes the pooled classified workforce non-monetary share ($73.7\%$, $N = 6,224$).
    \end{minipage}
\end{figure}

Moreover, as we can see in Figure~\ref{fig:work_values_by_occupation_income}, when we look at the breakdown of Work Values Profiles across ISCO-08 major groups, non-pecuniary work attachment is pervasive across occupations, regardless of whether they are ``blue-collar'' or ``white-collar.'' Indeed, intrinsic attachment is present in 38\% of Professionals and 34\% of Managers as well as in 31\% of Services and Sales workers and 33\% of Craft and Related Trades workers. The prevalence of non-monetary attachment to work is also nearly \textit{invariant} across income levels, with intrinsic shares at 32\% for low-income people, 33\% for those that are middle-income, and 34\% for those that are high-income. These numbers are 25\%, 23\%, and 21\%, respectively, for community-oriented workers, and 19\%, 15\%, and 19\% for status-driven workers.\footnote{An interesting note, however, is that higher education appears to be positively associated with the likelihood of a person having an Intrinsic orientation ($\beta = +0.060, p < 0.01$), while part-time employment positively predicts Community orientation ($\beta = +0.041, p < 0.01$) and negatively predicts Transactional orientation ($\beta = -0.056, p < 0.01$).}

Given this paper's focus on studying how workers may respond to the AI economic transition, it is also important to understand whether these work values profiles correspond with a heightened perceived belief that AI will replace a person's job over the next ten years. For descriptive purposes, responses are grouped into a binary indicator distinguishing between High Threat (\textit{Certain, Very likely}, or \textit{Somewhat Likely}) and Low Threat (\textit{Unlikely} or \textit{Very Unlikely}). For regression specifications, the item is reverse-coded onto a continuous 1--7 scale ($1=\text{Lowest threat}$, $7=\text{Highest threat}$) and standardized into a $z$-score ($\mu=0, \sigma=1$). Table~\ref{tab:ai_threat_tier} shows the subjective perception of AI's displacement threat by occupational category. In direct contrast with prior waves of technological change, which tended to concentrate risk among routine manual workers, people in white-collar occupations are most likely to report concern about AI-related job replacement (40\%) relative to manual and service workers (29\%). This finding is striking due to its complete alignment with recent observational evidence on AI usage and potential task automation throughout economic activities \citep{Eloundou2024, IscenkoEtAl2026}. In other words, the same occupations where preliminary evidence suggests automation might occur are exactly the places where people most perceive the risk of automation.

\begin{table}[!htbp]
    \centering
    \begin{threeparttable}
        \caption{\textbf{Subjective AI Job Replacement Threat by Occupational Category}}
        \label{tab:ai_threat_tier}
        \footnotesize
        \renewcommand{\arraystretch}{1.08}
        \setlength{\tabcolsep}{0pt}
        \begin{tabular*}{\textwidth}{@{\extracolsep{\fill}}lcccc@{}}
            \toprule
            \textbf{Occupational Category} & 
            \textbf{High Threat} & 
            \textbf{Low Threat} & 
            \textbf{Missing} & 
            \textbf{Total ($N$)} \\
            \midrule
            White-Collar Occupational Tier               & 2,082 (40.4\%) & 2,934 (57.0\%) & 135 (2.6\%)  & 5,151 \\
            Manual \& Service Occupational Tier          & 564 (28.8\%)   & 1,343 (68.7\%) & 48 (2.5\%)   & 1,955 \\
            Not in Workforce                             & 0 (0.0\%)      & 0 (0.0\%)      & 1,128 (100\%)& 1,128 \\
            \midrule
            \textbf{Active Workforce ($N=7,106$)}        & \textbf{2,646 (37.2\%)} & \textbf{4,277 (60.2\%)} & \textbf{183 (2.6\%)} & \textbf{7,106} \\
            \textbf{Total Sample ($N=8,234$)}            & \textbf{2,646 (32.1\%)} & \textbf{4,277 (51.9\%)} & \textbf{1,311 (15.9\%)} & \textbf{8,234} \\
            \bottomrule
        \end{tabular*}
        \begin{tablenotes}[flushleft]
            \footnotesize
            \item \textit{Notes:} High Threat means respondents report that AI/automation is certain, very likely, or somewhat likely to replace their job within 10 years. Occupational tiers come from the survey's 10-category employment classification (\texttt{employment1}): the ``White-Collar'' tier is composed of professional/technical, managerial/administrative, clerical, sales, and farm-proprietor roles ($N=5,151$), while the ``Manual \& Service'' tier is composed of service, skilled, semi-skilled, unskilled manual, and farm-worker roles ($N=1,955$).
        \end{tablenotes}
    \end{threeparttable}
\end{table}

In Figure~\ref{fig:ai_threat_by_work_values}, I look at attitudes toward AI across the four Work Values Profiles ($N=6,224$), including assessing people's perceptions about their job's risk of automation (a four-item Personal Threat Sub-Index averaging the 10-year likelihood that AI will replace or significantly change one's job and AI's expected impact on job quality and job numbers in one's industry) as well as the broader societal impacts of AI (a five-item Societal Threat Sub-Index averaging AI's expected impact on national job quality, national job numbers, the national economy, the country generally, and national politics). I find that transactionally motivated workers report the highest level of anxiety on the overall index and regarding AI's societal impacts, whereas community- and status-driven workers report greater concern about AI's risk to their own job. Status-driven workers express high levels of concern about their job's susceptibility to automation while also demonstrating less concern about broader societal impacts. As we will discuss later, however, these groups react very differently to the threat of automation, through both their political behavior and demand for work. Across the UK and US samples, workers view AI as worse for society than for themselves. We find the opposite in Denmark and, especially, India, where active workplace AI use is associated with a $+0.28\text{ SD}$ perception of their own job's automation but a $-0.17\text{ SD}$ perception of societal threats.

To capture \textit{how much} workers value their employment beyond the compensation they directly receive ($T^*$ in Equation~\ref{eq:reservation_transfer}), active workers completed an iterative auction-style sequence asking what percentage of their current salary (from 60\% to 120\% in 10-percentage-point increments, with benefits preserved) they would require as a guaranteed basic income to voluntarily quit their job, alongside a post-displacement time-use module. Finally, to understand how automation threats and work values moderate socio-political attitudes (summarized alongside all survey variables in Appendix Table~\ref{tab:variable_definitions}), the survey captures two cross-national political outcomes across all four countries---\textit{Perceived Negative AI Political Impact} (expectations of societal harm to the national political system over the next decade; $z$-score, $N=6,487$ full workforce; $N=5,688$ classified) and \textit{Right-Wing or Populist Party Preference} (binary vote intention for national right-wing or populist parties; $N=6,018$ full; $N=5,295$ classified)---supplemented in the United Kingdom survey module ($N=1,667$ full workforce; $N=1,498$ classified) by three additional attitudinal batteries capturing \textit{Economic Right Ideology Placement} (a standardized composite mean across four redistribution and corporate power grids) as well as institutional alienation.\footnote{Specifically, the supplementary UK module also includes \textit{Political System Riggedness} (agreement that the economic and political system is rigged to serve the rich and powerful; $z$-score, $N=1,667$ full workforce; $N=1,498$ classified) and \textit{Institutional Chaos Preference} (agreement with dismantling or burning down broken institutions rather than preserving them; $z$-score, $N=1,667$ full; $N=1,498$ classified); full question wordings and coding rules are reported in Appendix Table~\ref{tab:variable_definitions}.}

In the next section, we begin this paper's core analysis. Since all Work Values Profiles are mutually exclusive among classified respondents ($N=6,224$), all profile regressions in Sections~\ref{sec:results_reservation} and \ref{sec:results_political} employ a \textit{cell-means specification}. In this model, the general constant is omitted, and one indicator is included for each profile in addition to a profile-specific AI threat slope ($\text{AI Threat} \mid \text{Profile}$). Panel~A reports the covariate-adjusted mean for each Work Values Profile at mean AI threat, with demographic controls and country indicators mean-centered in each estimation sample. Panel~B, meanwhile, reports the within-profile AI threat slope. Panels~C and D provide Wald test results of differences between work profiles, which are mathematically identical to coefficients from a regression with Transactional as the omitted baseline. To better understand age-related dynamics, each model is presented separately for younger ($\le 41$) and older ($> 41$) workers (the split is at the sample's median age). I note, however, that despite our inclusion of a battery of controls designed to account for observable correlates, coefficient values are descriptive and should not be interpreted as offering causal evidence. 

\section{Measuring the Value of Work}
\label{sec:results_reservation}

We have so far established that the majority of people in this study's sample value their work due to factors beyond financial compensation. Yet a key question is to gauge \textit{how much} people value their work. To explore this, I now discuss a module in which survey participants were asked about the percentage of their salary they would be willing to accept as a guaranteed income in order to voluntarily leave their job. This auction-style question was asked sequentially, with an initial offering of 60\% of their current earnings, increasing sequentially up to 120\% in 10 percentage point increments. This module is intended to directly measure this paper's first hypothesis: that we should expect workers who derive a significant amount of non-pecuniary value from their work (be it through status, intrinsic meaning, or community) will exhibit a higher attachment to continued work, relative to transactional workers ($H_1$). The results of this are presented below in Figure~\ref{fig:reservation_wage_curve}.

\begin{figure}[!htbp]
    \centering
    \includegraphics[width=0.88\linewidth]{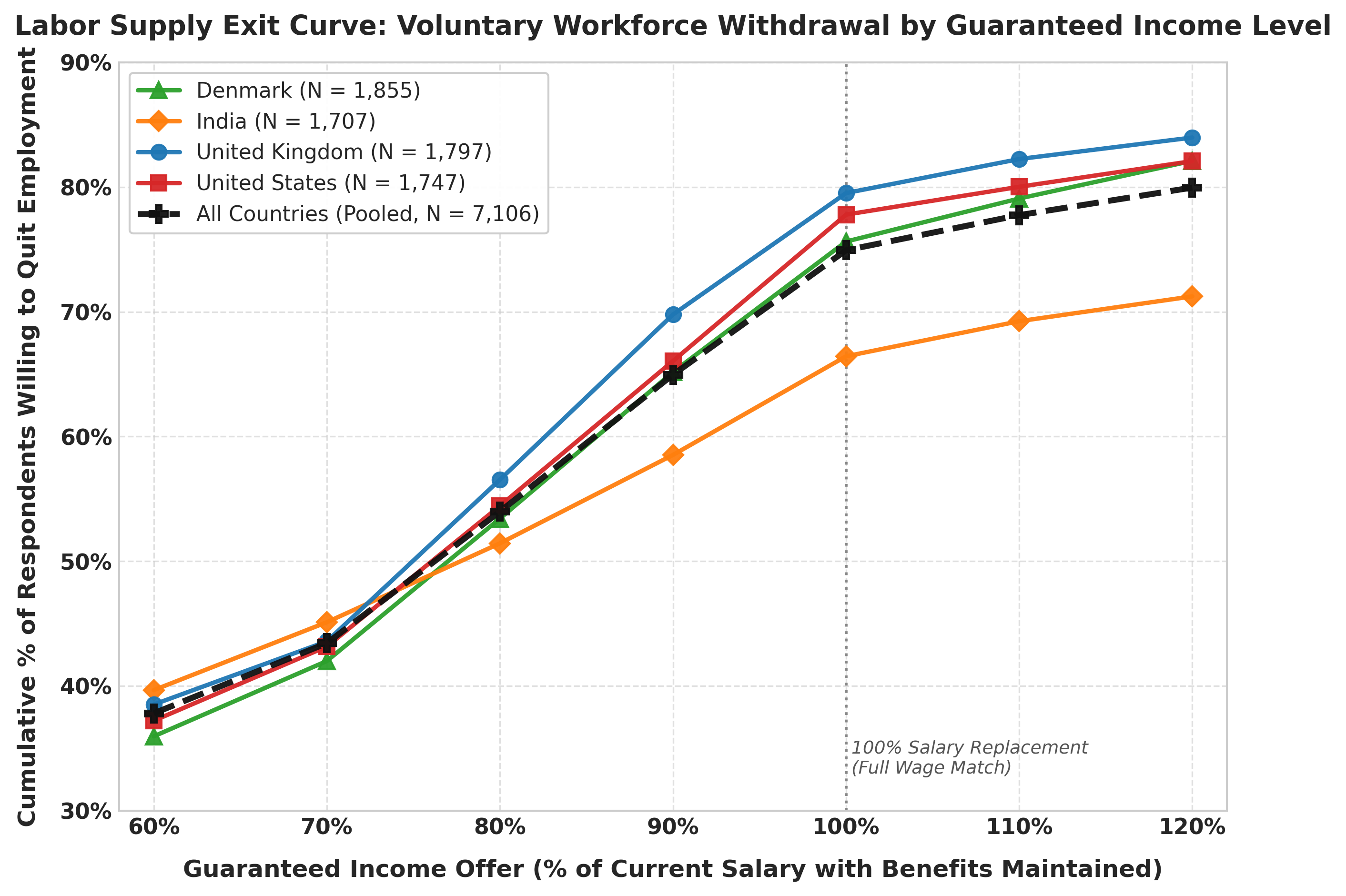}
    \caption{\textbf{Voluntary Workforce Withdrawal by Guaranteed Income Threshold}}
    \label{fig:reservation_wage_curve}
    \vspace{0.10cm}
    \begin{minipage}{\linewidth}
        \footnotesize 
        \textit{Notes:} Lines display the total share of survey respondents who are willing to voluntarily exit paid employment at each sequential guaranteed income threshold (60\% to 120\% of current salary, preserving healthcare, pension, and paid leave benefits). Non-quitters who refused all offers up to 120\% salary (Denmark: 17.9\%, India: 28.8\%, UK: 16.0\%, US: 17.9\%, Pooled: 20.0\%) remain in the denominator (Total $N=7,106$). The vertical dashed line marks full wage parity (100\% replacement).
    \end{minipage}
\end{figure}

The data paints a complex picture of the ways in which people value their work. By 80\% wage replacement a slight majority of people across countries would choose to voluntarily leave their job. By 90\% wage replacement, this increases to around 65\%. And around 75\% of workers say they would choose to leave their job if they were offered 100\% substitution of their current wage. Yet, curiously, 20\% of respondents ($N=1,424$) refused to quit working, even when offered 120\% of their current wage, suggesting around one fifth of workers have a particularly deep attachment to their employment. For the rest of the sample ($N=5,682$), the average reservation income required to voluntarily leave employment sat at 76\% of each worker's current salary, and the median sat at 70\%.\footnote{
The survey auction module stops at 120\% of a person's current salary. As a consequence, we do not observe preferences beyond the threshold of 120\%, and we cannot know for certain if or when the right-censored 20\% of respondents ($N=1,424$) would ultimately agree to voluntarily leave their job. If we conservatively assume that those respondents would exit at the next hypothetical auction increment of 130\% of salary, the active-workforce average reservation wage increases by 10.85 percentage points from 76\% ($\text{SD}=17.85$) to 87\% ($\text{SD}=26.91$). The active-workforce median rises from 70.0\% to 80.0\% (or to a mean of 85\% if top-coded strictly at the observed 120\% ceiling). Meanwhile, if we instead assume those holdouts require 200\% of their current salary (double their pay) to exit employment, the average rises by 25 percentage points to 100.75\% ($\text{SD}=52.19$; median $= 80.0\%$).} There are also slight outliers across the sample, with UK respondents generally most keen to leave their job (perhaps due to the country's particularly pessimistic view of its economy). In the online-only India sample, where work ranks highest as a source of personal identity ($+90$ net importance), nearly 29\% of workers refused to quit their job even at 120\% of their current wage.

There are two interesting implications of this. The first of these is that, despite the fact that a majority of workers are heavily driven by non-monetary factors in their work, most people say they would choose to quit work altogether if their current income were guaranteed. Second, for a small and tenacious minority of workers, 100--120\% permanent income substitution would not be enough for them to leave their job. Yet what do these findings actually mean in practice? For instance, a person might choose to leave their job when given some level of guaranteed income, on the basis that they could then pursue other meaningful unpaid work-related activities. Moreover, are expectations about future expected wage growth shaping the decision to stay or leave? And finally, do particular forms of non-pecuniary work value (e.g.\ community or status) moderate the decision to continue working?

\begin{figure}[!htbp]
    \centering
    \includegraphics[width=0.90\linewidth]{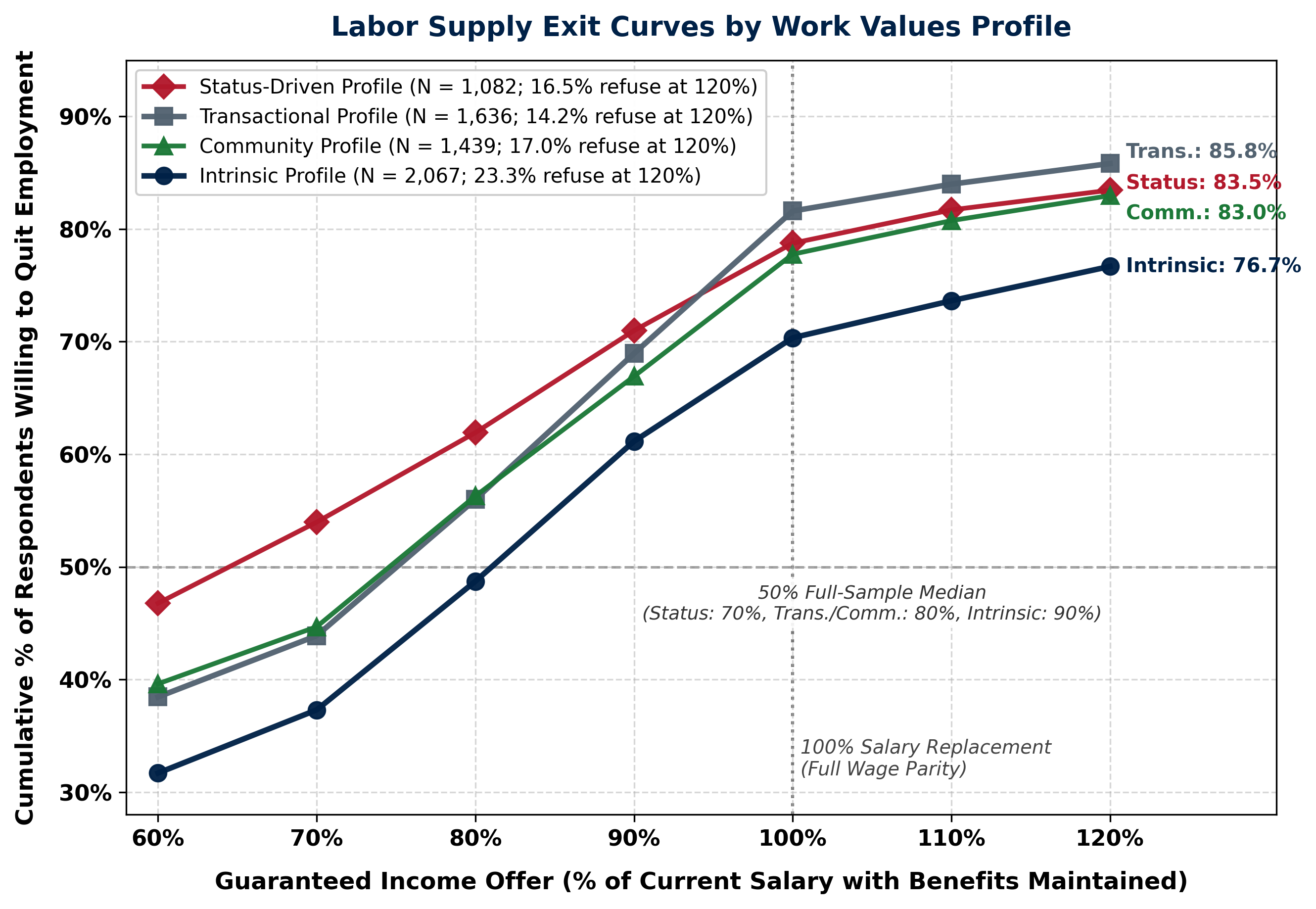}
    \caption{\textbf{Labor Supply Exit Curves by Work Values Profile}}
    \label{fig:reservation_wage_curve_profiles}
    \vspace{0.10cm}
    \begin{minipage}{\linewidth}
        \footnotesize 
        \textit{Notes:} Lines plot the cumulative percentage of classified respondents ($N=6,224$) willing to voluntarily exit paid employment at each sequential guaranteed income threshold (60\% to 120\% of current salary, with benefits preserved) across the four Work Values Profiles, keeping non-quitters who refused all offers up to 120\% in the denominator (see Appendix Figure~\ref{fig:app_reservation_curve_regions} for the regional disaggregation between the US--UK--Denmark subsample and the online India sample). The horizontal dashed line at 50\% identifies the exact full-sample median reservation threshold for each group (Status-Driven: 70\%, Transactional and Community: 80\%, Intrinsic: 90\%).
    \end{minipage}
\end{figure}

\begin{figure}[!htbp]
    \centering
    \includegraphics[width=0.92\linewidth]{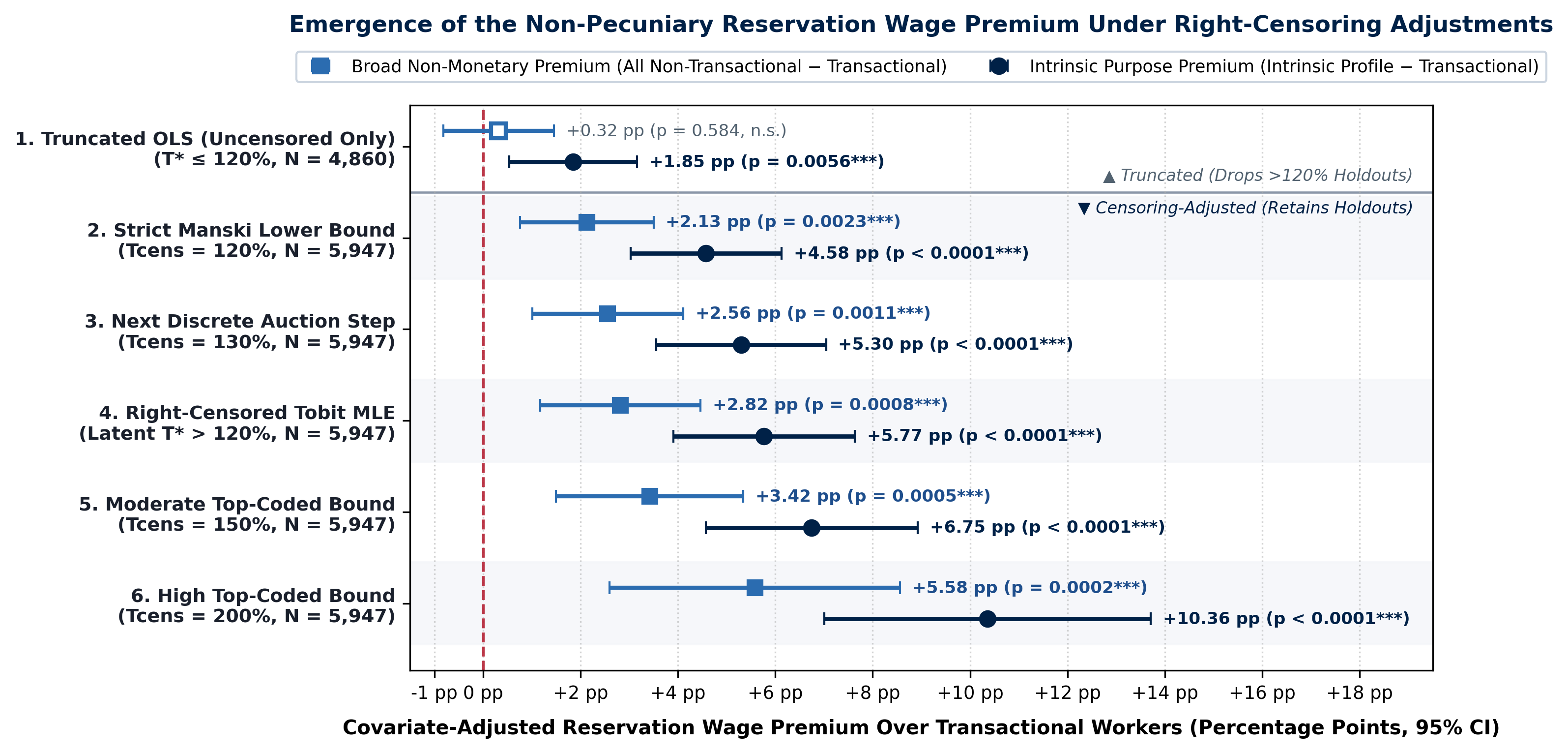}
    \caption{\textbf{Non-Pecuniary Reservation Wage Premium Under Right-Censoring Adjustments}}
    \label{fig:non_monetary_premium_censoring}
    \vspace{0.10cm}
    \begin{minipage}{\linewidth}
        \footnotesize 
        \textit{Notes:} Point estimates and 95\% confidence intervals for the covariate-adjusted Broad Non-Monetary Premium ($\text{Non-Transactional} - \text{Transactional}$, blue squares) and Intrinsic Purpose Premium ($\text{Intrinsic} - \text{Transactional}$, navy circles) across the baseline Truncated OLS specification ($T^* \le 120\%, N = 4,860$; open square denotes $p > 0.10$) and five right-censoring-adjusted specifications across the full classified sample ($N = 5,947$: top-coded bounds $T_{\text{cens}} \in \{120\%, 130\%, 150\%, 200\%\}$ and parametric Right-Censored Tobit MLE). All models control for the full Model 3 conditioning set and country fixed effects (Table~\ref{tab:reservation_regressions} and Appendix Table~\ref{tab:app_censoring_regressions}). The top horizontal axis and parenthetical callouts report the equivalent annual dollar premium evaluated at the approximate US median full-time salary of $\$60,000$/year \citep{BLS2025}. $^{***} p < 0.01$.
    \end{minipage}
\end{figure}

First, let's consider whether work values profiles shape the minimum wage replacement at which a person would be willing to leave their employment. Descriptively, we find that workers holding an Intrinsic Profile do indeed demand the highest level of wage replacement across work value profiles: among respondents who accepted an offer at or below 120\% ($N=5,086$), the median reservation wage sits at 80\% of salary for Intrinsic workers (mean $= 78\%$), compared to 70\% for both Transactional (mean $= 77\%$) and Community (mean $= 76\%$) profiles, and 60\% for Status-Driven profiles (mean $= 73\%$). Since less than half of respondents in any profile refused to quit at 120\%, we can also calculate the exact median across \textit{all} classified respondents ($N=6,224$) by retaining those who said ``no'' even at 120\% in the upper tail of each group's distribution. When this group is included, every profile's median does indeed shift upward by exactly 10 percentage points while preserving the same relationship between groups. These results are plotted in Figure~\ref{fig:reservation_wage_curve_profiles} (with the side-by-side regional disaggregation between the US--UK--Denmark subsample and the online India sample presented in Appendix Figure~\ref{fig:app_reservation_curve_regions}). In these initial descriptive comparisons, I highlight medians alongside means for two reasons. The first is that reservation wages are generated on a 10-percentage-point grid (60\% to 120\% of current salary). Second, since about one-fifth of active workers refuse all offers up to the 120\% ceiling---well below 50\% in every profile---the exact full-sample median is point-identified without dropping holdouts or imposing an arbitrary top-code value. By contrast, if we were to estimate conditional means, this would entail either sample truncation (Table~\ref{tab:reservation_regressions}) or right-censoring and Tobit maximum likelihood estimation (Figure~\ref{fig:non_monetary_premium_censoring} and Appendix Table~\ref{tab:app_censoring_regressions}).

To more fully interrogate these interactions, I look at the association between non-pecuniary work attachments and the percentage demanded for wage replacement, once we control for earnings, occupation, and demographic characteristics. The dependent variable here is our measure of ``uncensored reservation income''---that is, the percentage of current salary required to voluntarily quit working, again, ranging from 60 to 120 percent.\footnote{Following sequential listwise deletion of right-censored respondents ($N=1,668$, which includes $244$ of the $1,128$ non-working respondents), remaining non-working respondents lacking an occupational tier ($N=884$), remaining unclassified ``Other'' respondents ($N=596$), and missing control data ($N=226$), the primary classified estimation sample comprises $N=4,860$ active workers.} Table~\ref{tab:reservation_regressions} shows the resulting cell-means OLS estimates across three specifications.

\begin{table}[!htbp]
    \centering
    \begin{threeparttable}
        \caption{\textbf{Regression Models of Reservation Income on Work Values, Occupational Category, and AI Threat (Cell-Means Specification with Direct Contrasts)}}
        \label{tab:reservation_regressions}
        \footnotesize
        \setlength{\tabcolsep}{0pt}
        \renewcommand{\arraystretch}{1.04}
        \begin{tabular*}{\textwidth}{@{\extracolsep{\fill}}lccc@{}}
            \toprule
            \textbf{Independent Variable / Contrast} & 
            \textbf{(1) Simple} & 
            \textbf{(2) + Demog.} & 
            \textbf{(3) + Country FEs} \\
            \midrule
            \multicolumn{4}{@{}l}{\textit{Panel A: Adjusted profile means (at covariate means)}} \\
            \quad Transactional Profile                      & 75.83 (0.49)        & 75.68 (0.49)        & 75.79 (0.49)        \\
            \quad Intrinsic Profile                          & 77.59 (0.45)        & 77.60 (0.45)        & 77.64 (0.45)        \\
            \quad Community Profile                          & 76.00 (0.52)        & 76.06 (0.52)        & 76.00 (0.52)        \\
            \quad Status-Driven Profile                      & 73.55 (0.64)        & 73.73 (0.64)        & 73.62 (0.65)        \\
            \midrule
            \multicolumn{4}{@{}l}{\textit{Panel B: AI threat slope within profile}} \\
            \quad AI Threat $\mid$ Transactional             & $-$2.32*** (0.51)   & $-$2.14*** (0.52)   & $-$2.09*** (0.52)   \\
            \quad AI Threat $\mid$ Intrinsic                 & $-$3.62*** (0.46)   & $-$3.46*** (0.46)   & $-$3.44*** (0.46)   \\
            \quad AI Threat $\mid$ Community                 & $-$4.24*** (0.52)   & $-$3.98*** (0.53)   & $-$3.99*** (0.53)   \\
            \quad AI Threat $\mid$ Status-Driven             & $-$2.68*** (0.63)   & $-$2.58*** (0.63)   & $-$2.64*** (0.63)   \\
            \midrule
            \multicolumn{4}{@{}l}{\textit{Panel C: Direct contrasts in levels}} \\
            \quad Transactional $-$ Intrinsic                & $-$1.76*** (0.66)   & $-$1.92*** (0.66)   & $-$1.85*** (0.67)   \\
            \quad Transactional $-$ Community                & $-$0.17 (0.71)      & $-$0.39 (0.71)      & $-$0.21 (0.72)      \\
            \quad Transactional $-$ Status-Driven            & 2.28*** (0.80)      & 1.94** (0.81)       & 2.17*** (0.82)      \\
            \quad Transactional $-$ Non-Transactional        & $-$0.25 (0.57)      & $-$0.47 (0.57)      & $-$0.32 (0.58)      \\
            \midrule
            \multicolumn{4}{@{}l}{\textit{Panel D: Direct contrasts in AI threat slopes}} \\
            \quad Trans.\ $-$ Intrinsic (slope)              & 1.29* (0.68)        & 1.32* (0.68)        & 1.35** (0.68)       \\
            \quad Trans.\ $-$ Community (slope)              & 1.91*** (0.73)      & 1.84** (0.73)       & 1.90*** (0.73)      \\
            \quad Trans.\ $-$ Status-Driven (slope)          & 0.36 (0.81)         & 0.44 (0.81)         & 0.55 (0.81)         \\
            \quad Trans.\ $-$ Non-Transactional (slope)      & 1.26** (0.59)       & 1.27** (0.59)       & 1.33** (0.59)       \\
            \midrule
            \multicolumn{4}{@{}l}{\textit{Controls (mean-centered)}} \\
            \quad White-Collar Occupational Tier             & 1.24** (0.59)       & 0.93 (0.61)         & 1.04* (0.61)        \\
            \quad AI Threat $\times$ White-Collar Tier       & 0.57 (0.56)         & 0.56 (0.55)         & 0.51 (0.56)         \\
            \quad Middle Income Tercile                      & 0.08 (0.64)         & $-$0.91 (0.67)      & $-$0.94 (0.68)      \\
            \quad High Income Tercile                        & 0.07 (0.69)         & $-$1.10 (0.77)      & $-$1.15 (0.78)      \\
            \quad Age (years)                                & ---                 & 0.05** (0.02)       & 0.05** (0.02)       \\
            \quad Female ($1 = \text{Female}$)               & ---                 & 1.26** (0.52)       & 1.26** (0.52)       \\
            \quad Higher Education (Bachelor's+)             & ---                 & $-$0.08 (0.56)      & 0.08 (0.59)         \\
            \quad Part-Time Employment ($1 = \text{Yes}$)    & ---                 & $-$3.44*** (0.66)   & $-$3.27*** (0.66)   \\
            \midrule
            \textbf{Country Fixed Effects}                   & No                  & No                  & Yes (4 FEs)         \\
            \textbf{Observations ($N$)}                      & 4,860               & 4,860               & 4,860               \\
            \textbf{$R^2$}                                   & 0.04                & 0.05                & 0.05                \\
            \bottomrule
        \end{tabular*}
        \begin{tablenotes}[flushleft]
            \scriptsize
            \item \textit{Notes:} OLS cell-means regression estimates with standard errors in parentheses (* $p < 0.10$, ** $p < 0.05$, *** $p < 0.01$). Dependent variable is Reservation Income (percentage of current salary required to voluntarily quit working, 60\%--120\%). Right-censored observations ($N=1,668$), remaining non-working respondents ($N=884$), remaining unclassified ``Other'' respondents ($N=596$), and missing control data ($N=226$) are excluded ($N=4,860$). AI threat, controls, and country indicators are mean-centered, so Panel~A reports covariate-adjusted profile means at mean AI threat and Panel~B the within-profile AI threat slope. Panels~C--D report Wald tests of direct differences; ``Non-Transactional'' is the sample-size-weighted average of Intrinsic, Community, and Status-Driven.
        \end{tablenotes}
    \end{threeparttable}
\end{table}

The regression results in Table~\ref{tab:reservation_regressions} provide strong support for this paper's first hypothesis (\textit{Hypothesis 1}) among intrinsically motivated workers, while suggesting interesting heterogeneity across non-pecuniary profiles. Specifically, \textit{Hypothesis 1} predicted that in direct comparisons with transactionally motivated workers, workers who place a significant importance on the non-monetary value of labor will require higher wages to voluntarily exit employment. In Panel~A (Model 3, controlling for household income terciles, occupational tier, educational attainment, age, gender, part-time status, and country fixed effects), we can see that workers who hold an \textit{Intrinsic} Profile demand the highest reservation wage to voluntarily quit working at 78\%  ($\text{SE}=0.45$), followed by \textit{Community}-oriented workers and Transactional workers, both at around 76\%. Curiously, however, \textit{Hypothesis 1} is not supported on the current-job exit margin for \textit{Status-Driven} workers, who again held the lowest reservation wage at 74\%  ($\text{SE}=0.65$). 

When we compare direct Wald contrasts in Panel~C, we find results in line with \textit{Hypothesis 1}. Specifically, the data suggests that workers with a \textit{Transactional} profile require a statistically significant 1.76 to 1.92 percentage points lower wage replacement than \textit{Intrinsic} workers in order to voluntarily leave work.\footnote{Specifically, $\text{Transactional} - \text{Intrinsic} = -1.85\text{ pp}$ ($\text{SE}=0.67, p < 0.01$) in Model 3; $-1.92\text{ pp}$ ($\text{SE}=0.66, p < 0.01$) in Model 2; and $-1.76\text{ pp}$ ($\text{SE}=0.66, p < 0.01$) in Model 1.} When we constrict our analysis just to the US, UK, and Denmark ($N=3,837$), excluding India as an outlier partly due to its online sample, this figure increases to between 2.01 and 2.15 percentage points ($p < 0.01$, with a standard error of $0.72$ across Models 1--3; and $2.33\text{ percentage points}$, $\text{SE}=0.73, p<0.01$, in the additive specification without threat interactions). In contrast to \textit{Hypothesis 1}, however, we find that \textit{Status-Driven} workers held the lowest reservation income in the sample, with a $2.17\text{ percentage point}$ ($p<0.01$) lower reservation wage when compared with \textit{Transactional} workers. It is meanwhile associated with a wage requirement to leave work that is $4.02\text{ percentage points}$ below \textit{Intrinsic} workers ($p<0.01$).

It is important to note, however, that the estimates in Table~\ref{tab:reservation_regressions} are substantially \textit{downward-biased} because restricting the estimation sample to respondents who accepted an offer at or below 120\% ($N=4,860$) has the effect of truncating the dependent variable \citep{Killingsworth1985}. As shown in Figure~\ref{fig:reservation_wage_curve} and Figure~\ref{fig:reservation_wage_curve_profiles}, and as previously discussed, preferences around wage replacement for voluntary workforce withdrawal are highly elastic \textit{below} $100\%$ wage replacement, with a midpoint arc elasticity of $\varepsilon = 1.27$. This elasticity declines considerably afterwards, however, with an arc elasticity of $\varepsilon = 0.34$ at and after $100\%$ wage replacement, amounting to a $74\%$ collapse in elasticity. Again, $25\%$ of the active workforce refuses to quit at $100\%$ wage replacement, and $80\%$ of this group ($20\%$ of the active workforce, $N=1,424$) continues to refuse to leave work at $120\%$ salary. To better understand this group of ``holdouts,'' Appendix Table~\ref{tab:app_reservation_descriptives} and Appendix Table~\ref{tab:app_censoring_regressions} provide additional analysis relevant to this paper's hypotheses. Specifically, I note that the rate of ``holding out'' at $120\%$ wage substitution is considerably higher for groups that place a high non-monetary value on their work. For instance, the rate of ``holding out'' for \textit{Intrinsic} workers sits at $23\%$ ($N=482$), at $17\%$ for \textit{Community}-oriented workers ($N=245$), and at $17\%$ for \textit{Status}-motivated workers ($N=179$). The share of ``hold outs'' is meanwhile lowest for \textit{Transactional} workers at $14\%$ ($N=232$).\footnote{Two-proportion $z$-tests against the \textit{Transactional} baseline ($14.18\%$) confirm that these refusal rate differences are statistically significant for \textit{Intrinsic} ($+9.14\text{ pp}, \text{SE}=1.27, z=7.00, p<0.01$) and \textit{Community} ($+2.85\text{ pp}, \text{SE}=1.31, z=2.17, p=0.03$) workers and marginally significant for \textit{Status-Driven} workers ($+2.36\text{ pp}, \text{SE}=1.42, z=1.68, p=0.09$), while \textit{Intrinsic} refusal also significantly exceeds both \textit{Community} and \textit{Status-Driven} refusal ($p<0.01$); see Appendix Table~\ref{tab:app_reservation_descriptives} (Panel~A) and Appendix Table~\ref{tab:app_censoring_regressions} (Column~2) for full standard errors, $z$-statistics, and covariate-adjusted linear probability model estimates.} This shows that listwise deletion eliminates the most work-attached quarter of the \textit{Intrinsic}-profile distribution but only $14\%$ of \textit{Transactional} profiles. As a result, the OLS estimates presented earlier represent a clear lower bound of the wage premium demanded by workers with an \textit{Intrinsic} Work Profile.

Yet what happens if we eliminate this distortion in our model? Figure~\ref{fig:non_monetary_premium_censoring} (alongside Appendix Figure~\ref{fig:app_censoring_robustness} and Appendix Table~\ref{tab:app_censoring_regressions}) shows that, if we correct for the removal of ``hold outs'' from our sample with complete covariates ($N=5,947$), we see a substantial increase in the estimated reservation wage premium. Even under the strictest Manski-style lower bound estimate, where we assign ``hold outs'' to $120\%$ (the ceiling that they already rejected), the statistically significant \textit{Intrinsic} premium more than doubles, relative to \textit{Transactional} profile workers, from $+1.85\text{ pp}$ ($p<0.01$) to \textbf{$+4.58$} pp ($87\%$ vs.\ $83\%, \text{SE}=0.79, t=5.81, p<0.01$; see Appendix Table~\ref{tab:app_reservation_descriptives}, Panel~B, and Appendix Table~\ref{tab:app_censoring_regressions}, Columns~3--6, for full adjusted means, standard errors, and overall sample standard deviations across all censoring specifications). Assigning ``hold outs'' to the next item in the guaranteed-wage auction interval ($130\%$) increases this statistically significant premium further to $+5.30$  pp ($\text{SE}=0.89, p<0.01$). As an additional test, I used a Right-Censored Tobit MLE model, which treats any observation above $120\%$ as right-censored, but does not substitute in arbitrary values for replacement reservation wages. Under this model, the statistically significant \textit{Intrinsic} premium increases further to $+5.77\text{ pp}$ ($\text{SE}=0.95, p<0.01$, with $\hat{\sigma}=27.09$), while \textit{Community}-oriented worker status is associated with a premium of $+1.74\text{ pp}$ ($\text{SE}=1.04, p<0.10$). 

As a final test, let's explore a scenario where ``hold outs'' would have required even higher levels of wage replacement (e.g.\ $150\%$ or $200\%$ of salary) if they were given the option. Figure~\ref{fig:non_monetary_premium_censoring} (and Panel~B of Appendix Figure~\ref{fig:app_censoring_robustness}) unsurprisingly shows that the \textit{Intrinsic} premium increases further to $+6.75\text{ pp}$ ($\text{SE}=1.11, p<0.01$) at $150\%$ and \textbf{$+10.36$} pp ($\text{SE}=1.71, p<0.01$) at $200\%$. Meanwhile, the broader contrast between workers who value the non-monetary benefits of work and those that do not increases considerably. Although this gap was initially small and statistically non-significant in our truncated OLS model ($+0.32\text{ pp}, p=0.58$), under these censoring-adjusted specifications we see a statistically significant non-pecuniary premium emerge, rising with each auction threshold (see footnote).\footnote{The results from this test are: $+2.13\text{ pp}, \text{SE}=0.70, p<0.01$ at the $120\%$ bound; $+2.56\text{ pp}, \text{SE}=0.79, p<0.01$ at $130\%$; $+2.82\text{ pp}, \text{SE}=0.84, p<0.01$ in Tobit MLE; $+3.42\text{ pp}, \text{SE}=0.98, p<0.01$ at $150\%$; and $+5.58\text{ pp}, \text{SE}=1.52, p<0.01$ at $200\%$. } More work is needed to better understand the exact magnitude of these premiums, yet the approach we have taken so far provides evidence for the existence of a meaningful non-pecuniary premium, particularly for intrinsically motivated workers.

Although these premium estimates may sound small---and indeed, many of them are---they are more considerable than it might seem. In our four-country pooled regression (Table~\ref{tab:reservation_regressions}, Model 3, $N=4,860$ across the US, UK, Denmark, and India), even the most conservative \textit{Intrinsic} premium of $+1.85\text{ percentage points}$ ($p<0.01$) is larger in magnitude than moving from the bottom third to the top third of a respondent's national household income distribution across the four surveyed countries ($-1.15\text{ pp}, \text{SE}=0.78, p>0.10$, holding country fixed effects constant). Taking the approximate median full-time annual salary of $\$60,000$ in the United States \citep{BLS2025}, the \textit{Intrinsic} premium amounts to $\$1,110$ per year in the US. Moreover, once we include ``hold outs'' across the full classified sample ($N=5,947$, Figure~\ref{fig:non_monetary_premium_censoring} and Appendix Table~\ref{tab:app_censoring_regressions}), the premium amounts to $\$2,748$ per year when top-coded at $120\%$ ($+4.58\text{ pp}$), $\$3,462$ per year in our Right-Censored Tobit model ($+5.77\text{ pp}$), and reaches $\$6,216$ per year under the most extreme $200\%$ top-coded version of the model ($+10.36\text{ pp}$; with the broad non-monetary premium across all non-transactional workers reaching $\$3,348$ per year at $+5.58\text{ pp}$).

Why do Status-Driven workers accept a lower guaranteed basic income threshold than Intrinsic workers? A separate survey module asked respondents how they would spend their time if made redundant and supported by a basic income. Respondents selected across nine items, visualized in Figure~\ref{fig:post_work_time_use_profiles} (and aggregated into three composite domains in Appendix Figure~\ref{fig:app_time_use_domains}). The results directly test and appear to support \textit{Hypothesis 2}---that workers who value their employment due to status, intrinsic, or communal motivations will pursue further work or prosocial activities rather than pure leisure when receiving income transfers. \textit{Transactional} workers appeared to primarily treat basic income as an opportunity to retire into leisure, with $79\%$ choosing at least one leisure option and only $11\%$ seeking another full-time job. Meanwhile, a majority of \textit{Intrinsic} workers said they would pivot toward unpaid prosocial and creative engagement. And $55\%$ of \textit{Status-Driven} workers state that they would use the basic income to continue in paid work or entrepreneurship. \textit{Community}-oriented workers blended high desire for family and civic engagement with moderate interest in leisure and re-entry into work. As shown in Appendix Figure~\ref{fig:app_time_use_domains}, all of these cross-profile divergences across the three composite domains remain statistically significant ($p < 0.01$) in multivariate linear probability models controlling for age, gender, education, part-time status, household income terciles, occupational tier, subjective AI threat, and country fixed effects ($N = 5,947$).\footnote{In our cell-means OLS model, with the full control set, we find that all three non-pecuniary Work Values profiles are more likely to pursue continued paid work or entrepreneurship, with a gap of $+5.44\text{ pp}, \text{SE}=1.48, p<0.01$ when we pool all non-monetary profiles together. All three non-pecuniary profiles share similar rates of intent to re-enter into paid work. \textit{Intrinsic} and \textit{Community} oriented workers are, meanwhile, much more likely to enter into unpaid volunteering or family care. Transactional workers are much more likely than other cohorts to functionally retire into leisure, rest, or travel ($76.22\%$ of the sample). } In line with \textit{Hypothesis 2} and consistent with \citet{Shayo2009}, status-seeking workers appear to regard a state-guaranteed basic income not as a substitute for working, but instead as an income floor that can help subsidize transitions into higher-prestige employment or business ownership. However, because covariate-adjusted intentions to pursue paid work or entrepreneurship are similar across \textit{Status-Driven} ($49\%$), \textit{Community} ($47\%$), and \textit{Intrinsic} ($47\%$) workers, re-employment intentions alone only partially account for why status-driven workers accept the lowest transfer threshold in \textit{Hypothesis 1} ($74\%$ vs.\ $78\%$ for \textit{Intrinsic})---suggesting that status-oriented workers also feel less attached to their \textit{current} role and more confident in switching jobs (as shown below).

\begin{figure}[!htbp]
    \centering
    \includegraphics[width=0.92\linewidth]{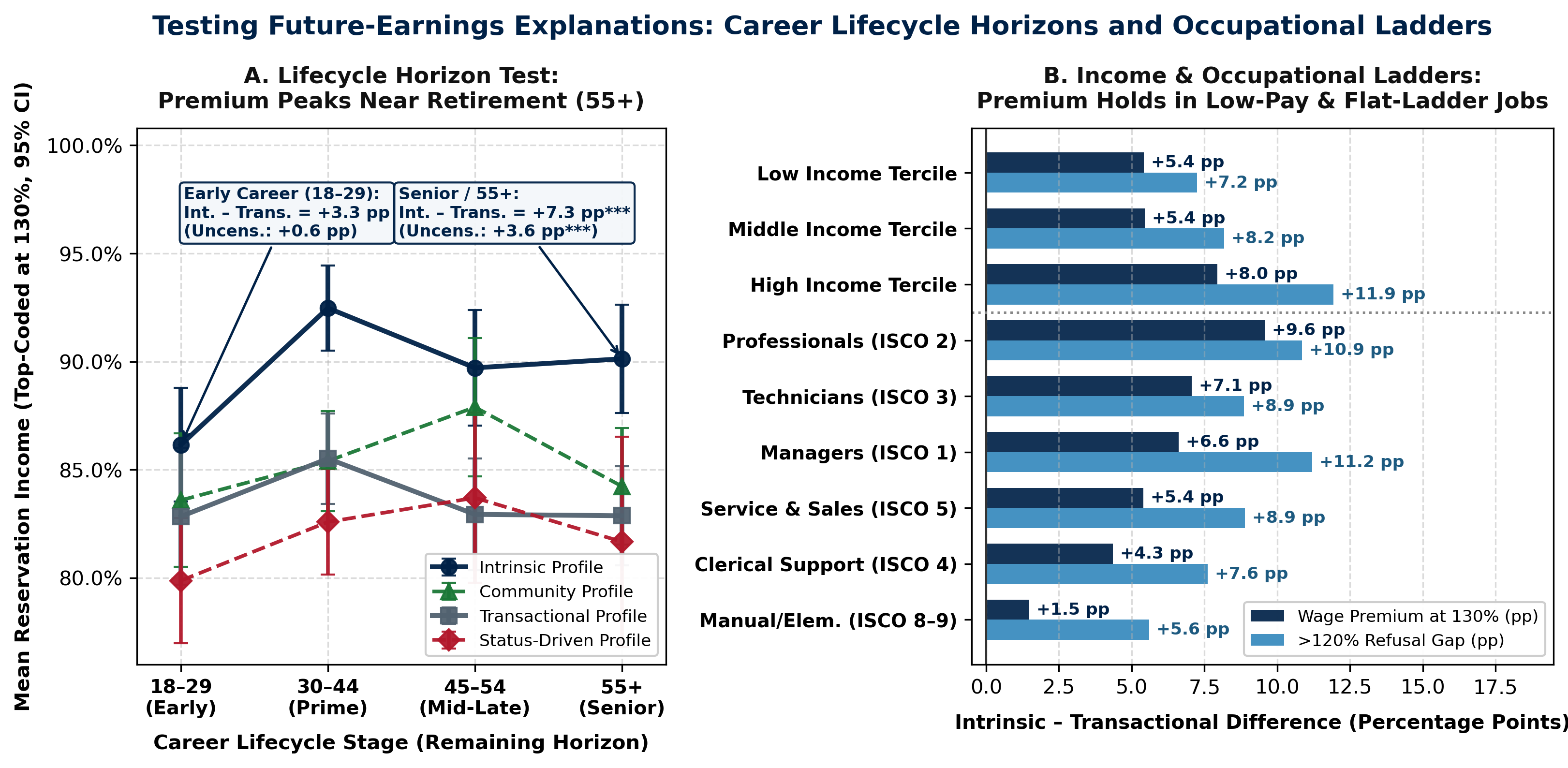}
    \caption{\textbf{Testing Future-Earnings Explanations: Career Lifecycle Horizons and Occupational Ladders}}
    \label{fig:future_earnings_expectations}
    \vspace{0.10cm}
    \begin{minipage}{\linewidth}
        \footnotesize 
        \textit{Notes:} Panel A plots mean reservation income (top-coded at $130\%$ with 95\% confidence intervals) across four age brackets ($N=5,938$). In contrast to a standard future-earnings-growth hypothesis, which would suggest the highest threshold among early-career 18--29 year-olds with the longest remaining career horizon, we see an opposite trend. Overall reservation wages are lowest among 18--29 year-olds (rising into mid-career at $130\%$ top-coding and rising across all four age brackets). Meanwhile, the \textit{Intrinsic} premium peaks among senior workers aged $55+$. Panel B displays this (at $130\%$ top-coding) and the $>120\%$ refusal rate gap across household income terciles and ISCO-08 occupational major groups. It shows that, although the Intrinsic premium appears largest among upward-mobile professionals, managers, and technicians, significant premiums also hold for low-income workers and in other occupations. For the companion test across outside re-employment confidence, see Appendix Figure~\ref{fig:app_reemployment_confidence}.
    \end{minipage}
\end{figure}

Finally, a natural question one might ask is whether higher reservation wages---particularly the refusal of one-fifth of respondents to quit even at 120\% of their current salary---might be explained simply by an expectation for future earnings growth. Indeed, a worker who expects a steep future pay increase or promotion might reject any income guarantee tied to their \textit{current} wage, since this would entail losing out on future higher earnings. As we see in Figure~\ref{fig:future_earnings_expectations} (and Appendix Figure~\ref{fig:app_reemployment_confidence}), while steeper career progression ladders in professional occupations may account for part of the cross-occupational variation in reservation wages, three independent empirical tests in our data suggest that future-earnings expectations alone cannot explain the non-pecuniary attachment identified in this paper.

First, on the lifecycle horizon margin (Figure~\ref{fig:future_earnings_expectations}, Panel~A), we would expect reservation wages to be highest among young and early-career workers who have three to four decades of wage growth and career advancement ahead of them. We would meanwhile expect reservation wages to be flattest among senior workers who are approaching retirement. Instead, early-career workers appear to have the lowest reservation wages. Indeed, average reservation wages rise across age cohorts, from $73\%$ among 18--29 year-olds to $76\%$ among 30--44 year-olds, $77\%$ among 45--54 year-olds, and $78\%$ among workers aged 55+ (consistent with the positive linear age slope of $+0.05\text{ pp}$ per year, $p<0.05$, in Table~\ref{tab:reservation_regressions}). When right-censored holdouts are top-coded at $130\%$ (as plotted in Figure~\ref{fig:future_earnings_expectations}, Panel~A), the trajectory is concave rather than strictly linear---rising sharply from $83\%$ among 18--29 year-olds to $87\%$ among 30--44 year-olds before leveling off at $86\%$ among 45--54 year-olds and $85\%$ among workers aged 55+.\footnote{This slight plateau in unadjusted $130\%$ top-coded means after age 44 reflects cross-national age composition in part. The online India sample (which has the highest $>120\%$ refusal rate at $28.7\%$) accounts for $36.4\%$ of 18--29 year-olds and $26.6\%$ of 30--44 year-olds, compared to $18.3\%$ of 45--54 year-olds and just $9.5\%$ of workers aged 55+. Within both the uncensored and $130\%$ top-coded series, however, 18--29 year-olds exhibit the lowest reservation wages of any age group.} Moreover, under both measures, the \textit{Intrinsic} premium over \textit{Transactional} workers is smallest and statistically non-significant among early-career 18--29 year-olds and largest among senior workers aged 55+ (where $20\%$ of \textit{Intrinsic} workers refuse $120\%$ salary vs.\ $12\%$ of \textit{Transactional} workers), potentially because older workers have had more time to form non-pecuniary attachments with their areas of employment, or work in a more autonomous and senior role.

\begin{figure}[!htbp]
    \centering
    \includegraphics[width=0.88\linewidth]{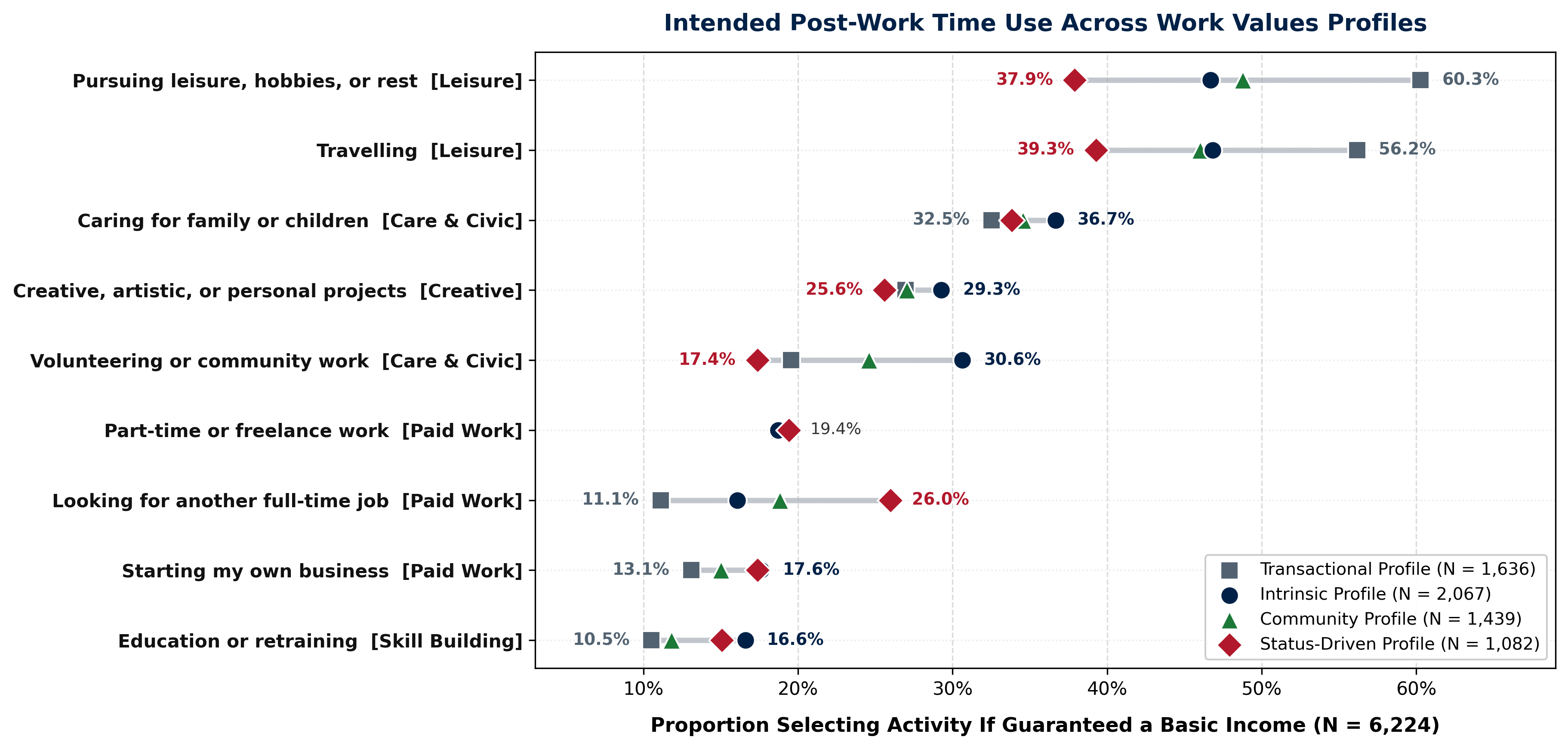}
    \caption{\textbf{Intended Time Use Under a Guaranteed Basic Income Across All Four Work Values Profiles ($N=6,224$)}}
    \label{fig:post_work_time_use_profiles}
    \vspace{0.10cm}
    \begin{minipage}{\linewidth}
        \footnotesize 
        \textit{Notes:} Points display the exact proportion of respondents in each Work Values Profile ($N=6,224$) selecting each of the nine individual post-displacement time-use activities if guaranteed a basic income; horizontal gray bars span the range across profiles. Respondents could select multiple activities. Composite domain totals, 95\% confidence intervals, and multivariate cell-means regression contrasts ($N=5,947$) are presented in Appendix Figure~\ref{fig:app_time_use_domains}.
    \end{minipage}
\end{figure}

Second, across income tiers and occupational ladders (Figure~\ref{fig:future_earnings_expectations}, Panel~B), upward-mobile white-collar groups do exhibit the largest \textit{Intrinsic} premiums (at $130\%$ top-coding)---including among Professionals, Technicians, and Managers. Meanwhile, the \textit{Intrinsic} contrast is small and statistically non-significant among Manual and Elementary workers. One potential explanation of this is that steeper career progression prospects and greater task autonomy in high-skill occupations may contribute in part to the magnitude of the \textit{Intrinsic} premium. Future wage expectations may indeed still play a role; yet, I argue they cannot fully account for the pattern observed in the data. Indeed,  $>120\%$ refusal rates are virtually identical across low-income and high-income workers ($20\%$ vs.\ $19\%$). Moreover, the data still identifies statistically significant \textit{Intrinsic} wage premiums and refusal gaps from the lowest household income tercile ($+7.25\text{ pp}$ refusal gap, $p<0.01$), the middle income tercile ($+8.17\text{ pp}$ refusal gap, $p<0.01$), and in Services and Sales jobs and in Clerical Support. 

Third, when we directly measure respondents' confidence in their outside labor market and earnings prospects (plotted in Appendix Figure~\ref{fig:app_reemployment_confidence}), we see that re-employment confidence is associated with a significantly \textit{lower} reservation wage ($\beta = -0.79\text{ pp per SD}, \text{SE}=0.29, p<0.01$ uncensored; $\beta = -0.96\text{ pp per SD}, \text{SE}=0.39, p=0.01$ at $130\%$), perhaps because workers who are confident they can easily secure comparable outside pay (led by Status-Driven workers at $z = +0.43\text{ SD}$) are more willing to accept a basic income while changing jobs. Yet this seems to suggest workers who feel most confident in their professional opportunities (a possible proxy for future wage growth) demand lower wage replacement, not higher.\footnote{Moreover, controlling directly for outside re-employment confidence leaves the covariate-adjusted \textit{Intrinsic} premium over \textit{Transactional} workers completely intact and slightly larger ($\text{Transactional} - \text{Intrinsic} = -1.99\text{ pp}, \text{SE}=0.68, p<0.01$ uncensored; $-5.47\text{ pp}, \text{SE}=0.90, p<0.01$ at $130\%$ top-coding, $N=5,850$).}

Another interesting finding from Panel~B of Table~\ref{tab:reservation_regressions} is that as perceptions of AI's likelihood of replacing one's job increase, the wage replacement demanded by workers decreases. This is true across all four profiles in the sample. Panel~D, in particular, shows that this negative association with perceptions of the AI threat to jobs is significantly steeper among Community and Intrinsic workers than among Transactional ones.\footnote{Specifically, in Table~\ref{tab:reservation_regressions} (Model 3), a one-standard-deviation increase in perceived AI job replacement threat is associated with a $-3.99\text{ pp}$ ($\text{SE}=0.53, p<0.01$) reduction in reservation wages among \textit{Community} workers and a $-3.44\text{ pp}$ ($\text{SE}=0.46, p<0.01$) reduction among \textit{Intrinsic} workers, compared to $-2.09\text{ pp}$ ($\text{SE}=0.52, p<0.01$) among \textit{Transactional} workers (direct slope contrasts: $\text{Transactional} - \text{Community} = +1.90\text{ pp}, p<0.01$; $\text{Transactional} - \text{Intrinsic} = +1.35\text{ pp}, p=0.05$; and $\text{Transactional} - \text{Non-Transactional} = +1.33\text{ pp}, p=0.02$). Consequently, among workers who feel secure in their jobs ($z = -1.0\text{ SD}$ of AI threat), the uncensored \textit{Intrinsic} premium over \textit{Transactional} workers rises to $+3.20\text{ pp}$ ($81.08\%$ vs.\ $77.88\%, p<0.01$; and $+3.67\text{ pp}$ at the zero-threat floor, $p<0.01$).} More research is needed to understand the directionality of this relationship---for instance, it may be that more pessimistic people demand lower wages, or it may be that perceptions of AI's threat of job automation encourage someone to be conservative and accept a guaranteed basic income for less than they otherwise might. 

Finally, given the emergence of recent literature that suggests entry-level workers might be most exposed to AI-enabled automation \citep{BrynjolfssonChandarChen2025}, Table~\ref{tab:reservation_age_split} re-estimates Model 3 separately for respondents at or below the median age ($\le 41$ years, $N = 2,527$) versus older respondents ($> 41$ years, $N = 2,333$). The results in Table~\ref{tab:reservation_age_split} appear to support the pattern we observed earlier in Figure~\ref{fig:future_earnings_expectations}. Across both age cohorts, \textit{Intrinsic}-attached workers demanded the highest reservation wage (and \textit{Status-Driven} workers the lowest). Meanwhile, the attachment premium gap between \textit{Intrinsic} and \textit{Community}-oriented workers on the one hand and \textit{Transactional} workers on the other appears to shift with age. Among younger workers ($\le 41$), the baseline reservation wage gap between \textit{Intrinsic} and \textit{Transactional} workers is small and statistically non-significant ($-0.97\text{ pp}, p = 0.32$), yet this shifts for older workers. Younger \textit{Status-Driven} workers stand out, in particular, for their willingness to leave their current job at a lower basic income threshold to pursue new career opportunities. Moreover, in line with recent evidence on entry-level exposure to AI \citep{BrynjolfssonChandarChen2025}, younger workers ($\le 41$) in our sample also perceive themselves to be at substantially higher risk of AI replacing their job than older workers. This heightened perception of AI risk among younger workers holds across every Work Values Profile---led by younger \textit{Status-Driven} and \textit{Community} workers, followed by younger \textit{Intrinsic} and \textit{Transactional} ($38\%$ vs.\ $23\%$) workers. Of course, we cannot say for certain whether younger people are correct in their perception of AI risk---such belief may, for instance, be simply a consequence of greater AI awareness or broader anxieties among entry-level workers. Still, as Panels~B and~D of Table~\ref{tab:reservation_age_split} show, a higher perceived AI risk is associated with a lower demanded reservation wage across all profiles in both age groups; however, the steeper sensitivity of non-monetary workers relative to \textit{Transactional} workers is most pronounced among older respondents.

\begin{table}[!htbp]
    \centering
    \begin{threeparttable}
        \caption{\textbf{Reservation Income Regressions, Younger vs.\ Older Respondents}}
        \label{tab:reservation_age_split}
        \footnotesize
        \renewcommand{\arraystretch}{1.05}
        \setlength{\tabcolsep}{0pt}
        \begin{tabular*}{\textwidth}{@{\extracolsep{\fill}}lcc@{}}
            \toprule
            \textbf{Variable / Contrast} & \textbf{Younger ($\le 41$)} & \textbf{Older ($> 41$)} \\
            \midrule
            \multicolumn{3}{@{}l}{\textit{Panel A: Adjusted profile means (at covariate means)}} \\
            \quad Transactional                               & 75.49 (0.75)      & 76.10 (0.65)      \\
            \quad Intrinsic                                   & 76.47 (0.63)      & 78.96 (0.65)      \\
            \quad Community                                   & 74.63 (0.73)      & 77.43 (0.74)      \\
            \quad Status-Driven                               & 72.55 (0.77)      & 75.05 (1.09)      \\
            \midrule
            \multicolumn{3}{@{}l}{\textit{Panel B: AI threat slope within profile}} \\
            \quad AI Threat $\mid$ Transactional              & $-$3.07*** (0.77) & $-$1.40* (0.71)   \\
            \quad AI Threat $\mid$ Intrinsic                  & $-$4.02*** (0.64) & $-$2.72*** (0.72) \\
            \quad AI Threat $\mid$ Community                  & $-$3.27*** (0.78) & $-$4.45*** (0.78) \\
            \quad AI Threat $\mid$ Status-Driven              & $-$2.17*** (0.84) & $-$3.12*** (0.99) \\
            \midrule
            \multicolumn{3}{@{}l}{\textit{Panel C: Direct contrasts in levels}} \\
            \quad Transactional $-$ Intrinsic                 & $-$0.97 (0.97)    & $-$2.86*** (0.91) \\
            \quad Transactional $-$ Community                 & 0.86 (1.05)       & $-$1.33 (0.98)    \\
            \quad Transactional $-$ Status-Driven             & 2.95*** (1.08)    & 1.05 (1.29)       \\
            \quad Transactional $-$ Non-Transactional         & 0.73 (0.85)       & $-$1.57** (0.79)  \\
            \midrule
            \multicolumn{3}{@{}l}{\textit{Panel D: Direct contrasts in AI threat slopes}} \\
            \quad Transactional $-$ Intrinsic (slope)         & 0.95 (0.99)       & 1.33 (1.00)       \\
            \quad Transactional $-$ Community (slope)         & 0.20 (1.08)       & 3.06*** (1.04)    \\
            \quad Transactional $-$ Status-Driven (slope)     & $-$0.90 (1.13)    & 1.73 (1.21)       \\
            \quad Transactional $-$ Non-Transactional (slope) & 0.18 (0.87)       & 2.01** (0.85)     \\
            \midrule
            \multicolumn{3}{@{}l}{\textit{Selected Controls (mean-centered)}} \\
            \quad White-Collar Occupational Tier              & 1.93** (0.84)     & 0.16 (0.90)       \\
            \quad AI Threat $\times$ White-Collar Tier        & 0.03 (0.79)       & 0.59 (0.86)       \\
            \midrule
            \textbf{Observations ($N$)}                       & 2,527             & 2,333             \\
            \textbf{$R^2$}                                    & 0.06              & 0.05              \\
            \bottomrule
        \end{tabular*}
        \begin{tablenotes}[flushleft]
            \scriptsize
            \item \textit{Notes:} *** $p < 0.01$, ** $p < 0.05$, * $p < 0.10$. Standard errors in parentheses. Subgroup estimates split at the median age (41 years) of the pooled classified sample; each subgroup model re-estimates the full Model 3 specification of Table~\ref{tab:reservation_regressions} (including a linear, mean-centered age control, income terciles, gender, education, part-time status, and country indicators).
        \end{tablenotes}
    \end{threeparttable}
\end{table}

\section{The Politics of Work in the Age of AI}
\label{sec:results_political} 

In this section, we examine how work values and occupational standing moderate socio-political attitudes about technological displacement, testing \textit{Hypothesis 3}. As previously discussed in Section~\ref{sec:lit_review}, there is a robust literature that suggests material income losses---whether they be from trade or technological automation---can induce a political reaction in the form of movement toward anti-establishment populist parties  \citep{Im2019, Gingrich2019, KurerGallego2019, Kurer2020}. This shift is often explained on account of the fact that negative economic shocks can catalyze losses to real income, which in turn can create dissatisfaction and resentment of the political establishment \citep{Margalit2011, AutorDornHanson2013, EvansTilley2017, BicchiKuoGallego2024}. Yet, as we discussed in Section~\ref{sec:lit_review}, a puzzle exists. Specifically, income transfers and welfare generosity did not appear to effectively moderate the political backlash to prior moments of disruptive technological change \citep{Gingrich2019}. This then begs a question about what is kindling the populist shift among workers exposed to automation. One answer, articulated in the literature, is loss of ``status,'' which income transfers may not compensate for \citep{GidronHall2017, Kurer2020}. 

I now test this hypothesis, along with a broader exploration of how the non-monetary value of work shapes political attitudes and the reaction to technological change, with a particular emphasis on the UK, where I was able to access a broader sample of More in Common panel data. To do this, I have generated cell-means OLS and logistic regression models, linking subjective perceptions of AI's threat to respondents' own jobs over the next ten years ($z$-scores) and our four Work Values Profiles, alongside five political dependent variables. Two of these political outcomes were fielded across all four countries in our survey (Denmark, India, the UK, and the US). These are perceptions of AI's negative political impact ($N = 5{,}688$ classified workers) and right-wing or populist party vote intention ($N = 5{,}295$). The remaining three outcomes come from an additional UK-only survey module ($N = 1{,}498$ classified workers), capturing whether respondents feel the political and economic system is ``rigged,'' whether they want to preserve institutions or dismantle them, and their self-described left--right economic ideology.\footnote{Since the four continuous political dependent variables (Columns 1, 2, 3, and 5) and the continuous AI threat predictor are measured on different raw survey scales, each continuous variable is standardized into a $z$-score ($\mu = 0, \sigma = 1$) prior to OLS estimation; the OLS regression coefficients ($\beta$) and adjusted cell means are expressed directly in standard deviation ($\text{SD}$) units to allow direct comparison across outcomes (while Column 4 is a binary indicator estimated via logistic regression in log-odds).} Table~\ref{tab:political_regressions_profiles} reports the results of this analysis. 

Three findings emerge from this analysis, presented in Table~\ref{tab:political_regressions_profiles}. First, in Column 1 (our measure of perceptions about AI's political impact, coded $1 = \text{Very positive}$ to $5 = \text{Very negative}$ and standardized into a $z$-score across all four countries), we can observe the marked divergences in Work Values Profiles in adjusted levels (Panel~A) and AI threat slopes (Panel~B). For instance, \textit{Transactional} workers exhibited the highest expectation that AI will harm their country's political system ($+0.19\text{ SD}$), compared to $-0.02\text{ SD}$ among \textit{Intrinsic} workers, $-0.03\text{ SD}$ among \textit{Community} workers, and $-0.16\text{ SD}$ among \textit{Status-Driven} workers. This amounts to a pooled non-monetary-vs-\textit{Transactional} contrast of $\beta = +0.25\text{ SD}, p < 0.01$ in Panel~C. Second, across the UK-only modules, \textit{Transactional} workers express a higher preference for dismantling existing political institutions than \textit{Intrinsic} workers at the baseline as well as modestly higher perceptions that the system is rigged. Third, in Column 5, which captures economic ideological placement in the UK sample (coded $1 = \text{Left / Pro-redistribution}$ to $5 = \text{Right / Pro-market}$), \textit{Transactional} workers appear significantly further to the economic left in baseline levels, which is potentially consistent with this group's possible prioritization of material compensation. Meanwhile, higher levels of belief that AI will replace one's job are associated with an economically leftward shift toward pro-redistribution policies among \textit{Community} ($\beta = -0.12\text{ SD}, p < 0.05$) and \textit{Status-Driven} workers. Of course, we don't know the directionality of this relationship. 

\begin{table}[!htbp]
\centering
    \begin{threeparttable}
        \caption{\textbf{Regression of Political Attitudes on AI Threat by Work Values Profile}}
        \label{tab:political_regressions_profiles}
        \scriptsize
        \renewcommand{\arraystretch}{1.05}
        \setlength{\tabcolsep}{0pt}
        \begin{tabular*}{\textwidth}{@{\extracolsep{\fill}}lccccc@{}}
            \toprule
            \textbf{Variable / Contrast} &
            \textbf{(1) AI Imp.} &
            \textbf{(2) Rigged} &
            \textbf{(3) Chaos} &
            \textbf{(4) Pop.\ Vote} &
            \textbf{(5) Econ.\ R} \\
            & \textit{(4 Countries)} & \textit{(UK Only)} & \textit{(UK Only)} & \textit{(4 Countries)} & \textit{(UK Only)} \\
            \midrule
            \multicolumn{6}{@{}l}{\textit{Panel A: Adjusted profile means (at covariate means)}} \\
            \quad Transactional Profile                & 0.19 (0.02)       & 0.08 (0.05)       & 0.06 (0.04)       & 0.03 (0.06)       & $-$0.11 (0.05)    \\
            \quad Intrinsic Profile                    & $-$0.02 (0.02)    & $-$0.03 (0.04)    & $-$0.12 (0.04)    & 0.04 (0.05)       & 0.05 (0.04)       \\
            \quad Community Profile                    & $-$0.03 (0.03)    & 0.01 (0.05)       & 0.04 (0.05)       & 0.03 (0.06)       & 0.07 (0.05)       \\
            \quad Status-Driven Profile                & $-$0.16 (0.03)    & $-$0.09 (0.08)    & 0.02 (0.08)       & 0.01 (0.07)       & 0.11 (0.08)       \\
            \midrule
            \multicolumn{6}{@{}l}{\textit{Panel B: AI threat slope within profile}} \\
            \quad AI Threat $\mid$ Transactional       & $-$0.08*** (0.03) & 0.03 (0.05)       & 0.07 (0.05)       & 0.03 (0.06)       & $-$0.02 (0.05)    \\
            \quad AI Threat $\mid$ Intrinsic           & $-$0.19*** (0.02) & 0.01 (0.05)       & 0.02 (0.05)       & 0.07 (0.05)       & $-$0.06 (0.05)    \\
            \quad AI Threat $\mid$ Community           & $-$0.22*** (0.03) & 0.03 (0.06)       & 0.08 (0.06)       & $-$0.09 (0.06)    & $-$0.12** (0.06)  \\
            \quad AI Threat $\mid$ Status-Driven       & $-$0.30*** (0.03) & 0.01 (0.08)       & 0.07 (0.08)       & 0.10 (0.07)       & $-$0.14* (0.08)   \\
            \midrule
            \multicolumn{6}{@{}l}{\textit{Panel C: Direct contrasts in levels}} \\
            \quad Transactional $-$ Intrinsic          & 0.21*** (0.03)    & 0.11* (0.06)      & 0.18*** (0.06)    & $-$0.01 (0.08)    & $-$0.15** (0.06)  \\
            \quad Transactional $-$ Community          & 0.22*** (0.04)    & 0.07 (0.07)       & 0.02 (0.07)       & 0.01 (0.08)       & $-$0.17** (0.07)  \\
            \quad Transactional $-$ Status-Driven      & 0.35*** (0.04)    & 0.17* (0.09)      & 0.04 (0.09)       & 0.02 (0.09)       & $-$0.22** (0.09)  \\
            \quad Transactional $-$ Non-Transactional  & 0.25*** (0.03)    & 0.10* (0.05)      & 0.10* (0.05)      & 0.00 (0.07)       & $-$0.17*** (0.06) \\
            \midrule
            \multicolumn{6}{@{}l}{\textit{Panel D: Direct contrasts in AI threat slopes}} \\
            \quad Trans.\ $-$ Intrinsic (slope)        & 0.11*** (0.03)    & 0.02 (0.07)       & 0.05 (0.07)       & $-$0.04 (0.08)    & 0.04 (0.07)       \\
            \quad Trans.\ $-$ Community (slope)        & 0.14*** (0.04)    & 0.01 (0.08)       & $-$0.01 (0.07)    & 0.12 (0.08)       & 0.10 (0.08)       \\
            \quad Trans.\ $-$ Status-Driven (slope)    & 0.23*** (0.04)    & 0.03 (0.10)       & 0.00 (0.09)       & $-$0.07 (0.09)    & 0.12 (0.10)       \\
            \quad Trans.\ $-$ Non-Trans.\ (slope)      & 0.15*** (0.03)    & 0.02 (0.06)       & 0.02 (0.06)       & 0.00 (0.07)       & 0.08 (0.06)       \\
            \midrule
            \multicolumn{6}{@{}l}{\textit{Demographic Covariates (mean-centered)}} \\
            \quad Age (years)                          & 0.01*** (0.00)    & 0.00 (0.00)       & $-$0.01*** (0.00) & 0.00** (0.00)     & 0.00** (0.00)     \\
            \quad Female                               & 0.03 (0.02)       & $-$0.02 (0.05)    & 0.08 (0.05)       & $-$0.22*** (0.06) & 0.03 (0.05)       \\
            \quad Higher Educ.\ (BA+)                  & $-$0.05* (0.03)   & $-$0.20*** (0.06) & $-$0.28*** (0.06) & $-$0.49*** (0.07) & 0.14** (0.06)     \\
            \quad Part-Time                            & 0.11*** (0.03)    & 0.00 (0.07)       & 0.02 (0.07)       & $-$0.18** (0.08)  & $-$0.01 (0.07)    \\
            \quad Middle Income                        & $-$0.00 (0.03)    & $-$0.06 (0.07)    & $-$0.03 (0.07)    & 0.20*** (0.08)    & 0.05 (0.07)       \\
            \quad High Income                          & $-$0.19*** (0.04) & $-$0.31*** (0.08) & $-$0.12 (0.08)    & 0.46*** (0.09)    & 0.29*** (0.08)    \\
            \midrule
            Country Fixed Effects                      & Yes (4 FEs)       & No (UK)           & No (UK)           & Yes (4 FEs)       & No (UK)           \\
            Observations ($N$)                         & 5,688             & 1,498             & 1,498             & 5,295             & 1,498             \\
            $R^2$ / Pseudo $R^2$                       & 0.16              & 0.04              & 0.05              & 0.04              & 0.04              \\
            \bottomrule
        \end{tabular*}
        \begin{tablenotes}[flushleft]
            \scriptsize
            \item \textit{Notes:} Standard errors in parentheses. * $p < 0.10$, ** $p < 0.05$, *** $p < 0.01$. Cell-means specification is used to restrict to the four classified Work Values Profiles (``Other'' excluded) with no general constant. AI threat, controls, and country indicators are mean-centered. Columns 1 and 4 are estimated across all four countries; Columns 2, 3, and 5 are estimated using the supplementary UK-only module. In columns 1, 2, 3, 5, the OLS model is on $z$-scores (so coefficients $\beta$ and adjusted means are in $\text{SD}$ units). In column 4 it is binary Logit (log-odds).
        \end{tablenotes}
    \end{threeparttable}
\end{table}

The negative slopes of perceptions of personal job-replacement likelihood in Column~1 of Table~\ref{tab:political_regressions_profiles} as well as the higher baseline pessimism of \textit{Transactional} workers might appear to run counter to \textit{Hypothesis 3}. Yet, when we look more closely at people's perceptions of their own personal AI replacement expectations, alongside their broader expectations of industry job impacts, the picture looks different. Indeed, we find that, among workers who expect AI to replace their job, roughly half of \textit{Intrinsic} ($44\%$), \textit{Community} ($48\%$), and \textit{Status-driven} ($54\%$) workers actually expect AI's impact on work availability in their industry to be \textit{positive}, compared to just $23\%$ of \textit{Transactional} workers ($55\%$ of whom expect negative industry job impacts). Moreover, many surveyed respondents that expect AI to replace their job still view AI's impacts on jobs in their industry as a whole as positive. As a consequence, it is important to narrow our analysis just to those that have a negative view of AI's impacts on work availability for \textit{both} themselves personally and their industry as a whole. And indeed, across all countries, a perception that AI will harm job availability in a person's industry is associated with a much more negative evaluation of AI across all Work Value Profiles. The marginal backlash, however, is steeper among all three non-pecuniary profiles than among Transactional workers.\footnote{While the overall magnitude of the slopes in Column 1 of Appendix Table~\ref{tab:app_industry_harm_politics} partly reflects shared evaluation scaling across the AI impact items, a uniform battery effect cannot account for the significant cross-profile slope divergences or the parallel effects on separately fielded institutional riggedness and redistribution batteries in Columns 2 and 5.} Moreover, among those anticipating severe industry job harm, \textit{Intrinsic} ($+0.98\text{ SD}$) and \textit{Community} ($+0.96\text{ SD}$) workers express stronger macro-political pessimism than \textit{Transactional} workers ($+0.84\text{ SD}$).

\begin{figure}[!htbp]
    \centering
    \includegraphics[width=0.95\linewidth]{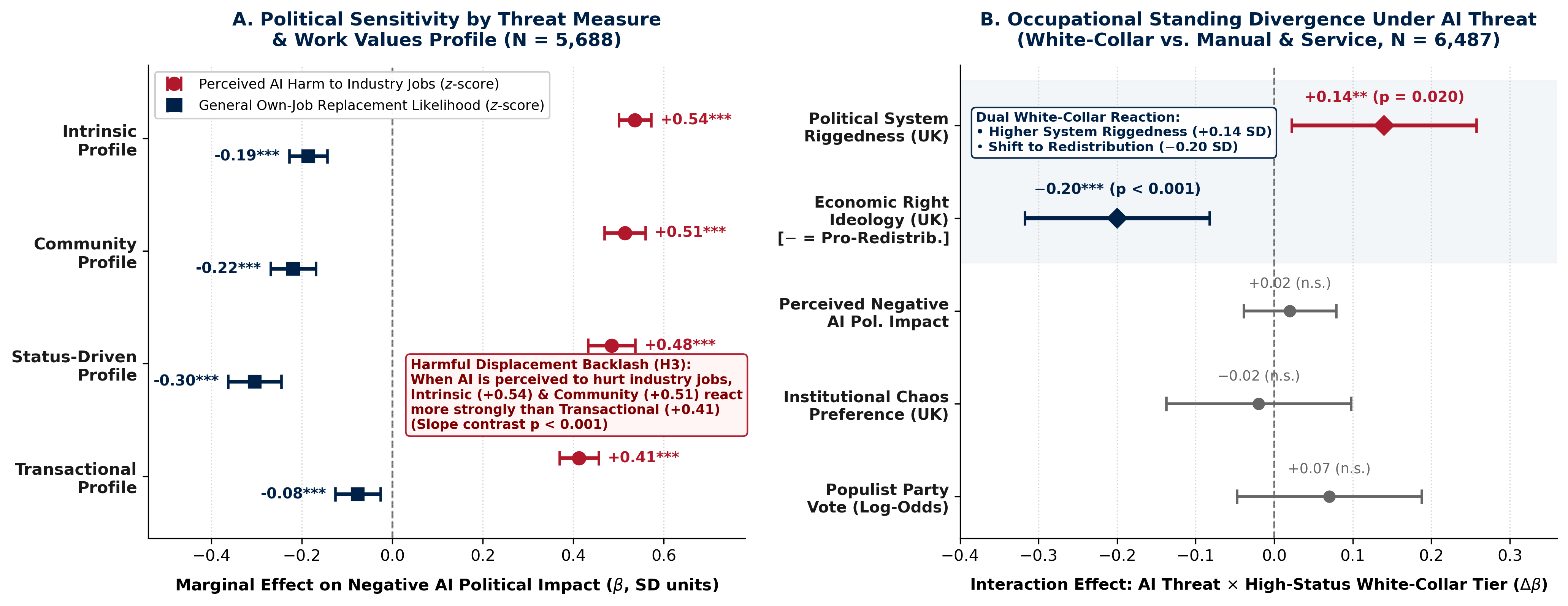}
    \caption{\textbf{Work Values, Occupational Category, and Political Backlash to AI Displacement}}
    \label{fig:political_profiles_and_tier}
    \vspace{0.10cm}
    \begin{minipage}{\linewidth}
        \footnotesize 
        \textit{Notes:} Panel A compares the covariate-adjusted marginal effect ($\beta$ and 95\% confidence intervals, in $\text{SD}$ units) on Perceived Negative AI Political Impact ($z$-score across all four countries, $N = 5{,}688$ / $5{,}722$) across the four Work Values Profiles when workplace AI exposure is measured by general own-job replacement likelihood (Table~\ref{tab:political_regressions_profiles}, Panel~B) versus perceived harm to job quality and job counts in the respondent's industry (Appendix Table~\ref{tab:app_industry_harm_politics}, Panel~B). Panel B plots the interaction coefficients ($\Delta\beta$ and 95\% confidence intervals) between standardized Subjective AI Threat and working in the White-Collar occupational tier (vs.\ Manual \& Service tier) across all five political dependent variables from Appendix Table~\ref{tab:political_regressions_tier} ($N = 1{,}667$ in the UK-only modules; $N = 6{,}487$ and $6{,}018$ in the four-country modules).
    \end{minipage}
\end{figure}

In this alternative specification using perceived industry-level job harm, an \textit{Intrinsic} work motivation is associated with more negative evaluations of AI's political impact across all four countries. Turning to the UK-only sample ($N = 1{,}493$; Appendix Table~\ref{tab:app_industry_harm_politics}), expecting AI to hurt jobs in one's industry is associated with a greater belief that the political and economic system is ``rigged.'' This is most pronounced with \textit{Status}-driven and \textit{Intrinsic}-motivated workers. It is also associated with heightened demand for left-wing economic redistribution among \textit{Intrinsic} workers and among \textit{Transactional} workers. In line with our theoretical expectations in Section~\ref{sec:theory}, workers who derive non-pecuniary meaning, community, and social status from their occupations appear to also be those who react most intensely when they perceive that technological change will genuinely damage employment in their sector. I insist again that this analysis is purely relational, not causal. We cannot say for certain the direction of any potential relationship, nor can we control for the battery of unobservables that we are unable to capture in data. 

Additional subgroup regressions reported in the Appendix show how these political patterns vary across age groups, household income levels, and occupational categories in practice. First, splitting the sample at the median age of 41 years (Appendix Table~\ref{tab:political_age_split}) suggests the same generational divide seen in our reservation wage analysis. Indeed, across all four countries, older workers ($> 41$) exhibit a substantially wider gap between \textit{Transactional} and non-monetary profiles in how personal AI exposure shapes their views of AI's political impact, while in the UK-only sample ($N = 1{,}498$), older \textit{Community}-oriented workers who fear AI will replace their job also express political beliefs that are more supportive of economic redistribution ($\beta = -0.18\text{ SD}, p < 0.05$), whereas younger \textit{Transactional} workers ($\le 41$) sit furthest to the economic left in baseline levels ($\beta = -0.22\text{ SD}, p < 0.01$). Second, estimating the four-country political models separately within low-, middle-, and high-income households (Appendix Table~\ref{tab:app_income_subgroups} and Figure~\ref{fig:app_income_subgroups}) and within white-collar versus manual and service occupations (Appendix Table~\ref{tab:app_occ_subgroups} and Figure~\ref{fig:app_threat_by_occ_status}) suggests that the attitudinal divide between \textit{Transactional} and non-monetary workers exists across occupations and income groups. In practice, the political divergence across Work Values profiles is just as large---and on the AI threat slope, even more pronounced---among high-income and white-collar workers, when compared with low-income and manual or service workers.

Finally, Figure~\ref{fig:occupation_marginal_effects} (with more extensive estimates reported in Appendix Table~\ref{tab:isco_granular_reg}) looks at the relationship between personal AI job replacement expectations and perceptions of AI's negative impact across each of the nine major ISCO-08 occupation groups. Since higher values on the dependent variable indicate more negative political evaluations of AI, negative coefficients mean that workers who report greater AI exposure in their own jobs also hold \textit{more favorable} (less pessimistic) views of AI's broader impact. As Figure~\ref{fig:occupation_marginal_effects} shows, this negative association holds broadly across seven of the nine major occupation groups. More research is needed to understand these dynamics, particularly work that can move beyond industry-level analysis of perceptions of AI risk, toward occupation or task level analysis.

\begin{figure}[!htbp]
    \centering
    \includegraphics[width=0.88\linewidth]{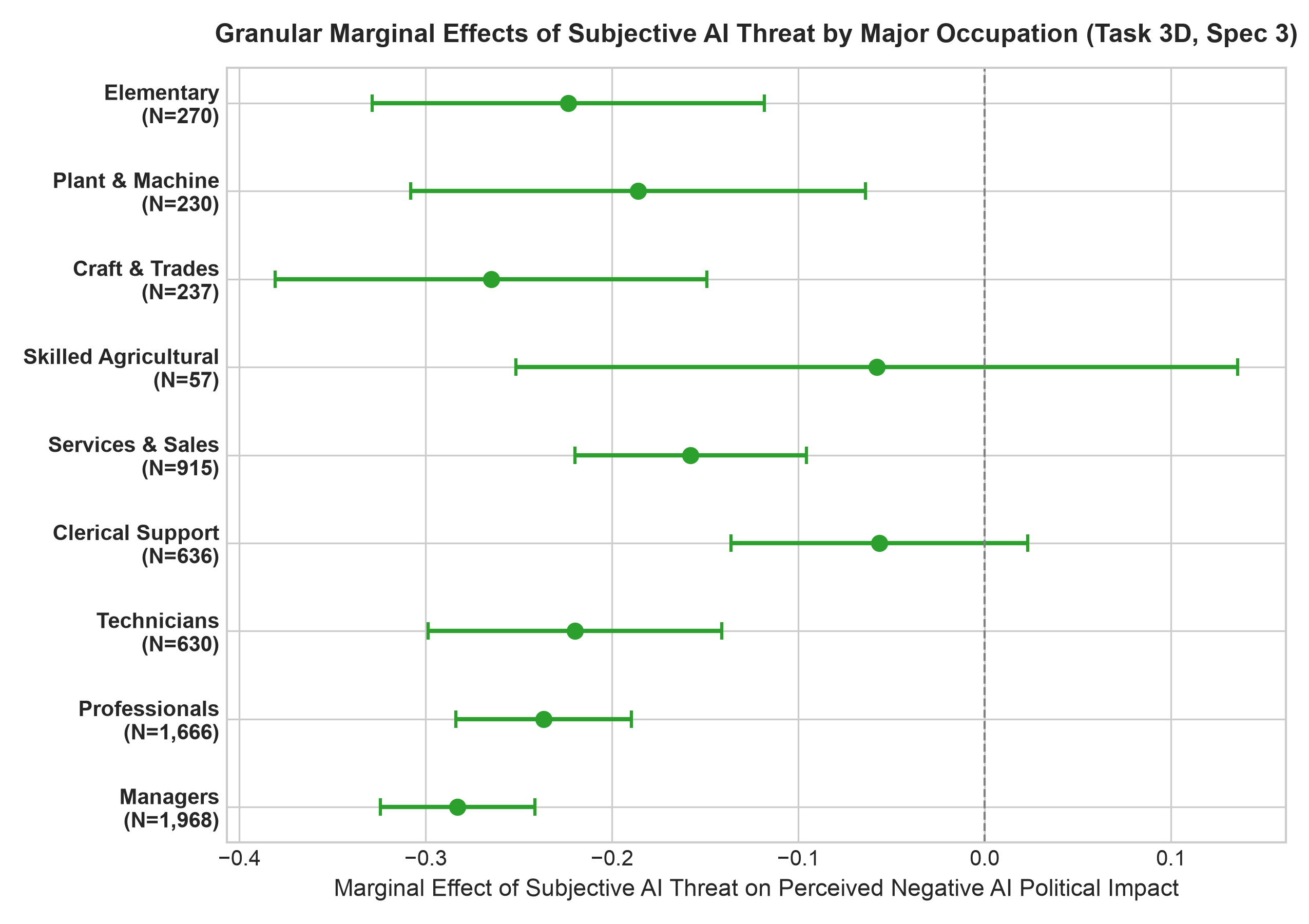}
    \caption{\textbf{Granular Marginal Effects of Subjective AI Threat on Perceived Negative AI Political Impact Across Major Occupation Groups}}
    \label{fig:occupation_marginal_effects}
    \vspace{0.10cm}
    \begin{minipage}{\linewidth}
        \footnotesize 
        \textit{Notes:} Marginal effects ($\beta$, in $\text{SD}$ units) and 95\% confidence intervals of standardized Subjective AI Threat ($z$-score) on Perceived Negative AI Political Impact ($1 = \text{Very positive}$ to $5 = \text{Very negative}$, $z$-score across all four countries) across ISCO-08 major occupation groups. The model controls for age, female gender, higher education, part-time status, and country fixed effects ($N = 6{,}480$). Full numerical point estimates, standard errors, and significance levels are reported in Appendix Table~\ref{tab:isco_granular_reg}.
    \end{minipage}
\end{figure}

\section{Discussion and Limitations}
\label{sec:discussion}

The analyses I have presented in this paper offer evidence meant to advance our understanding of what motivates non-monetary work attachment, how strong this pull is, and what its implications may be for the age of AI. I now evaluate our empirical results against the three primary hypotheses formalized in Section~\ref{sec:theory}. First, \textit{Hypothesis 1}---that workers who place significant importance on the non-monetary value they derive from their labor will require higher reservation wages to voluntarily exit employment than \textit{transactionally} motivated workers---appears to receive support, particularly for \textit{intrinsically}-motivated workers, though not across all non-monetary profiles uniformly. Indeed, in this paper's modelling, individuals with an \textit{Intrinsic} Profile exhibit a 10-percentage-point higher raw median reservation wage ($80\%$ vs.\ $70\%$) and require a statistically significant covariate-adjusted wage replacement premium of $1.76\text{ to }1.92\text{ percentage points}$ in our most conservative estimates and at $5.77\text{ percentage points}$ ($p < 0.01$) in less conservative estimates. One in four \textit{Intrinsic} workers ($23\%$ vs.\ $14\%$ of Transactional workers, $p < 0.01$) say they would refuse to leave their jobs at any price up to 120\% of current salary. Community-oriented workers ($\beta_C C_W$) similarly exhibit a higher extensive-margin refusal rate than Transactional workers ($17\%$ vs.\ $14\%, p = 0.03$). By contrast, \textit{Status-Driven} workers ($\beta_S S_W$) do not demand a higher reservation wage to leave their current roles, requiring a $60\%$ median wage replacement rate---the lowest threshold of any profile. Although belief in the non-pecuniary value of work is very common and appears to play a role in moderating occupational attachment, the precise expression of that non-pecuniary value can vary greatly, depending on the reasons a worker values their job outside of financial considerations. 

Second, \textit{Hypothesis 2}---that workers motivated by status, intrinsic, or communal values will pursue work-like or prosocial activities rather than pure leisure when receiving guaranteed income transfers outside their current job---again appears supported across all three non-monetary profiles and may help explain the Status-driven pattern in \textit{Hypothesis 1}. When survey respondents were asked how they would allocate their time if they could no longer work and instead received a guaranteed basic income, nearly four-fifths ($79\%$) of Transactional workers said they would pursue leisure activities (rest, hobbies, and travel). By contrast, Status-Driven ($55\%$), Community ($46\%$), and Intrinsic ($46\%$) workers were significantly more likely than Transactional workers ($39\%, p < 0.01$) to seek new paid work opportunities or start a business. Intrinsic ($57\%$) and Community ($51\%$) workers were also the most likely of any group to dedicate their time to unpaid volunteering or caregiving (vs.\ $45\%$ of Status-Driven and $46\%$ of Transactional workers, $p < 0.01$). For Status-Driven workers, a low reservation threshold in their current job may reflect a willingness to use guaranteed income partly to finance upward career mobility or entrepreneurship ($S_{NW} \ge S_W$), rather than detachment from work itself.

Third, \textit{Hypothesis 3} directly engages with a puzzle in political economy literature, supposing that workers who value the non-monetary components of their job will express greater concern about AI-related employment impacts and react more strongly in their labor-market and political attitudes when facing high displacement threat. This hypothesis receives potential support in both stated labor-market behavior and in political attitudes, though more research is needed to validate this finding. This paper shows that \textit{Status}-oriented and \textit{Community}-oriented workers report the highest level of concern about the threat of AI replacing their job (Figure~\ref{fig:ai_threat_by_work_values}).  Moreover, a perception of AI's replacement threat was associated with a steeper decline in respondents' reported reservation wages among \textit{Community} and \textit{Intrinsically}-motivated workers than among Transactional workers. When I looked at the relationship between perceptions of AI's job impacts and political attitudes (with a particular focus on the UK) (Table~\ref{tab:political_regressions_profiles}), we find that Transactional workers express more negative baseline views of AI's macro-political impact. Yet when we isolate for belief that AI will cause \textit{adverse} effects on both a person's job and their industry's job quality and job counts (Figure~\ref{fig:political_profiles_and_tier}, Panel~A, and Appendix Table~\ref{tab:app_industry_harm_politics}), we find clearer potential evidence to support \textit{Hypothesis 3}. Specifically, I show that \textit{Intrinsic}, \textit{Community}, and \textit{Status}-driven workers exhibit a significantly steeper negative political reaction than \textit{Transactional} workers. 

Several important methodological limitations warrant careful discussion. The key limitation of this study is its reliance on cross-sectional observational survey data, which cannot credibly establish causal relationships between perceptions of AI displacement threats, work values, and political attitudes. In particular, the Work Values constructed in this paper may themselves be endogenous and downstream of job quality, among other factors. Workers in rewarding, autonomous jobs may naturally prioritize intrinsic fulfillment, whereas workers in routinized or precarious roles may defensively adopt a transactional orientation toward work. Yet two empirical checks in Section~\ref{sec:results_reservation} should help temper some of this concern. When we control directly for standardized job satisfaction, we still find a distinct \textit{Intrinsic} reservation wage premium intact. Meanwhile, controlling for respondents' perceived confidence in finding an equally good outside job \textit{strengthens} the Intrinsic premium. Conceptually, therefore, Work Values Profiles are best interpreted not as fixed psychological traits, but instead as a worker's realized relationship within their current job---capturing part of what they stand to lose if that role is automated. Still, it remains possible that other unobservables also shape these relationships. Likewise, it is plausible that politically alienated or generally pessimistic individuals selectively perceive greater personal AI job threat, rather than AI threats causing political alienation. We cannot credibly determine the direction of this relationship within this paper's modeling. Similarly, I note that a worker who already holds anti-establishment beliefs might retrospectively rationalize those beliefs by emphasizing grievances against AI. Moreover, although the profile validation analysis in Section~\ref{sec:data} (Appendix Table~\ref{tab:app_profile_validation}) suggests that Work Values Profiles are not heavily correlated to income class and pre-existing political beliefs, this finding does not rule out the possibility that unobserved psychological dispositions (e.g. a disposition for anxiety, risk aversion, or general negativity bias) are driving higher perceived AI threat and more intense political backlash. 

A second methodological challenge is related to the exclusion of the $N = 2{,}010$ unclassified (``Other'') respondents—24\% of the total sample, comprising $N = 1{,}128$ non-working adults and $N = 882$ working respondents with four-way ``values'' ties. Although excluding this sample allows us to construct profile contrasts to compare active workers with clearly differentiated priorities, it also restricts us to $N = 6{,}224$ classified respondents. Relatedly, our Step 2 tie-breaking rule assigns the $N = 2{,}000$ respondents who tied at their maximum score (32\% of classified workers) to the empirically rarer domain. This then frequently places workers who rated both pay and a non-monetary dimension at the ceiling into a non-monetary profile. While this makes our contrasts against the \textit{Transactional} baseline conservative (since tied non-monetary workers also attach high importance to pay), continuous four-index PCA models---which remove the discrete categories and tie-breaking altogether---suggest both the positive Intrinsic effect and negative Status effect across the continuous workforce. Meanwhile, Appendix Table~\ref{tab:app_threshold_sensitivity} rebuilds the Work Values Profiles across five alternative cutoffs ($< 2.0$, $< 2.5$, $< 3.0$ baseline, $< 3.5$, and $\le 4.0$ [no minimum cutoff]) across 12 headline models. Over all five thresholds, the estimates remain virtually unchanged in both magnitude and significance, including the reservation income contrast and the AI threat slope contrast on perceived negative political impact.

Third, the evidence presented in the paper on reservation wages, while striking in its consistency, also warrants epistemic modesty. The iterative auction-style model captures stated willingness to quit at \textit{hypothetical} income replacement thresholds. As is common in economic surveys, we cannot capture revealed labor market behavior. No respondent actually left their job as a consequence of participating in the survey. Behavioral economics offers an abundance of evidence that preferences articulated in surveys can diverge significantly from real-world decisions, outside the bounds of a survey \citep{Hausman2012}. For instance, respondents may overstate their actual attachment to work when the choice is a costless survey, as opposed to a real-life decision with implications. So it can be difficult to truly substantively interpret the 20\% of respondents who refused to quit at 120\% of their current salary, and how they would actually behave in the real world. At the same time, excluding the 20\% of right-censored respondents who refused all offers up to 120\% of salary means that our baseline OLS estimates capture conservative lower-bound differences among uncensored compliers ($N = 4{,}860$), whereas retaining right-censored observations via our two-part hurdle model, top-coded bounds ($120\%$ to $200\%$), and parametric right-censored Tobit maximum likelihood estimation increases the estimated Intrinsic premium to $4.58\text{--}5.77\text{ percentage points}$ ($p < 0.01$; Appendix Figure~\ref{fig:app_censoring_robustness} and Appendix Table~\ref{tab:app_censoring_regressions}). 

\section{Conclusion}
\label{sec:conclusion}

This paper set out to better understand the ways in which people value their work, including the significance of employment's non-monetary value. To look at this question, I created an original cross-national survey of working-age adults across the United Kingdom, the United States, Denmark, and India, fielded by More in Common and examining the pecuniary, intrinsic, communal, and status-driven motivations for work attachment. I then traced how non-monetary workplace attachments shaped the guaranteed replacement income workers would demand to quit their job, how people would spend their time if they could no longer work, and whether the non-pecuniary value of work shapes the way people respond to AI. 

I highlight four findings in particular. First, $81\%$ of respondents across all occupations reported being satisfied with their jobs. Meanwhile, a majority of workers valued their roles for non-monetary reasons. Indeed, using a novel Work Values profile classification, I showed that one-third of workers say they are intrinsically motivated by work, finding their job valuable for its own sake. Meanwhile, one quarter of survey respondents value their job because of the community and social opportunities it brings them, and $17\%$ of respondents say they value their job because of the recognition and prestige it offers. On the other side, just over one-quarter of respondents valued their job principally for its pay and benefits. Furthermore, non-monetary work attachment can be found in near-equal measure across income levels and within both blue and white collar occupations. Second, most people say they would choose to leave their job if offered a full guaranteed income replacement. Yet $20\%$ of workers said they would stay in their job even if offered up to $120\%$ as guaranteed income. This share is consistent across the income distribution and occupational categories, but it is significantly higher for intrinsically motivated workers, who attached significant non-monetary value to their work. Third, non-monetary work attachment is associated with divergences in how people would spend their time if they received a guaranteed income but could not work in their current role. Specifically, transactionally-motivated workers said they would functionally retire, while people attaching non-monetary value to their job said they would volunteer or pursue further work or business opportunities. Finally, I find potential evidence that non-pecuniary work attachment plays a moderating role in shaping the political and economic backlash to AI; however, I note more research is needed to understand these dynamics. 

The findings from this paper appear to carry significant implications for how governments ought to design social and economic policies to manage the AI transition. While income transfer provisions---such as basic income or negative income tax programs---may be \textit{necessary} to preserve material livelihoods for workers who experience job displacement, they are unlikely to be \textit{sufficient}. Indeed, as this paper shows, workers appear to attach significant non-monetary value to their jobs, across the income distribution and occupational groups. Income transfers can fail entirely to compensate for this non-pecuniary value of work. And such a finding is entirely in line with an abundance of prior literature---cited in this paper---that points toward the socio-cultural and psychological salience of formal employment. Governments and researchers may need to adjust for these non-pecuniary aspects of work when conjuring ideas to support people during periods of economic change. Of course, if AI does indeed induce a major transformation of the labor market, people may find entirely new replacements for the non-monetary value they previously identified in their work. Such an outcome may require new systems for economic redistribution and benefit sharing. Regardless of whichever scenario of AI-enabled economic change the world experiences, disruptions to work may force society to wrestle with pressing questions about how to protect living standards, purpose, community, and dignity for workers as the world changes. 

\clearpage


\clearpage
\appendix
\section*{Appendix}
\suppressfloats[t]
\addcontentsline{toc}{section}{Appendix}
\setcounter{table}{0}
\renewcommand{\thetable}{A\arabic{table}}
\setcounter{figure}{0}
\renewcommand{\thefigure}{A\arabic{figure}}

\begin{figure}[!htbp]
    \centering
    \includegraphics[width=0.95\linewidth]{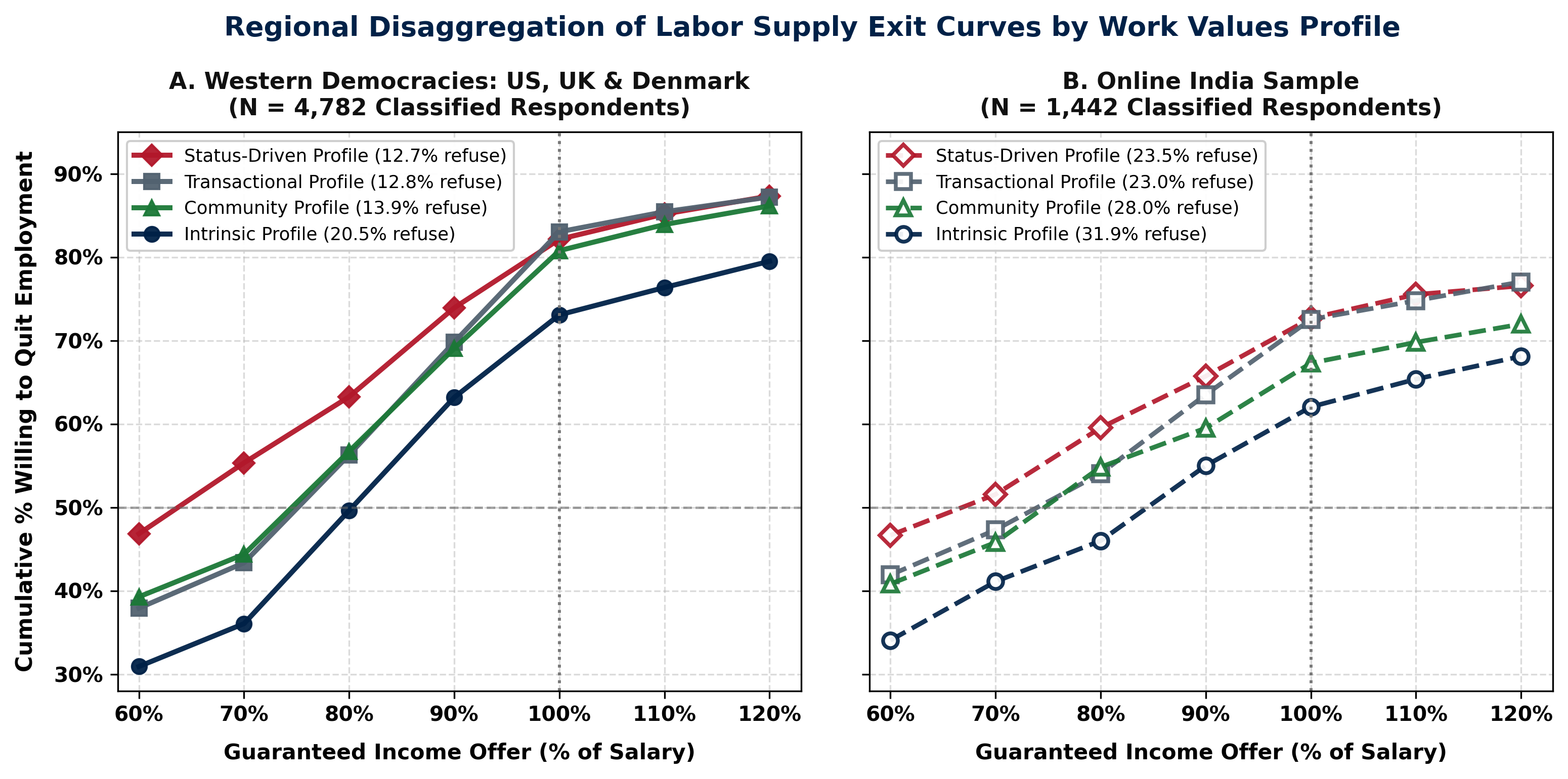}
    \caption{\textbf{Regional Divergences in Reservation Wage by Work Values}}
    \label{fig:app_reservation_curve_regions}
    \vspace{0.10cm}
    \begin{minipage}{\linewidth}
        \footnotesize
        \textit{Notes:} Panel A plots the share of respondents in the three Western democracies (US, UK, and Denmark pooled, $N = 4,782$) willing to voluntarily exit paid employment at each sequential guaranteed income threshold ($60\%$ to $120\%$ of current salary, with benefits preserved) across the four Work Values Profiles. Panel B plots the corresponding exit curves for the online India sample ($N = 1,442$). Within both regions, Intrinsic workers exhibit the lowest cumulative exit trajectory and the highest refusal rate at $120\%$ ($20.5\%$ vs.\ $12.8\%$ Transactional in US--UK--Denmark; $31.9\%$ vs.\ $23.0\%$ Transactional in India).
    \end{minipage}
\end{figure}

\begin{figure}[!htbp]
    \centering
    \includegraphics[width=0.96\linewidth]{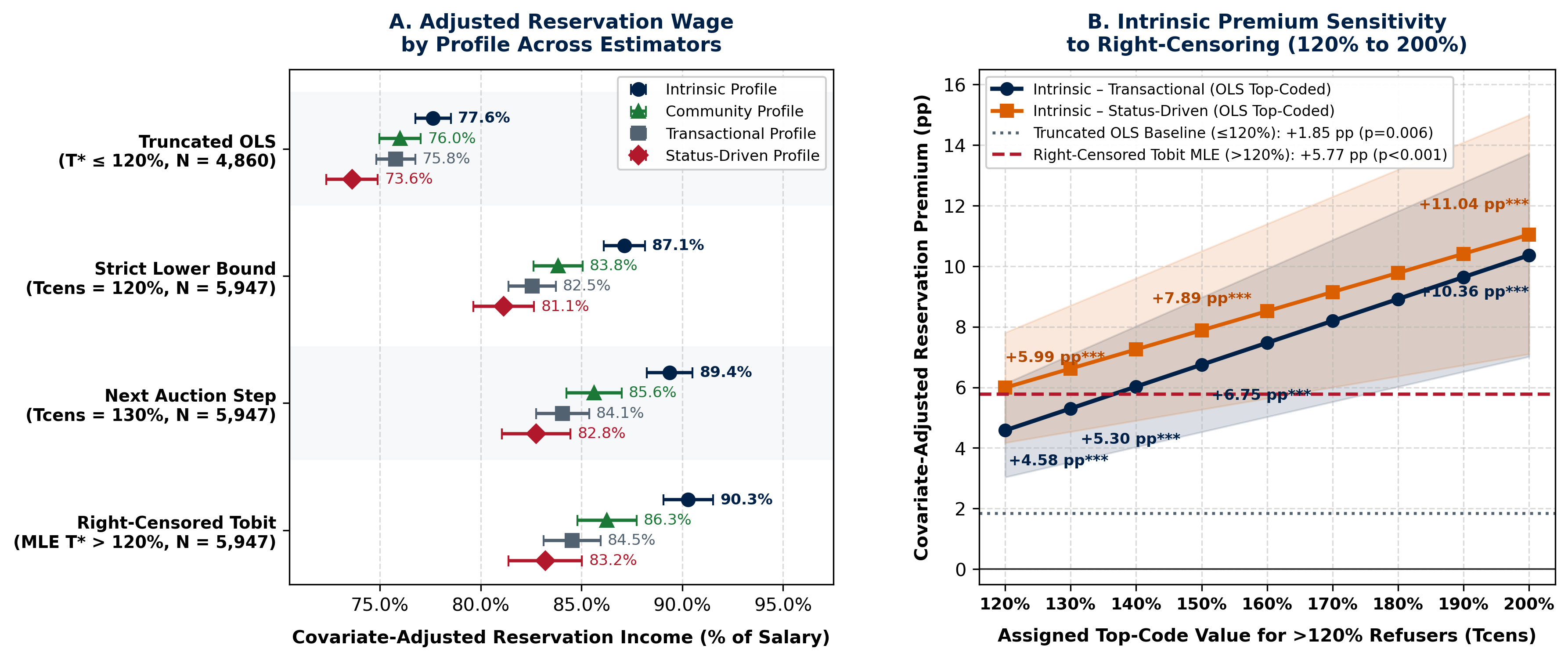}
    \caption{\textbf{Multi-Estimator Profile Means and Top-Coding Sensitivity Curve}}
    \label{fig:censoring_robustness}
    \label{fig:app_censoring_robustness}
    \vspace{0.10cm}
    \begin{minipage}{\linewidth}
        \footnotesize
        \textit{Notes:} Panel A plots covariate-adjusted reservation wage means and 95\% confidence intervals for each Work Values Profile across four specifications. In order, these are (1) the baseline Truncated OLS model restricted to uncensored compliers ($T^* \le 120\%, N = 4,860$), (2) the Manski-style lower bound model that assigns $>120\%$ refusers to $T_{\text{cens}} = 120\%$ ($N = 5,947$), (3) the next auction step model that assigns holdouts to $T_{\text{cens}} = 130\%$ ($N = 5,947$), and (4) a Right-Censored Tobit model estimated via maximum likelihood ($T^* > 120\%$ right-censored, $N = 5,947$). Panel B plots the covariate-adjusted \textit{Intrinsic} reservation wage premiums (and 95\% confidence bands) across top-coded thresholds from $T_{\text{cens}} = 120\%$ to $200\%$ in 10-percentage-point increments. This is then benchmarked against the Truncated OLS baseline ($+1.85\text{ pp}, p < 0.01$) and the Right-Censored Tobit MLE estimate ($+5.77\text{ pp}, p < 0.01$). Full regression tables across all specifications are reported in Appendix Table~\ref{tab:app_censoring_regressions}.
    \end{minipage}
\end{figure}

\begin{figure}[!htbp]
    \centering
    \includegraphics[width=0.88\linewidth]{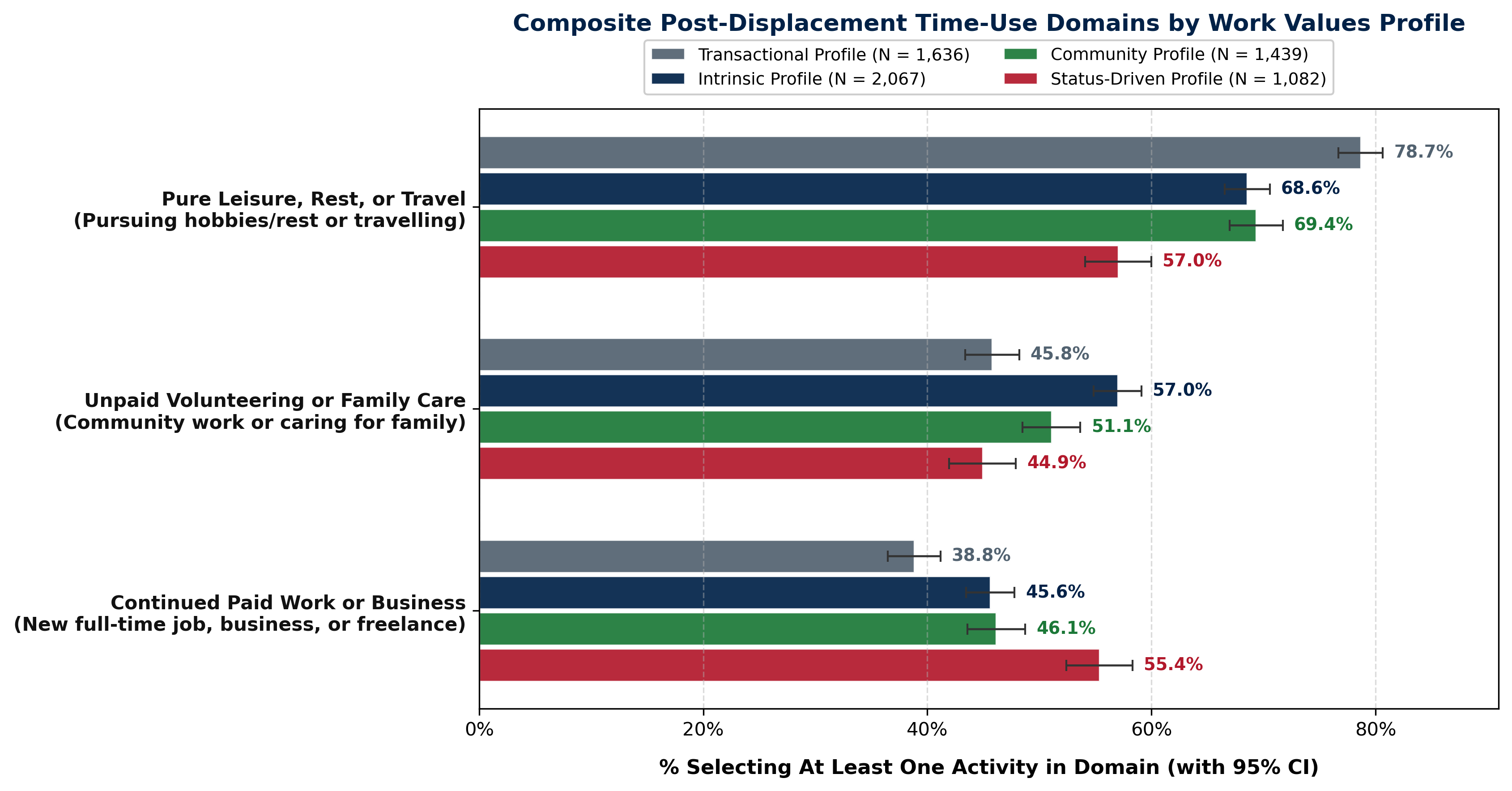}
    \caption{\textbf{Post-Displacement Time-Use by Work Values Profile ($N = 6,224$)}}
    \label{fig:app_time_use_domains}
    \vspace{0.10cm}
    \begin{minipage}{\linewidth}
        \footnotesize
        \textit{Notes:} Bars display the proportion of classified respondents ($N = 6,224$) in each Work Values Profile (with 95\% confidence intervals) selecting at least one activity within three composite post-displacement domains if guaranteed a basic income: (1) \textit{Pure Leisure, Rest, or Travel}, (2) \textit{Unpaid Volunteering or Family Care}, and (3) \textit{Continued Paid Work or Business}. Individual item breakdowns are plotted in Figure~\ref{fig:post_work_time_use_profiles}.
    \end{minipage}
\end{figure}

\begin{figure}[!htbp]
    \centering
    \includegraphics[width=0.88\linewidth]{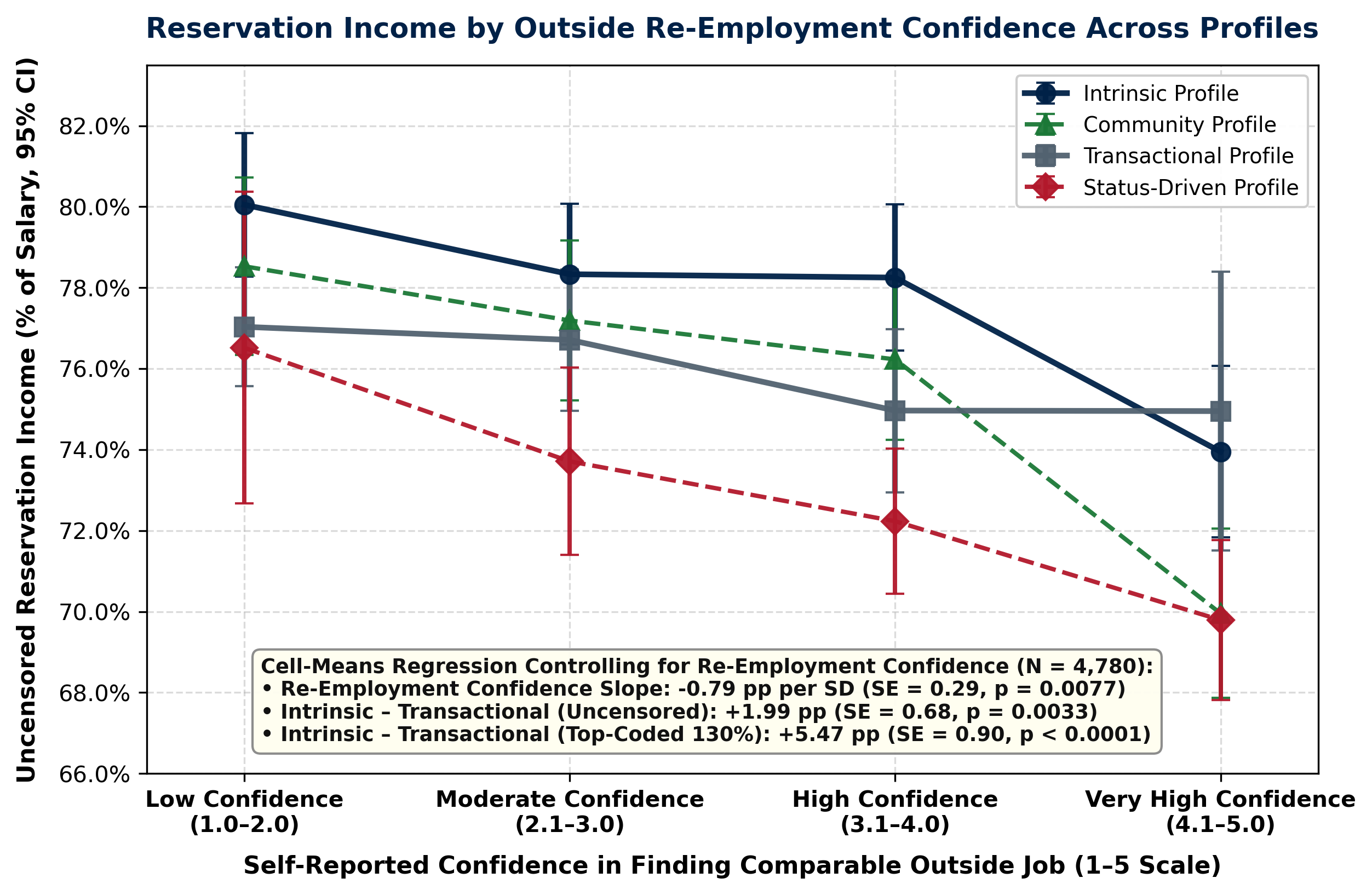}
    \caption{\textbf{Reservation Income by Outside Re-Employment Confidence Across Work Values Profiles}}
    \label{fig:app_reemployment_confidence}
    \vspace{0.10cm}
    \begin{minipage}{\linewidth}
        \footnotesize
        \textit{Notes:} Lines plot mean uncensored reservation income (and 95\% confidence intervals) across respondents' self-reported confidence in finding another comparable job (\texttt{lose\_job\_confident\_1}--\texttt{4}, 1--5 scale). 
    \end{minipage}
\end{figure}

\begin{figure}[!htbp]
    \centering
    \includegraphics[width=0.90\linewidth]{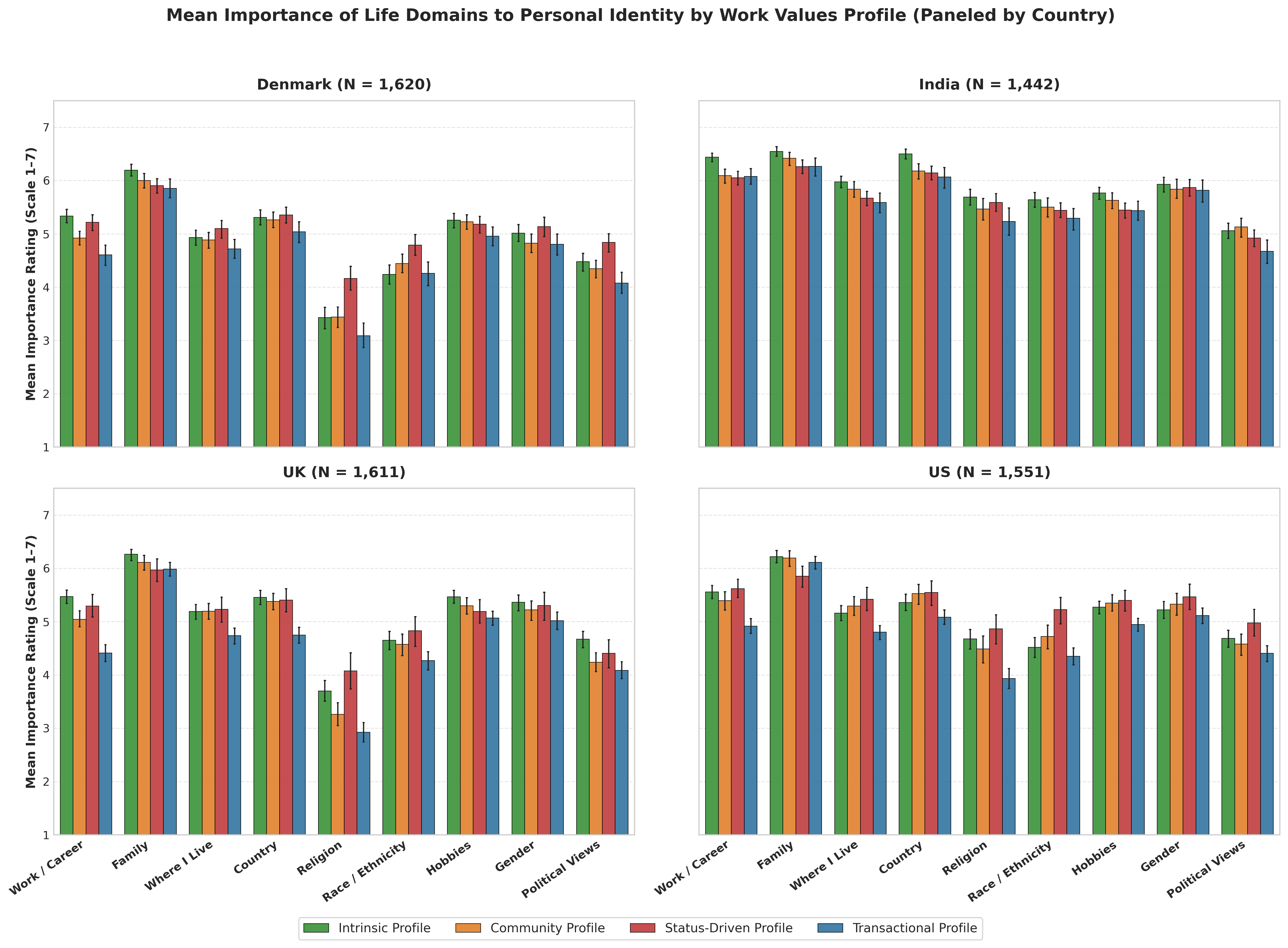}
    \caption{\textbf{Importance of Life Domains to Personal Identity by Work Values Profile}}
    \label{fig:app_important_to_identity}
    \vspace{0.10cm}
    \begin{minipage}{\linewidth}
        \footnotesize 
        \textit{Notes:} Bars display the average (mean) importance rating and 95\% confidence intervals ($1.96 \times \text{SE}$ whiskers) within each Work Values Profile on a 1--7 Likert scale (where 1 = ``Not important at all'' and 7 = ``Very important'') among classified respondents (``Other'' excluded; Total $N = 6,224$: Denmark $N = 1,620$, India $N = 1,442$, UK $N = 1,611$, US $N = 1,551$).
    \end{minipage}
\end{figure}

\begin{figure}[!htbp]
    \centering
    \includegraphics[width=0.85\linewidth]{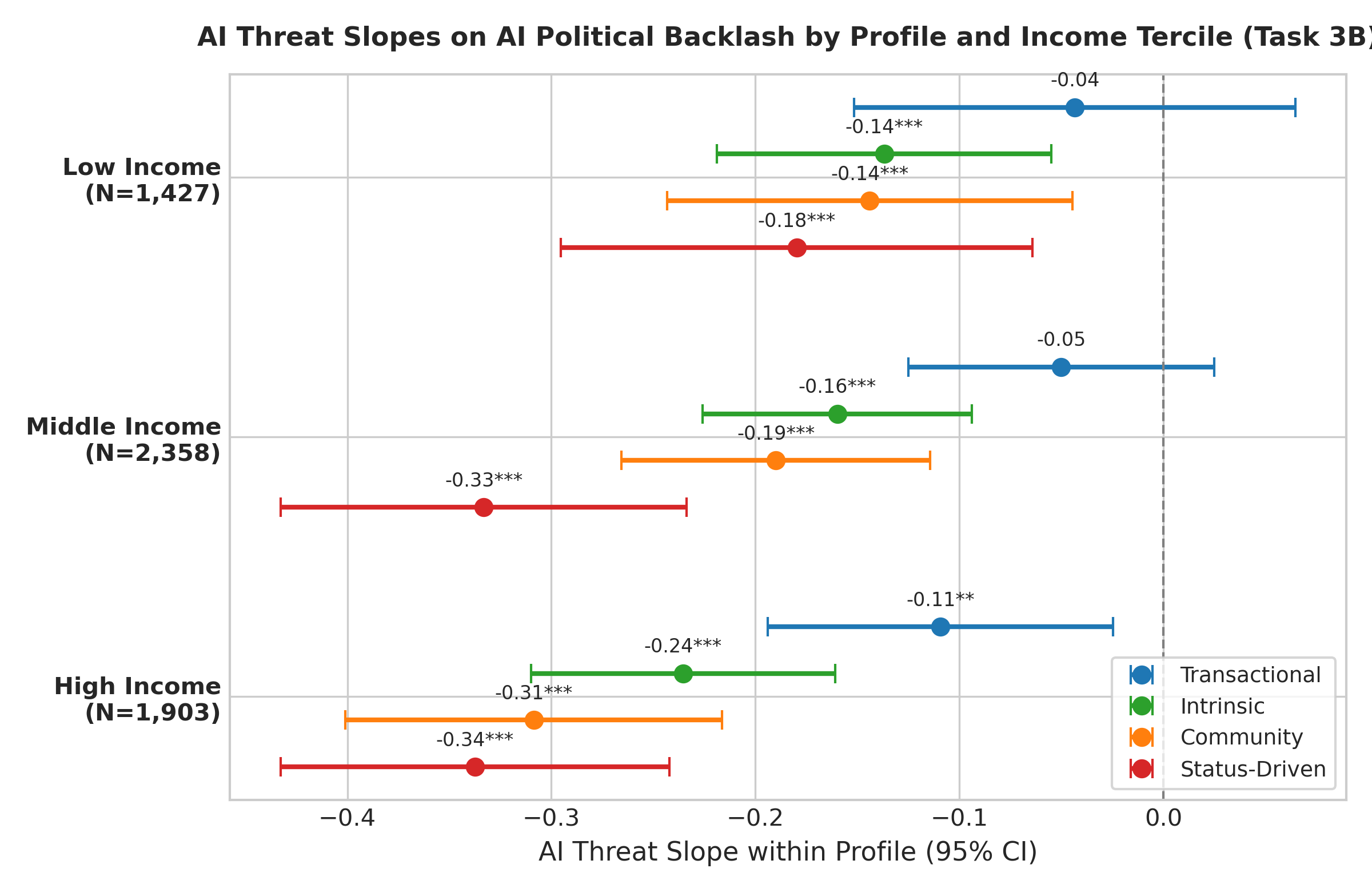}
    \caption{\textbf{AI Threat Slopes within Work Values Profiles across Income Terciles}}
    \label{fig:app_income_subgroups}
    \vspace{0.10cm}
    \begin{minipage}{\linewidth}
        \footnotesize 
        \textit{Notes:} Point estimates and 95\% confidence intervals for the AI threat slope within each Work Values Profile (cell-means specification; ``Other'' excluded) across low ($N=1,427$), middle ($N=2,358$), and high ($N=1,903$) household income terciles, corresponding to Panel~B of Table~\ref{tab:app_income_subgroups}.
    \end{minipage}
\end{figure}

\begin{figure}[!htbp]
    \centering
    \includegraphics[width=0.90\linewidth]{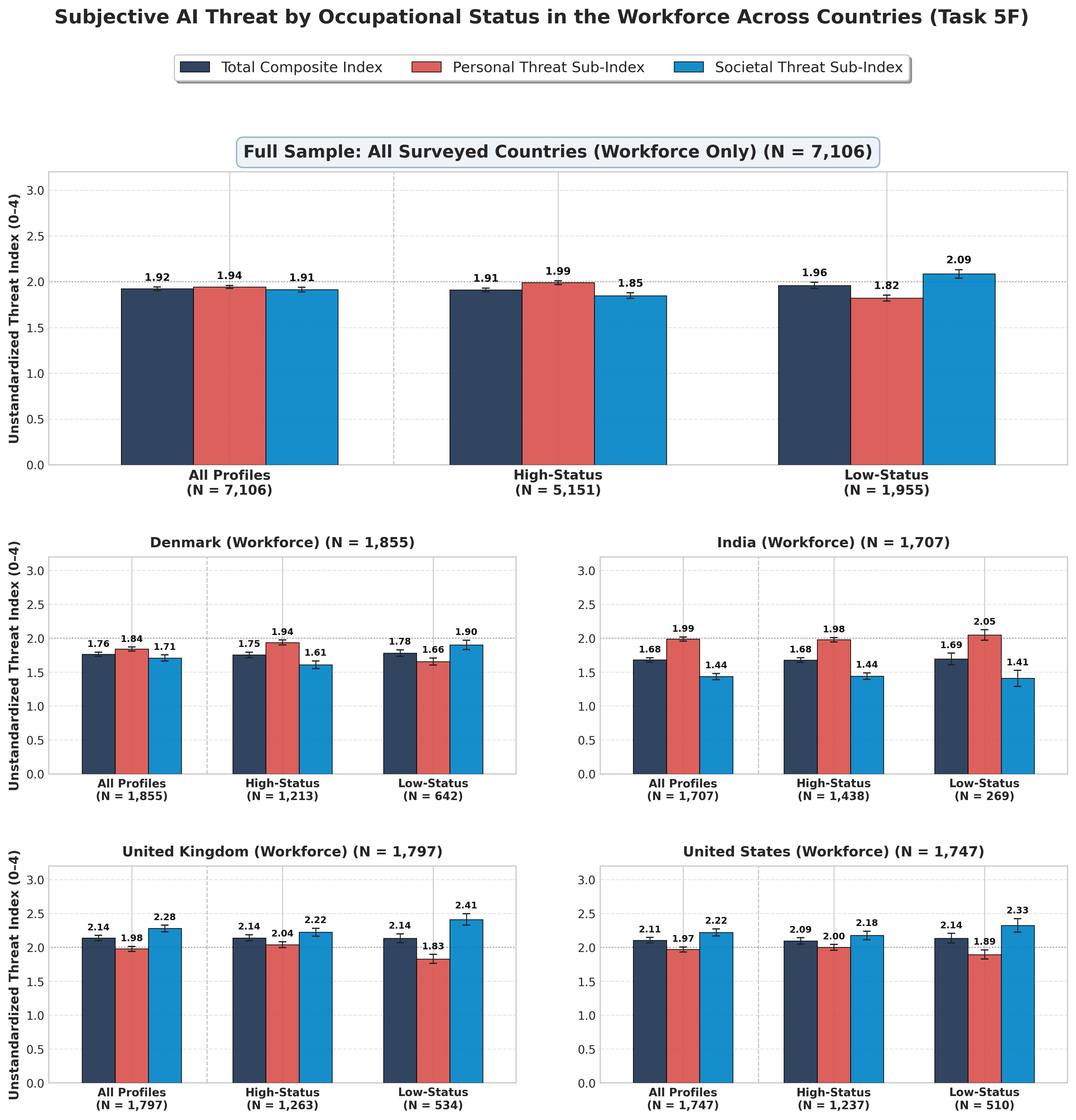}
    \caption{\textbf{Subjective AI Threat by Occupational Category in the Workforce Across Surveyed Countries}}
    \label{fig:app_threat_by_occ_status}
    \vspace{0.10cm}
    \begin{minipage}{\linewidth}
        \footnotesize 
        \textit{Notes:} Group means and 95\% confidence intervals (1.96 $\times$ SE whiskers) for unstandardized AI threat indices rescaled to a common [0, 4] metric where 0 represents minimum perceived threat, 2 represents the neutral scale midpoint, and 4 represents maximum perceived threat. The analytical sample is restricted exclusively to respondents actively participating in the workforce ($N = 7,106$). Within each panel, the leftmost grouping (`All Profiles') reports the pooled workforce average across all working respondents, followed by the White-Collar ($N = 5,151$) and Manual \& Service ($N = 1,955$) occupational tiers. Non-working respondents are excluded entirely.
    \end{minipage}
\end{figure}

\begin{figure}[!htbp]
    \centering
    \includegraphics[width=0.90\linewidth]{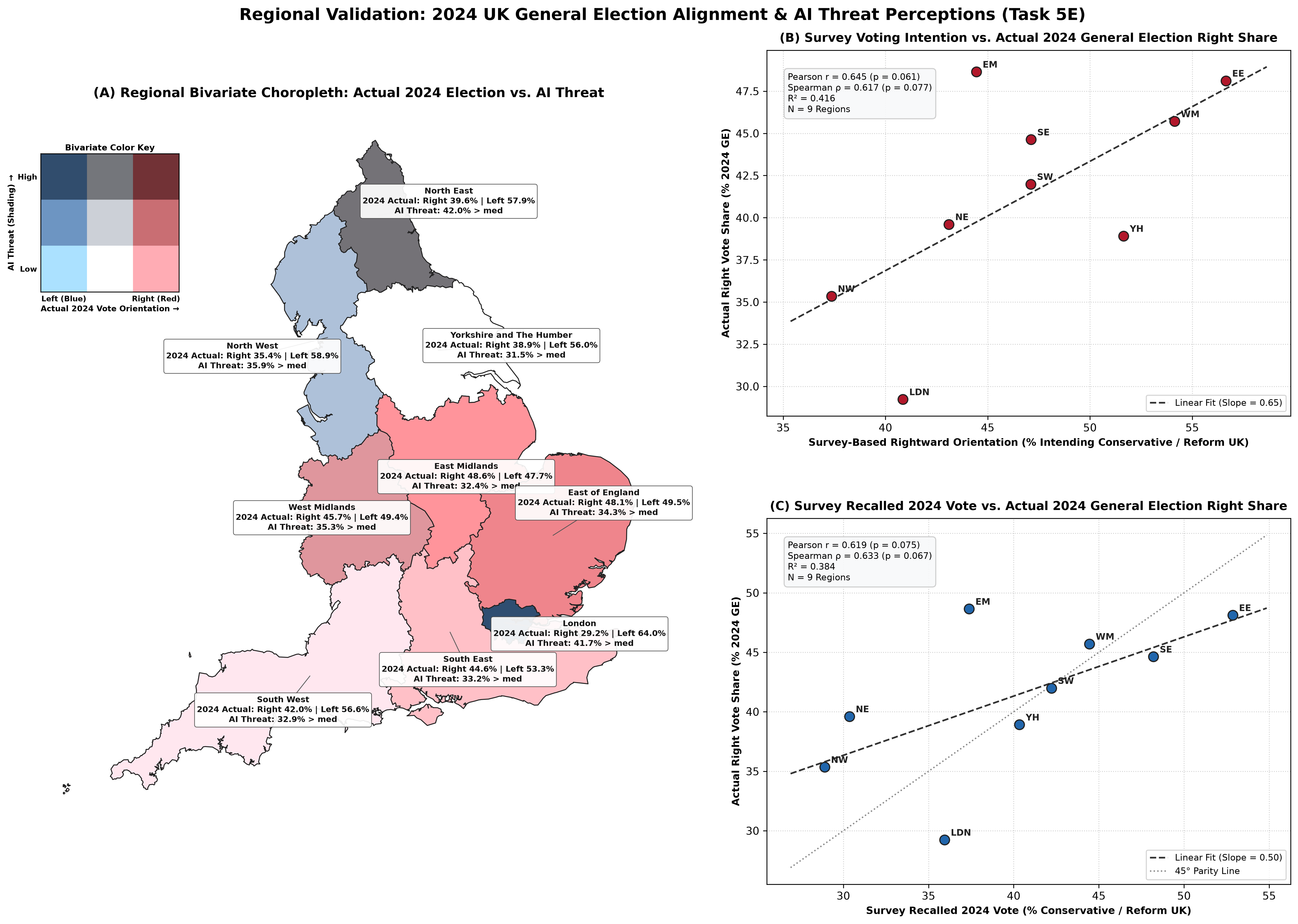}
    \caption{\textbf{Regional Validation: 2024 UK General Election Alignment and Bivariate Spatial Distribution with AI Threat Perceptions}}
    \label{fig:app_election_validation}
    \vspace{0.10cm}
    \begin{minipage}{\linewidth}
        \footnotesize 
        \textit{Notes:} Panel A presents a bivariate choropleth map across the 9 official Government Office Regions of England. Base color hue represents actual political orientation in the July 2024 General Election ($N = 24,288,102$ total votes cast across England): Blue denotes Left-oriented regions (Labour + Liberal Democrats + Green Party), Red denotes Right-oriented regions (Conservative + Reform UK). Color saturation/intensity represents the proportion of survey respondents ($N = 1,585$) scoring above the national median on subjective AI job replacement threat. Panels B and C benchmark regional survey political preferences against official election returns: Panel B compares regional voting intention for Right parties against actual election Right vote share ($r = 0.645$, $p = 0.06$); Panel C compares recalled 2024 vote choice against actual election returns ($r = 0.619$, $p = 0.08$) alongside a 45$^\circ$ parity line.
    \end{minipage}
\end{figure}

This appendix presents additional variable definitions, descriptive statistics, robustness checks, and figures to support the empirical analyses reported in the main text. Table~\ref{tab:variable_definitions} summarizes the definitions, coding rules, and sample coverage of all variables analyzed in the paper. Tables~\ref{tab:app_reservation_descriptives} and~\ref{tab:app_censoring_regressions} offer additional reservation income summary statistics, as well as right-censored refusal rates, and censoring-adjusted regression bounds. This is complemented by Figures~\ref{fig:app_reservation_curve_regions}--\ref{fig:app_reemployment_confidence}. Table~\ref{tab:app_profile_validation} reports the coefficients from the linear probability models that predict Work Values Profiles. It is supported by Figure~\ref{fig:app_important_to_identity} which looks at the importance of personal identity across life domains. Table~\ref{tab:app_industry_harm_politics} reports political attitude regressions using perceptions of AI's harm to jobs in a person's industry. Table~\ref{tab:political_regressions_tier} reports the results of this paper's model of political attitudes across white-collar versus manual and service occupations. Table~\ref{tab:political_age_split} reports the age-split political attitude models ($\le 41$ vs.\ $> 41$). Tables~\ref{tab:app_income_subgroups} and~\ref{tab:app_occ_subgroups} present the full subgroup regression estimates across household income terciles and occupational categories. It is complemented by Figures~\ref{fig:app_income_subgroups} and~\ref{fig:app_threat_by_occ_status}. Table~\ref{tab:isco_granular_reg} reports the full granular ISCO-08 interaction models. Figure~\ref{fig:app_election_validation} offers up external validation against actual 2024 UK General Election returns. And Table~\ref{tab:app_threshold_sensitivity} tests sensitivity across alternative Work Values Profile classification cutoffs ($< 2.0$ to $\le 4.0$). 

\begin{table}[!htbp]
    \centering
    \caption{\textbf{Survey Variables, Coding Rules, and Sample}}
    \label{tab:variable_definitions}
    \scriptsize
    \setlength{\tabcolsep}{0pt}
    \renewcommand{\arraystretch}{1.12}
    \begin{tabular*}{\textwidth}{@{\extracolsep{\fill}}>{\raggedright\arraybackslash}p{3.4cm}>{\raggedright\arraybackslash}p{5.3cm}>{\raggedright\arraybackslash}p{2.6cm}c@{}}
        \toprule
        \textbf{Variable / Index} & 
        \textbf{Survey Question \& Coding Rule} & 
        \textbf{Scale / Metric} & 
        \textbf{Coverage ($N$)} \\
        \midrule
        \multicolumn{4}{@{}l}{\textbf{Panel A: Core Independent Variables \& Moderators}} \\
        \addlinespace[2pt]
        Subjective AI Job Threat & Likelihood AI replaces respondent's job in 10 yrs (\texttt{ai\_replace\_job\_10years}); reverse-coded ($8 - x$) and standardized across sample. & Continuous $z$-score ($\mu=0, \sigma=1$); 1--7 raw & 4 Countries ($6,923$) \\
        \addlinespace[2.5pt]
        Work Values Profiles ($M, I_W, C_W, S_W$) & Mutually exclusive dominant domain on 1--4 job importance battery; ``Other'' ($N=2,010$) excluded. & 4 categorical indicators (Sum $= 1$) & 4 Countries ($6,224$) \\
        \addlinespace[2.5pt]
        White-Collar Occupational Tier & Binary indicator from 10-category employment item (\texttt{employment1}): $1$ = professional/technical, managerial/admin., clerical, sales, farm proprietors; $0$ = service and manual roles. & Binary (0 / 1) & 4 Countries ($7,106$) \\
        \midrule
        \multicolumn{4}{@{}l}{\textbf{Panel B: Behavioral Labor Supply Outcomes (Section \ref{sec:results_reservation})}} \\
        \addlinespace[2pt]
        Reservation Income ($T^*$) & Minimum guaranteed basic income (\% of current salary, benefits preserved) to quit job. & Continuous \% ($60\%$ to $120\%$) & 4 Countries ($6,566$ / $4,860^{\text{a}}$) \\
        \addlinespace[2.5pt]
        Refusal to Quit ($T^* > 120\%$) & Indicator $= 1$ if respondent refuses all basic income offers up to $120\%$ of salary. & Binary (0 / 1) & 4 Countries ($8,234$ / $6,224^{\text{a}}$) \\
        \addlinespace[2.5pt]
        Post-Displacement Time Use & Planned activities if made redundant with basic income; grouped into Leisure, Civic/Care, Continued Work. & Binary (0 / 1) per category & 4 Countries ($8,234$ / $6,224^{\text{a}}$) \\
        \midrule
        \multicolumn{4}{@{}l}{\textbf{Panel C: Political \& Ideological Outcomes (Section \ref{sec:results_political})}} \\
        \addlinespace[2pt]
        Perceived Negative AI Political Impact & Single-item expected impact of AI on national politics over next 10 yrs (\texttt{ai\_pos\_neg\_country\_1}, \texttt{ai\_pol\_neg}); full-sample $z$-score, higher = more negative. & Continuous $z$-score (1--5 raw) & 4 Countries ($6,487$ / $5,688^{\text{a}}$) \\
        \addlinespace[2.5pt]
        Right-Wing / Populist Party Vote & Binary indicator $= 1$ for national right-wing or populist party vote intention (\texttt{vi\_*}). & Binary (0 / 1); Logit & 4 Countries ($6,018$ / $5,295^{\text{a}}$) \\
        \addlinespace[2.5pt]
        Economic Right Ideology & Standardized mean of 4 economic redistribution/corporate regulation grids; higher = Right. & Continuous $z$-score (1--5 raw) & UK Module ($1,667$ / $1,498^{\text{a}}$) \\
        \addlinespace[2.5pt]
        Political System Riggedness$^{\text{b}}$ & Agreement that ``the economic and political system is rigged to serve the rich and powerful''; higher = more rigged. & Continuous $z$-score (1--5 raw) & UK Module ($1,667$ / $1,498^{\text{a}}$) \\
        \addlinespace[2.5pt]
        Institutional Chaos Preference$^{\text{b}}$ & Agreement with ``burning down'' or dismantling broken political institutions; higher = greater chaos preference. & Continuous $z$-score (1--5 raw) & UK Module ($1,667$ / $1,498^{\text{a}}$) \\
        \midrule
        \multicolumn{4}{@{}l}{\textbf{Panel D: Demographic Covariates \& Fixed Effects}} \\
        \addlinespace[2pt]
        Age, Female, Higher Educ., Part-Time, Income Terciles, Country FEs & Age (years); Female ($0/1$); Bachelor's+ ($0/1$); Part-Time ($0/1$); Low/Mid/High Household Income Terciles; Country indicators (DK, IN, UK, US). & Continuous / Binary indicators & 4 Countries ($8,234$) \\
        \bottomrule
    \end{tabular*}
    \par\vspace{0.12cm}
    \begin{minipage}{\textwidth}
        \scriptsize
        \textit{Notes:} $^{\text{a}}$ First number reports the total sample regression coverage; second number reports coverage among classified Work Values Profile respondents ($N = 6,224$, ``Other'' excluded). $^{\text{b}}$ Supplementary single-country attitudinal batteries fielded in the United Kingdom survey module.
    \end{minipage}
\end{table}

\begin{table}[!htbp]
    \centering
    \caption{\textbf{Reservation Income, Refusal Rates, and Robustness Checks}}
    \label{tab:app_reservation_descriptives}
    \renewcommand{\arraystretch}{1.08}
    \resizebox{\textwidth}{!}{%
        \begin{tabular}{@{}lccccccc@{}}
            \toprule
            \textbf{Subgroup / Estimator} &
            \textbf{Total $N$} &
            \textbf{Uncens.\ $N$} &
            \textbf{Refused at 120\%} &
            \textbf{Uncens.\ Mean} &
            \textbf{Uncens.\ Median} &
            \textbf{Contrast vs.\ Trans.} &
            \textbf{Contrast vs.\ Status} \\
            \midrule
            \multicolumn{8}{@{}l}{\textbf{Panel A: Uncensored Intensive Margin \& Extensive-Margin Refusal Rates ($T^* > 120\%$)}} \\
            \quad All Surveyed Respondents            & 8,234 & 6,566 & 1,668 (20.26\%) & 75.43\% & 70.0\% & --- & --- \\
            \quad Intrinsic Profile                   & 2,067 & 1,585 & 482 (23.32\%)   & 77.90\% & 80.0\% & $+9.14\text{ pp}^{***}$ (Refusal) & $+6.78\text{ pp}^{***}$ (Refusal) \\
            \quad Community Profile                   & 1,439 & 1,194 & 245 (17.03\%)   & 75.89\% & 70.0\% & $+2.85\text{ pp}^{**}$ (Refusal)  & $+0.49\text{ pp}$ (Refusal) \\
            \quad Transactional Profile               & 1,636 & 1,404 & 232 (14.18\%)   & 76.55\% & 70.0\% & [Reference] & $-2.36\text{ pp}^{*}$ (Refusal) \\
            \quad Status-Driven Profile               & 1,082 & 903   & 179 (16.54\%)   & 72.78\% & 60.0\% & $+2.36\text{ pp}^{*}$ (Refusal)   & [Reference] \\
            \quad Unclassified / Other (Excluded)     & 2,010 & 1,480 & 530 (26.37\%)   & 73.76\% & 70.0\% & --- & --- \\
            \addlinespace[2pt]
            \quad \textit{By Occupational Category:} & & & & & & & \\
            \quad \quad White-Collar Occupational Tier & 5,151 & 4,058 & 1,093 (21.22\%) & 75.92\% & 70.0\% & $+4.29\text{ pp}^{***}$ vs.\ Manual & --- \\
            \quad \quad Manual \& Service Occupational Tier & 1,955 & 1,624 & 331 (16.93\%) & 75.74\% & 70.0\% & [Reference] & --- \\
            \quad \quad Not in Workforce              & 1,128 & 884   & 244 (21.63\%)   & 72.58\% & 60.0\% & --- & --- \\
            \addlinespace[2pt]
            \quad \textit{By Country (Total $N = 8,234$):} & & & & & & & \\
            \quad \quad India ($N = 2,072$)           & 2,072 & 1,477 & 595 (28.72\%)   & ---     & ---    & --- & --- \\
            \quad \quad Denmark ($N = 2,032$)         & 2,032 & 1,666 & 366 (18.01\%)   & ---     & ---    & --- & --- \\
            \quad \quad United States ($N = 2,078$)   & 2,078 & 1,704 & 374 (18.00\%)   & ---     & ---    & --- & --- \\
            \quad \quad United Kingdom ($N = 2,052$)  & 2,052 & 1,718 & 334 (16.28\%)   & ---     & ---    & --- & --- \\
            \midrule
            \multicolumn{8}{@{}l}{\textbf{Panel B: Alternative Estimators of the Intrinsic Reservation Premium ($N = 4,860\text{ to }6,224$)}} \\
            \textbf{Estimation Specification} & \textbf{Intrinsic} & \textbf{Community} & \textbf{Transactional} & \textbf{Status-Driven} & \textbf{Int.\ $-$ Trans.} & \textbf{SE ($p$-value)} & \textbf{Int.\ $-$ Status} \\
            \midrule
            \quad 1. Truncated OLS at Mean Threat ($z = 0$)             & 77.64\% & 76.00\% & 75.79\% & 73.62\% & $+1.85\text{ pp}^{***}$ & 0.67 ($p < 0.01$) & $+4.02\text{ pp}^{***}$ \\
            \quad 2. Truncated OLS at Low Threat ($z = -1\text{ SD}$)   & 81.08\% & 79.99\% & 77.88\% & 76.26\% & $+3.20\text{ pp}^{***}$ & 0.88 ($p < 0.01$) & $+4.82\text{ pp}^{***}$ \\
            \quad 3. Truncated OLS at Zero Threat ($z = -1.35\text{ SD}$)& 82.28\% & 81.38\% & 78.61\% & 77.18\% & $+3.67\text{ pp}^{***}$ & 1.05 ($p < 0.01$) & $+5.10\text{ pp}^{***}$ \\
            \quad 4. Older ($>41$) at Mean Threat ($z = 0$)             & 78.96\% & 77.43\% & 76.10\% & 75.05\% & $+2.86\text{ pp}^{***}$ & 0.91 ($p < 0.01$) & $+3.91\text{ pp}^{***}$ \\
            \quad 5. Older ($>41$) at Low Threat ($z = -1\text{ SD}$)   & 81.68\% & 81.88\% & 77.50\% & 78.17\% & $+4.18\text{ pp}^{***}$ & 1.27 ($p < 0.01$) & $+3.51\text{ pp}^{*}$   \\
            \quad 6. Full Sample Top-Coded ($T_{\text{cens}} = 120\%$, Unadj.) & 87.81\% & 83.44\% & 82.45\% & 80.27\% & $+5.35\text{ pp}^{***}$ & 0.78 ($p < 0.01$) & $+7.54\text{ pp}^{***}$ \\
            \quad 7. Full Sample Top-Coded ($T_{\text{cens}} = 120\%$, Adj.)   & 87.13\% & 83.84\% & 82.55\% & 81.14\% & $+4.58\text{ pp}^{***}$ & 0.79 ($p < 0.01$) & $+5.99\text{ pp}^{***}$ \\
            \quad 8. Next Auction Step Bound ($T_{\text{cens}} = 130\%$, Adj.) & 89.38\% & 85.63\% & 84.07\% & 82.75\% & $+5.30\text{ pp}^{***}$ & 0.89 ($p < 0.01$) & $+6.62\text{ pp}^{***}$ \\
            \quad 9. Moderate Top-Coded Bound ($T_{\text{cens}} = 150\%$, Adj.)& 93.87\% & 89.21\% & 87.12\% & 85.98\% & $+6.75\text{ pp}^{***}$ & 1.11 ($p < 0.01$) & $+7.89\text{ pp}^{***}$ \\
            \quad 10. High Top-Coded Bound ($T_{\text{cens}} = 200\%$, Adj.)   & 105.10\%& 98.17\% & 94.73\% & 94.05\% & $+10.36\text{ pp}^{***}$& 1.71 ($p < 0.01$) & $+11.04\text{ pp}^{***}$\\
            \quad 11. Right-Censored Tobit MLE ($T^* > 120\%$, Adj.)           & 90.30\% & 86.27\% & 84.54\% & 83.20\% & $+5.77\text{ pp}^{***}$ & 0.95 ($p < 0.01$) & $+7.10\text{ pp}^{***}$ \\
            \bottomrule
        \end{tabular}%
    }
    \par\vspace{0.12cm}
    \begin{minipage}{\textwidth}
        \scriptsize
        \textit{Notes:} $^{*} p < 0.10$, $^{**} p < 0.05$, $^{***} p < 0.01$. Panel A analyzes the total profile counts and occupational category counts by decomposing them into uncensored respondents that name a finite reservation income ($N = 6{,}566$, overall $\text{SD} = 18.42$) and right-censored respondents that refused all offers up to $120\%$ of salary ($N = 1{,}668$).  Panel B offers alternative estimators of the reservation wage across profiles: Rows 1--5 evaluate the covariate-adjusted cell-means regression (Model 3, Table~\ref{tab:reservation_regressions} and Table~\ref{tab:reservation_age_split}) across levels of Subjective AI Threat ($z$); Rows 6--10 show right-censored observations across the full classified sample ($N = 5{,}947$ with complete covariates) via top-coded bounds $T_{\text{cens}} \in \{120\%, 130\%, 150\%, 200\%\}$; Row 11 estimates a Right-Censored Tobit model via maximum likelihood ($\hat{\sigma} = 27.09$) treating refusers as right-censored at $120\%$. Full cell-means regression estimates across all censoring-adjusted specifications are reported in Table~\ref{tab:app_censoring_regressions} and visualized in Figure~\ref{fig:app_censoring_robustness}.
    \end{minipage}
\end{table}

\begin{table}[!htbp]
    \centering
    \caption{\textbf{Censoring-Adjusted Regression Models of Reservation Income}}
    \label{tab:app_censoring_regressions}
    \renewcommand{\arraystretch}{1.08}
    \resizebox{\textwidth}{!}{%
        \begin{tabular}{@{}lcccccc@{}}
            \toprule
            \textbf{Variable / Contrast} &
            \textbf{(1) Trunc.\ OLS} &
            \textbf{(2) Refused LPM} &
            \textbf{(3) Bound 120\%} &
            \textbf{(4) Step 130\%} &
            \textbf{(5) Bound 200\%} &
            \textbf{(6) Tobit MLE} \\
            & ($T^* \le 120\%$) & ($\Pr(T^* > 120\%)$) & ($T_{\text{cens}} = 120\%$) & ($T_{\text{cens}} = 130\%$) & ($T_{\text{cens}} = 200\%$) & (Cens.\ $>120\%$) \\
            \midrule
            \multicolumn{7}{@{}l}{\textit{Panel A: Adjusted profile means (at covariate means)}} \\
            \quad Transactional Profile                & 75.79 (0.49)       & 15.23 (1.00)       & 82.55 (0.60)       & 84.07 (0.67)       & 94.73 (1.30)       & 84.54 (0.72)       \\
            \quad Intrinsic Profile                    & 77.64 (0.45)       & 22.46 (0.86)       & 87.13 (0.52)       & 89.38 (0.58)       & 105.10 (1.12)      & 90.30 (0.63)       \\
            \quad Community Profile                    & 76.00 (0.52)       & 17.91 (1.03)       & 83.84 (0.62)       & 85.63 (0.70)       & 98.17 (1.34)       & 86.27 (0.75)       \\
            \quad Status-Driven Profile                & 73.62 (0.65)       & 16.14 (1.29)       & 81.14 (0.77)       & 82.75 (0.87)       & 94.05 (1.67)       & 83.20 (0.93)       \\
            \midrule
            \multicolumn{7}{@{}l}{\textit{Panel B: AI threat slope within profile}} \\
            \quad AI Threat $\mid$ Transactional       & $-$2.09*** (0.52)  & $-$2.11** (1.04)   & $-$2.67*** (0.63)  & $-$2.88*** (0.70)  & $-$4.36*** (1.36)  & $-$3.04*** (0.75)  \\
            \quad AI Threat $\mid$ Intrinsic           & $-$3.44*** (0.46)  & $-$4.46*** (0.89)  & $-$4.53*** (0.53)  & $-$4.98*** (0.60)  & $-$8.10*** (1.15)  & $-$5.39*** (0.65)  \\
            \quad AI Threat $\mid$ Community           & $-$3.99*** (0.53)  & $-$6.20*** (1.06)  & $-$6.00*** (0.63)  & $-$6.62*** (0.72)  & $-$10.96*** (1.38) & $-$7.06*** (0.77)  \\
            \quad AI Threat $\mid$ Status-Driven       & $-$2.64*** (0.63)  & $-$4.22*** (1.25)  & $-$4.11*** (0.75)  & $-$4.54*** (0.84)  & $-$7.49*** (1.62)  & $-$4.81*** (0.90)  \\
            \midrule
            \multicolumn{7}{@{}l}{\textit{Panel C: Direct contrasts in levels}} \\
            \quad Transactional $-$ Intrinsic          & $-$1.85*** (0.67)  & $-$7.23*** (1.31)  & $-$4.58*** (0.79)  & $-$5.30*** (0.89)  & $-$10.36*** (1.71) & $-$5.77*** (0.95)  \\
            \quad Transactional $-$ Community          & $-$0.21 (0.72)     & $-$2.68* (1.44)    & $-$1.29 (0.86)     & $-$1.56 (0.97)     & $-$3.43* (1.87)    & $-$1.74* (1.04)    \\
            \quad Transactional $-$ Status-Driven      & 2.17*** (0.82)     & $-$0.91 (1.65)     & 1.41 (0.99)        & 1.32 (1.11)        & 0.68 (2.14)        & 1.34 (1.19)        \\
            \quad Transactional $-$ Non-Transactional  & $-$0.32 (0.58)     & $-$4.31*** (1.16)  & $-$2.13*** (0.70)  & $-$2.56*** (0.79)  & $-$5.58*** (1.52)  & $-$2.82*** (0.84)  \\
            \midrule
            \multicolumn{7}{@{}l}{\textit{Panel D: Direct contrasts in AI threat slopes}} \\
            \quad Transactional $-$ Intrinsic (slope)  & 1.35** (0.68)      & 2.35* (1.35)       & 1.86** (0.81)      & 2.09** (0.91)      & 3.74** (1.75)      & 2.35** (0.98)      \\
            \quad Transactional $-$ Community (slope)  & 1.90*** (0.73)     & 4.09*** (1.46)     & 3.33*** (0.88)     & 3.74*** (0.99)     & 6.60*** (1.90)     & 4.02*** (1.06)     \\
            \quad Transactional $-$ Status-Driven (slope)& 0.55 (0.81)      & 2.11 (1.61)        & 1.44 (0.97)        & 1.65 (1.09)        & 3.13 (2.10)        & 1.77 (1.17)        \\
            \quad Transactional $-$ Non-Trans.\ (slope)& 1.33** (0.59)      & 2.83** (1.18)      & 2.22*** (0.71)     & 2.50*** (0.80)     & 4.48*** (1.54)     & 2.73*** (0.86)     \\
            \midrule
            Demographic \& Income Controls             & Yes                & Yes                & Yes                & Yes                & Yes                & Yes                \\
            Country Fixed Effects                      & Yes (4 FEs)        & Yes (4 FEs)        & Yes (4 FEs)        & Yes (4 FEs)        & Yes (4 FEs)        & Yes (4 FEs)        \\
            Observations ($N$)                         & 4,860              & 5,947              & 5,947              & 5,947              & 5,947              & 5,947              \\
            $R^2$ / $\hat{\sigma}$                     & 0.05               & 0.04               & 0.06               & 0.06               & 0.05               & $\hat{\sigma}=27.09$\\
            \bottomrule
        \end{tabular}%
    }
    \par\vspace{0.12cm}
    \begin{minipage}{\textwidth}
        \scriptsize
        \textit{Notes:} * $p < 0.10$, ** $p < 0.05$, *** $p < 0.01$. Standard errors in parentheses. All models include the full Model 3 conditioning set of Table~\ref{tab:reservation_regressions} (mean-centered age, female gender, higher education, part-time status, household income terciles, white-collar occupational tier $\times$ AI threat, and country fixed effects). Column 1 shows the baseline Truncated OLS model ($T^* \le 120\%, N = 4,860$). Column 2 offers an estimate of the extensive-margin Linear Probability Model (in percentage points, $0$ to $100$) for refusing to exit employment even at $120\%$ of current salary ($T^* > 120\%, N = 5,947$). Columns 3--5 estimate full-sample OLS models top-coding right-censored refusers at $T_{\text{cens}} = 120\%$ (strict lower bound), $130\%$ (next discrete auction step), and $200\%$ ($N = 5,947$). Column 6 estimates a parametric Right-Censored Tobit model via maximum likelihood treating refusers ($N = 1,087$) as right-censored above $120\%$.
    \end{minipage}
\end{table}

\begin{table}[!htbp]
    \centering
    \caption{\textbf{Predicting Work Values Profiles by Demographics and Politics}}
    \label{tab:app_profile_validation}
    \renewcommand{\arraystretch}{1.08}
    \resizebox{\textwidth}{!}{%
        \begin{tabular}{@{}lcccccccc@{}}
            \toprule
            & \multicolumn{2}{c}{\textbf{Transactional}} & \multicolumn{2}{c}{\textbf{Status-Driven}} & \multicolumn{2}{c}{\textbf{Intrinsic}} & \multicolumn{2}{c}{\textbf{Community}} \\
            \cmidrule(lr){2-3} \cmidrule(lr){4-5} \cmidrule(lr){6-7} \cmidrule(lr){8-9}
            \textbf{Predictor} & \textbf{(1) Demog} & \textbf{(2) + Polit} & \textbf{(3) Demog} & \textbf{(4) + Polit} & \textbf{(5) Demog} & \textbf{(6) + Polit} & \textbf{(7) Demog} & \textbf{(8) + Polit} \\
            \midrule
            White-Collar Occupational Tier & 0.000 (0.014) & 0.005 (0.015) & 0.002 (0.012) & 0.001 (0.013) & 0.025* (0.015) & 0.029* (0.016) & $-$0.028** (0.014) & $-$0.035** (0.015) \\
            Middle Income                 & 0.021 (0.015) & 0.011 (0.016) & $-$0.025** (0.013) & $-$0.027* (0.014) & $-$0.002 (0.016) & 0.008 (0.018) & 0.007 (0.015) & 0.008 (0.016) \\
            High Income                   & $-$0.001 (0.017) & $-$0.006 (0.018) & 0.018 (0.015) & 0.016 (0.016) & $-$0.018 (0.019) & $-$0.006 (0.020) & 0.001 (0.016) & $-$0.004 (0.018) \\
            Higher Education (Bachelor+)  & $-$0.028** (0.013) & $-$0.028* (0.014) & $-$0.011 (0.011) & $-$0.017 (0.012) & 0.060*** (0.014) & 0.057*** (0.015) & $-$0.020 (0.013) & $-$0.012 (0.013) \\
            Part-Time Employment          & $-$0.056*** (0.014) & $-$0.055*** (0.015) & 0.007 (0.013) & $-$0.005 (0.014) & 0.009 (0.016) & 0.014 (0.017) & 0.041*** (0.015) & 0.046*** (0.016) \\
            Age                           & 0.002*** (0.000) & 0.002*** (0.000) & $-$0.003*** (0.000) & $-$0.003*** (0.000) & 0.000 (0.000) & 0.000 (0.000) & 0.001 (0.000) & 0.001 (0.000) \\
            Female                        & 0.007 (0.011) & 0.010 (0.012) & $-$0.016 (0.010) & $-$0.012 (0.011) & 0.021* (0.013) & 0.019 (0.013) & $-$0.013 (0.011) & $-$0.017 (0.012) \\
            Right-Wing / Populist Vote    &               & 0.002 (0.012) &               & 0.002 (0.011) &               & 0.000 (0.013) &               & $-$0.004 (0.012) \\
            Denmark Fixed Effect          & 0.123*** (0.023) & 0.115*** (0.025) & 0.352*** (0.021) & 0.370*** (0.024) & 0.246*** (0.026) & 0.227*** (0.029) & 0.279*** (0.023) & 0.289*** (0.026) \\
            India Fixed Effect            & 0.104*** (0.023) & 0.102*** (0.026) & 0.385*** (0.022) & 0.406*** (0.025) & 0.276*** (0.027) & 0.250*** (0.030) & 0.235*** (0.023) & 0.242*** (0.027) \\
            UK Fixed Effect               & 0.248*** (0.025) & 0.240*** (0.027) & 0.244*** (0.021) & 0.265*** (0.024) & 0.280*** (0.028) & 0.263*** (0.031) & 0.228*** (0.025) & 0.232*** (0.027) \\
            US Fixed Effect               & 0.301*** (0.026) & 0.302*** (0.029) & 0.247*** (0.022) & 0.272*** (0.025) & 0.271*** (0.029) & 0.242*** (0.031) & 0.181*** (0.025) & 0.184*** (0.028) \\
            \midrule
            Observations ($N$)            & 6,082 & 5,391 & 6,082 & 5,391 & 6,082 & 5,391 & 6,082 & 5,391 \\
            $R^2$                         & 0.050 & 0.051 & 0.047 & 0.047 & 0.007 & 0.007 & 0.011 & 0.013 \\
            \bottomrule
        \end{tabular}%
    }
    \par\vspace{0.12cm}
    \begin{minipage}{\textwidth}
        \scriptsize
        \textit{Notes:} Linear probability models (OLS with heteroskedasticity-robust standard errors in parentheses). Dependent variables are mutually exclusive binary indicators for each work values profile among classified active-workforce respondents (``Other'' excluded), so the four outcomes sum to one. * $p < 0.10$, ** $p < 0.05$, *** $p < 0.01$.
    \end{minipage}
\end{table}

\begin{table}[!htbp]
    \centering
    \begin{threeparttable}
        \caption{\textbf{Political Attitudes Regressed on Perceived AI Harm to Industry Jobs by Work Values Profile (Cell-Means Specification)}}
        \label{tab:app_industry_harm_politics}
        \scriptsize
        \setlength{\tabcolsep}{0pt}
        \renewcommand{\arraystretch}{1.05}
        \begin{tabular*}{\textwidth}{@{\extracolsep{\fill}}lccccc@{}}
            \toprule
            \textbf{Variable / Contrast} &
            \textbf{(1) AI Imp.} &
            \textbf{(2) Rigged} &
            \textbf{(3) Chaos} &
            \textbf{(4) Pop.\ Vote} &
            \textbf{(5) Econ.\ R} \\
            & \textit{(4 Countries)} & \textit{(UK Only)} & \textit{(UK Only)} & \textit{(4 Countries)} & \textit{(UK Only)} \\
            \midrule
            \multicolumn{6}{@{}l}{\textit{Panel A: Adjusted profile means (at mean perceived industry job harm)}} \\
            \quad Transactional Profile                & 0.09 (0.02)       & 0.03 (0.05)       & 0.03 (0.05)       & 0.07 (0.06)       & $-$0.07 (0.05)    \\
            \quad Intrinsic Profile                    & 0.03 (0.02)       & 0.01 (0.04)       & $-$0.09 (0.04)    & 0.03 (0.05)       & 0.01 (0.04)       \\
            \quad Community Profile                    & 0.01 (0.02)       & 0.02 (0.05)       & 0.04 (0.05)       & 0.02 (0.06)       & 0.06 (0.05)       \\
            \quad Status-Driven Profile                & $-$0.18 (0.03)    & $-$0.05 (0.08)    & 0.04 (0.08)       & 0.02 (0.07)       & 0.06 (0.08)       \\
            \midrule
            \multicolumn{6}{@{}l}{\textit{Panel B: Perceived AI industry job harm slope within profile ($z$-score)}} \\
            \quad Industry Harm $\mid$ Transactional   & 0.41*** (0.02)    & 0.21*** (0.05)    & 0.10** (0.05)     & $-$0.06 (0.06)    & $-$0.13*** (0.05) \\
            \quad Industry Harm $\mid$ Intrinsic       & 0.54*** (0.02)    & 0.23*** (0.04)    & 0.14*** (0.04)    & $-$0.20*** (0.05) & $-$0.14*** (0.04) \\
            \quad Industry Harm $\mid$ Community       & 0.51*** (0.02)    & 0.15*** (0.05)    & 0.09* (0.06)      & $-$0.05 (0.06)    & $-$0.07 (0.06)    \\
            \quad Industry Harm $\mid$ Status-Driven   & 0.48*** (0.03)    & 0.30*** (0.07)    & 0.00 (0.08)       & $-$0.14** (0.07)  & $-$0.10 (0.08)    \\
            \midrule
            \multicolumn{6}{@{}l}{\textit{Panel C: Direct contrasts in levels}} \\
            \quad Transactional $-$ Intrinsic          & 0.06** (0.03)     & 0.02 (0.06)       & 0.12* (0.06)      & 0.04 (0.08)       & $-$0.08 (0.06)    \\
            \quad Transactional $-$ Community          & 0.08** (0.03)     & 0.01 (0.07)       & $-$0.02 (0.07)    & 0.05 (0.08)       & $-$0.13* (0.07)   \\
            \quad Transactional $-$ Status-Driven      & 0.28*** (0.03)    & 0.08 (0.09)       & $-$0.01 (0.09)    & 0.05 (0.09)       & $-$0.13 (0.09)    \\
            \quad Transactional $-$ Non-Transactional  & 0.12*** (0.03)    & 0.02 (0.05)       & 0.05 (0.06)       & 0.05 (0.07)       & $-$0.10* (0.06)   \\
            \midrule
            \multicolumn{6}{@{}l}{\textit{Panel D: Direct contrasts in industry job harm slopes}} \\
            \quad Trans.\ $-$ Intrinsic (slope)        & $-$0.12*** (0.03) & $-$0.02 (0.06)    & $-$0.04 (0.06)    & 0.14* (0.08)      & 0.01 (0.06)       \\
            \quad Trans.\ $-$ Community (slope)        & $-$0.10*** (0.03) & 0.06 (0.07)       & 0.01 (0.07)       & $-$0.01 (0.08)    & $-$0.06 (0.07)    \\
            \quad Trans.\ $-$ Status-Driven (slope)    & $-$0.07** (0.03)  & $-$0.09 (0.09)    & 0.10 (0.09)       & 0.08 (0.09)       & $-$0.03 (0.09)    \\
            \quad Trans.\ $-$ Non-Trans.\ (slope)      & $-$0.10*** (0.03) & $-$0.01 (0.06)    & 0.00 (0.06)       & 0.08 (0.07)       & $-$0.02 (0.06)    \\
            \midrule
            Demographic \& Income Controls             & Yes               & Yes               & Yes               & Yes               & Yes               \\
            Country Fixed Effects                      & Yes (4 FEs)       & No (UK)           & No (UK)           & Yes (4 FEs)       & No (UK)           \\
            Observations ($N$)                         & 5,722             & 1,493             & 1,493             & 5,296             & 1,493             \\
            $R^2$ / Pseudo $R^2$                       & 0.34              & 0.08              & 0.06              & 0.04              & 0.05              \\
            \bottomrule
        \end{tabular*}
        \begin{tablenotes}[flushleft]
            \scriptsize
            \item \textit{Notes:} Standard errors in parentheses. * $p < 0.10$, ** $p < 0.05$, *** $p < 0.01$. Cell-means specification identical to Table~\ref{tab:political_regressions_profiles}, substituting standardized Perceived AI Harm to Industry Jobs (mean of \texttt{ai\_pos\_neg\_3} [job quality in my industry] and \texttt{ai\_pos\_neg\_4} [number of jobs in my industry], coded $1=\text{Very positive}$ to $5=\text{Very negative}$, standardized $z$-score) as the workplace threat measure. Negative slope contrasts ($\text{Transactional} - \text{Non-Transactional} = -0.10^{***}$) indicate that Intrinsic ($+0.54^{***}$), Community ($+0.51^{***}$), and Status-Driven ($+0.48^{***}$) workers exhibit significantly steeper negative political reactions per unit of perceived industry job harm than Transactional workers ($+0.41^{***}$).
        \end{tablenotes}
    \end{threeparttable}
\end{table}

\begin{table}[!htbp]
    \centering
    \begin{threeparttable}
        \caption{\textbf{Regression Models of Political Attitudes on AI Threat Interacted with White-Collar Occupational Category}}
        \label{tab:political_regressions_tier}
        \scriptsize
        \setlength{\tabcolsep}{0pt}
        \renewcommand{\arraystretch}{1.08}
        \begin{tabular*}{\textwidth}{@{\extracolsep{\fill}}lccccc@{}}
            \toprule
            \textbf{Variable} &
            \textbf{(1) AI Imp.} &
            \textbf{(2) Rigged} &
            \textbf{(3) Chaos} &
            \textbf{(4) Pop.\ Vote} &
            \textbf{(5) Econ.\ R} \\
            & \textit{(4 Countries)} & \textit{(UK Only)} & \textit{(UK Only)} & \textit{(4 Countries)} & \textit{(UK Only)} \\
            \midrule
            \multicolumn{6}{@{}l@{}}{\textbf{Main Effects (Omitted: Manual \& Service Tier)}} \\
            \hspace{0.45em}AI Threat ($z$-score)               & $-$0.23*** (0.02) & $-$0.10** (0.05)  & 0.09* (0.05)      & $-$0.03 (0.05)    & 0.06 (0.05)       \\
            \hspace{0.45em}White-Collar Occupational Tier          & 0.00 (0.03)       & $-$0.04 (0.06)    & $-$0.06 (0.06)    & $-$0.07 (0.07)    & 0.07 (0.06)       \\
            \addlinespace
            \multicolumn{6}{@{}l@{}}{\textbf{Interaction: AI Threat $\times$ White-Collar Tier}} \\
            \hspace{0.45em}Threat $\times$ White-Collar Tier       & 0.02 (0.03)       & 0.14** (0.06)     & $-$0.02 (0.06)    & 0.07 (0.06)       & $-$0.20*** (0.06) \\
            \addlinespace
            \multicolumn{6}{@{}l@{}}{\textbf{Demographic Covariates}} \\
            \hspace{0.45em}Age (years)                         & 0.01*** (0.00)    & 0.00 (0.00)       & $-$0.01*** (0.00) & 0.01*** (0.00)    & 0.00* (0.00)      \\
            \hspace{0.45em}Female                              & 0.04* (0.02)      & $-$0.01 (0.05)    & 0.11** (0.05)     & $-$0.17*** (0.06) & 0.00 (0.05)       \\
            \hspace{0.45em}Higher Educ.\ (BA+)                 & $-$0.05* (0.03)   & $-$0.21*** (0.06) & $-$0.29*** (0.06) & $-$0.45*** (0.06) & 0.09* (0.06)      \\
            \hspace{0.45em}Part-Time                           & 0.11*** (0.03)    & $-$0.01 (0.06)    & $-$0.03 (0.06)    & $-$0.17** (0.07)  & $-$0.01 (0.06)    \\
            \hspace{0.45em}Middle Income                       & $-$0.01 (0.03)    & $-$0.05 (0.07)    & $-$0.05 (0.07)    & 0.22*** (0.07)    & 0.06 (0.07)       \\
            \hspace{0.45em}High Income                         & $-$0.22*** (0.04) & $-$0.31*** (0.08) & $-$0.13 (0.08)    & 0.46*** (0.08)    & 0.28*** (0.08)    \\
            \addlinespace
            Constant                                  & Removed\tnote{a}  & 0.17 (0.11)       & 0.46*** (0.11)    & Removed\tnote{a}  & $-$0.34*** (0.11) \\
            \midrule
            Country Fixed Effects                     & Yes (4 FEs)       & No (UK)            & No (UK)            & Yes (4 FEs)        & No (UK)            \\
            Observations ($N$)                        & 6,487             & 1,667              & 1,667              & 6,018              & 1,667              \\
            $R^2$ / Pseudo $R^2$                      & 0.15              & 0.05               & 0.04               & 0.03               & 0.04               \\
            \bottomrule
        \end{tabular*}
        \begin{tablenotes}[flushleft]
            \scriptsize
            \item \textit{Notes:} Standard errors in parentheses. * $p < 0.10$, ** $p < 0.05$, *** $p < 0.01$. Occupational tiers reflect the binary White-Collar vs.\ Manual \& Service classification (\texttt{employment1}) defined in Table~\ref{tab:variable_definitions}. Columns 1, 2, 3, 5: OLS on $z$-scores (in $\text{SD}$ units); Column 4: binary Logit (log-odds). \item[a] Country fixed effects enter directly with the model constant removed.
        \end{tablenotes}
    \end{threeparttable}
\end{table}

\begin{table}[!htbp]
    \centering
    \caption{\textbf{Age-Split Extension: Political Attitudes on AI Threat by Work Values Profile, Younger ($\le 41$) vs.\ Older ($> 41$) Respondents}}
    \label{tab:political_age_split}
    \renewcommand{\arraystretch}{1.08}
    \resizebox{\textwidth}{!}{%
        \begin{tabular}{@{}lcccccccccc@{}}
            \toprule
            & \multicolumn{2}{c}{\textbf{(1) AI Neg.\ Impact}} & \multicolumn{2}{c}{\textbf{(2) System Rigged}} & \multicolumn{2}{c}{\textbf{(3) Inst.\ Chaos}} & \multicolumn{2}{c}{\textbf{(4) Populist Vote}} & \multicolumn{2}{c}{\textbf{(5) Econ.\ Right}} \\
            & \multicolumn{2}{c}{\textit{(4 Countries)}} & \multicolumn{2}{c}{\textit{(UK Only)}} & \multicolumn{2}{c}{\textit{(UK Only)}} & \multicolumn{2}{c}{\textit{(4 Countries)}} & \multicolumn{2}{c}{\textit{(UK Only)}} \\
            \cmidrule(lr){2-3} \cmidrule(lr){4-5} \cmidrule(lr){6-7} \cmidrule(lr){8-9} \cmidrule(lr){10-11}
            \textbf{Variable / Contrast} & $\le 41$ & $> 41$ & $\le 41$ & $> 41$ & $\le 41$ & $> 41$ & $\le 41$ & $> 41$ & $\le 41$ & $> 41$ \\
            \midrule
            \multicolumn{11}{@{}l}{\textit{Panel A: Adjusted profile means (at covariate means)}} \\
            Transactional & 0.02 (0.04) & 0.35 (0.03) & 0.00 (0.08) & 0.12 (0.05) & 0.08 (0.08) & 0.02 (0.05) & $-$0.00 (0.09) & 0.05 (0.08) & $-$0.20 (0.07) & $-$0.05 (0.06) \\
            Intrinsic     & $-$0.17 (0.03) & 0.15 (0.03) & $-$0.07 (0.07) & 0.01 (0.06) & $-$0.04 (0.06) & $-$0.19 (0.06) & 0.17 (0.07) & $-$0.10 (0.07) & $-$0.00 (0.06) & 0.09 (0.07) \\
            Community     & $-$0.15 (0.04) & 0.11 (0.04) & $-$0.05 (0.08) & 0.06 (0.07) & 0.00 (0.08) & 0.08 (0.07) & $-$0.04 (0.08) & 0.11 (0.08) & 0.05 (0.08) & 0.07 (0.07) \\
            Status-Driven & $-$0.29 (0.04) & $-$0.02 (0.05) & $-$0.13 (0.11) & $-$0.05 (0.12) & 0.12 (0.10) & $-$0.08 (0.12) & 0.02 (0.09) & 0.02 (0.12) & 0.04 (0.10) & 0.17 (0.13) \\
            \midrule
            \multicolumn{11}{@{}l}{\textit{Panel B: AI threat slope within profile}} \\
            AI Threat $\mid$ Trans.  & $-$0.17*** (0.04) & 0.02 (0.03) & $-$0.04 (0.08) & 0.07 (0.06) & 0.07 (0.08) & 0.07 (0.06) & 0.13 (0.09) & $-$0.04 (0.08) & 0.03 (0.07) & $-$0.04 (0.07) \\
            AI Threat $\mid$ Intr.   & $-$0.22*** (0.03) & $-$0.13*** (0.03) & $-$0.05 (0.07) & 0.09 (0.07) & $-$0.03 (0.07) & 0.06 (0.07) & 0.08 (0.07) & $-$0.00 (0.08) & 0.00 (0.06) & $-$0.13 (0.08) \\
            AI Threat $\mid$ Comm.   & $-$0.32*** (0.04) & $-$0.11*** (0.04) & $-$0.01 (0.09) & 0.06 (0.07) & 0.11 (0.09) & 0.08 (0.08) & $-$0.01 (0.09) & $-$0.16* (0.09) & $-$0.09 (0.08) & $-$0.18** (0.08) \\
            AI Threat $\mid$ Status  & $-$0.31*** (0.04) & $-$0.30*** (0.05) & $-$0.09 (0.11) & 0.16 (0.13) & $-$0.01 (0.11) & 0.19 (0.13) & 0.12 (0.09) & 0.05 (0.11) & $-$0.09 (0.10) & $-$0.18 (0.14) \\
            \midrule
            \multicolumn{11}{@{}l}{\textit{Panel C: Direct contrasts in levels}} \\
            Trans.\ $-$ Intrinsic   & 0.19*** (0.05) & 0.20*** (0.04) & 0.07 (0.10) & 0.11 (0.08) & 0.12 (0.10) & 0.21** (0.08) & $-$0.17 (0.11) & 0.15 (0.11) & $-$0.20** (0.09) & $-$0.14 (0.09) \\
            Trans.\ $-$ Community   & 0.17*** (0.05) & 0.24*** (0.05) & 0.05 (0.12) & 0.05 (0.09) & 0.08 (0.11) & $-$0.06 (0.09) & 0.04 (0.12) & $-$0.06 (0.11) & $-$0.25** (0.10) & $-$0.12 (0.10) \\
            Trans.\ $-$ Status      & 0.32*** (0.05) & 0.36*** (0.06) & 0.13 (0.13) & 0.16 (0.13) & $-$0.04 (0.13) & 0.09 (0.13) & $-$0.02 (0.12) & 0.03 (0.15) & $-$0.24** (0.12) & $-$0.22 (0.14) \\
            Trans.\ $-$ Non-Trans.  & 0.22*** (0.04) & 0.24*** (0.04) & 0.08 (0.09) & 0.09 (0.07) & 0.07 (0.09) & 0.09 (0.07) & $-$0.06 (0.10) & 0.06 (0.09) & $-$0.22*** (0.08) & $-$0.14* (0.08) \\
            \midrule
            \multicolumn{11}{@{}l}{\textit{Panel D: Direct contrasts in AI threat slopes}} \\
            Trans.\ $-$ Intr.\ (slope)  & 0.05 (0.05) & 0.14*** (0.05) & 0.00 (0.11) & $-$0.02 (0.09) & 0.10 (0.11) & 0.00 (0.09) & 0.04 (0.11) & $-$0.04 (0.11) & 0.03 (0.10) & 0.09 (0.10) \\
            Trans.\ $-$ Comm.\ (slope)  & 0.16*** (0.05) & 0.13*** (0.05) & $-$0.04 (0.12) & 0.01 (0.09) & $-$0.04 (0.12) & $-$0.01 (0.10) & 0.14 (0.12) & 0.12 (0.12) & 0.12 (0.11) & 0.14 (0.11) \\
            Trans.\ $-$ Status (slope)  & 0.14** (0.06) & 0.32*** (0.06) & 0.05 (0.14) & $-$0.09 (0.14) & 0.07 (0.14) & $-$0.13 (0.14) & 0.01 (0.13) & $-$0.08 (0.14) & 0.13 (0.13) & 0.14 (0.15) \\
            Trans.\ $-$ Non-Tr.\ (slope)& 0.11** (0.04) & 0.17*** (0.04) & $-$0.00 (0.10) & $-$0.02 (0.08) & 0.05 (0.09) & $-$0.02 (0.08) & 0.06 (0.10) & 0.01 (0.10) & 0.08 (0.09) & 0.11 (0.08) \\
            \midrule
            Observations ($N$)       & 3,033 & 2,655 & 664 & 834 & 664 & 834 & 2,787 & 2,508 & 664 & 834 \\
            $R^2$ / Pseudo $R^2$     & 0.15  & 0.14  & 0.06 & 0.05 & 0.07 & 0.07 & 0.05  & 0.03  & 0.04 & 0.05 \\
            \bottomrule
        \end{tabular}%
    }
    \par\vspace{0.12cm}
    \begin{minipage}{\textwidth}
        \scriptsize
        \textit{Notes:} *** $p < 0.01$, ** $p < 0.05$, * $p < 0.10$. Standard errors in parentheses. Subgroup estimates split at the median age (41) of the pooled classified sample; each subgroup model re-estimates the full specification of Table~\ref{tab:political_regressions_profiles} within the age group.
    \end{minipage}
\end{table}

\begin{table}[!htbp]
    \centering
    \begin{threeparttable}
        \caption{\textbf{Income Tercile Subgroup Regressions: Perceived Negative AI Political Impact ($z$-score, Cell-Means Specification)}}
        \label{tab:app_income_subgroups}
        \footnotesize
        \setlength{\tabcolsep}{0pt}
        \renewcommand{\arraystretch}{1.05}
        \begin{tabular*}{\textwidth}{@{\extracolsep{\fill}}lccc@{}}
            \toprule
            \textbf{Variable / Contrast} &
            \textbf{(1) Low Income} &
            \textbf{(2) Middle Income} &
            \textbf{(3) High Income} \\
            \midrule
            \multicolumn{4}{@{}l}{\textit{Panel A: Adjusted profile means (at covariate means)}} \\
            \quad Transactional Profile                & 0.31 (0.05)       & 0.22 (0.04)       & 0.04 (0.04)       \\
            \quad Intrinsic Profile                    & 0.05 (0.04)       & 0.08 (0.03)       & $-$0.20 (0.04)    \\
            \quad Community Profile                    & 0.03 (0.05)       & 0.05 (0.04)       & $-$0.16 (0.05)    \\
            \quad Status-Driven Profile                & $-$0.11 (0.06)    & $-$0.04 (0.05)    & $-$0.33 (0.05)    \\
            \midrule
            \multicolumn{4}{@{}l}{\textit{Panel B: AI threat slope within profile}} \\
            \quad AI Threat $\mid$ Transactional       & $-$0.04 (0.06)    & $-$0.05 (0.04)    & $-$0.11** (0.04)  \\
            \quad AI Threat $\mid$ Intrinsic           & $-$0.14*** (0.04) & $-$0.16*** (0.03) & $-$0.24*** (0.04) \\
            \quad AI Threat $\mid$ Community           & $-$0.14*** (0.05) & $-$0.19*** (0.04) & $-$0.31*** (0.05) \\
            \quad AI Threat $\mid$ Status-Driven       & $-$0.18*** (0.06) & $-$0.33*** (0.05) & $-$0.34*** (0.05) \\
            \midrule
            \multicolumn{4}{@{}l}{\textit{Panel C: Direct contrasts in levels}} \\
            \quad Transactional $-$ Intrinsic          & 0.27*** (0.07)    & 0.14*** (0.05)    & 0.25*** (0.06)    \\
            \quad Transactional $-$ Community          & 0.28*** (0.07)    & 0.18*** (0.05)    & 0.20*** (0.06)    \\
            \quad Transactional $-$ Status-Driven      & 0.42*** (0.08)    & 0.26*** (0.06)    & 0.37*** (0.07)    \\
            \quad Transactional $-$ Non-Transactional  & 0.31*** (0.06)    & 0.18*** (0.04)    & 0.27*** (0.05)    \\
            \midrule
            \multicolumn{4}{@{}l}{\textit{Panel D: Direct contrasts in AI threat slopes}} \\
            \quad Transactional $-$ Intrinsic (slope)  & 0.09 (0.07)       & 0.11** (0.05)     & 0.13** (0.06)     \\
            \quad Transactional $-$ Community (slope)  & 0.10 (0.07)       & 0.14*** (0.05)    & 0.20*** (0.06)    \\
            \quad Transactional $-$ Status-Driven (slope)& 0.14* (0.08)    & 0.28*** (0.06)    & 0.23*** (0.06)    \\
            \quad Transactional $-$ Non-Trans.\ (slope)& 0.11* (0.06)      & 0.16*** (0.04)    & 0.17*** (0.05)    \\
            \midrule
            \multicolumn{4}{@{}l}{\textit{Controls (mean-centered)}} \\
            \quad Age                                  & 0.00** (0.00)     & 0.00*** (0.00)    & 0.01*** (0.00)    \\
            \quad Female                               & $-$0.00 (0.05)    & 0.02 (0.04)       & 0.06 (0.04)       \\
            \quad Higher Education                     & 0.06 (0.06)       & $-$0.03 (0.04)    & $-$0.18*** (0.05) \\
            \quad Part-Time Employment                 & 0.17*** (0.05)    & 0.11** (0.05)     & $-$0.07 (0.08)    \\
            \midrule
            Country Fixed Effects                      & Yes               & Yes               & Yes               \\
            Observations ($N$)                         & 1,427             & 2,358             & 1,903             \\
            $R^2$                                      & 0.15              & 0.13              & 0.18              \\
            \bottomrule
        \end{tabular*}
        \begin{tablenotes}[flushleft]
            \scriptsize
            \item \textit{Notes:} OLS cell-means estimates from independent regressions within each household income tercile among classified respondents (``Other'' excluded). Dependent variable is Perceived Negative AI Political Impact ($z$-score, 4 countries). * $p < 0.10$, ** $p < 0.05$, *** $p < 0.01$.
        \end{tablenotes}
    \end{threeparttable}
\end{table}

\begin{table}[!htbp]
    \centering
    \begin{threeparttable}
        \caption{\textbf{Occupational Category Subgroup Regressions: Perceived Negative AI Political Impact ($z$-score, Cell-Means Specification)}}
        \label{tab:app_occ_subgroups}
        \footnotesize
        \setlength{\tabcolsep}{0pt}
        \renewcommand{\arraystretch}{1.05}
        \begin{tabular*}{\textwidth}{@{\extracolsep{\fill}}lcc@{}}
            \toprule
            \textbf{Variable / Contrast} &
            \textbf{(1) Manual \& Service Tier} &
            \textbf{(2) White-Collar Tier} \\
            \midrule
            \multicolumn{3}{@{}l}{\textit{Panel A: Adjusted profile means (at covariate means)}} \\
            \quad Transactional Profile                & 0.25 (0.05)       & 0.16 (0.03)       \\
            \quad Intrinsic Profile                    & 0.09 (0.04)       & $-$0.07 (0.02)    \\
            \quad Community Profile                    & 0.12 (0.05)       & $-$0.09 (0.03)    \\
            \quad Status-Driven Profile                & $-$0.07 (0.06)    & $-$0.19 (0.04)    \\
            \midrule
            \multicolumn{3}{@{}l}{\textit{Panel B: AI threat slope within profile}} \\
            \quad AI Threat $\mid$ Transactional       & $-$0.16*** (0.05) & $-$0.04 (0.03)    \\
            \quad AI Threat $\mid$ Intrinsic           & $-$0.19*** (0.04) & $-$0.18*** (0.03) \\
            \quad AI Threat $\mid$ Community           & $-$0.20*** (0.04) & $-$0.22*** (0.03) \\
            \quad AI Threat $\mid$ Status-Driven       & $-$0.30*** (0.06) & $-$0.30*** (0.04) \\
            \midrule
            \multicolumn{3}{@{}l}{\textit{Panel C: Direct contrasts in levels}} \\
            \quad Transactional $-$ Intrinsic          & 0.16** (0.06)     & 0.23*** (0.04)    \\
            \quad Transactional $-$ Community          & 0.13** (0.06)     & 0.25*** (0.04)    \\
            \quad Transactional $-$ Status-Driven      & 0.32*** (0.08)    & 0.36*** (0.05)    \\
            \quad Transactional $-$ Non-Transactional  & 0.19*** (0.05)    & 0.27*** (0.03)    \\
            \midrule
            \multicolumn{3}{@{}l}{\textit{Panel D: Direct contrasts in AI threat slopes}} \\
            \quad Transactional $-$ Intrinsic (slope)  & 0.03 (0.06)       & 0.14*** (0.04)    \\
            \quad Transactional $-$ Community (slope)  & 0.04 (0.06)       & 0.18*** (0.04)    \\
            \quad Transactional $-$ Status-Driven (slope)& 0.15** (0.07)   & 0.26*** (0.05)    \\
            \quad Transactional $-$ Non-Trans.\ (slope)& 0.06 (0.05)       & 0.18*** (0.03)    \\
            \midrule
            \multicolumn{3}{@{}l}{\textit{Controls (mean-centered)}} \\
            \quad Age                                  & 0.00* (0.00)      & 0.01*** (0.00)    \\
            \quad Female                               & 0.03 (0.05)       & 0.04 (0.03)       \\
            \quad Higher Education                     & $-$0.10* (0.06)   & $-$0.04 (0.03)    \\
            \quad Part-Time Employment                 & 0.13** (0.05)     & 0.09** (0.04)     \\
            \quad Middle Income                        & 0.05 (0.05)       & $-$0.04 (0.04)    \\
            \quad High Income                          & $-$0.07 (0.08)    & $-$0.24*** (0.04) \\
            \midrule
            Country Fixed Effects                      & Yes               & Yes               \\
            Observations ($N$)                         & 1,567             & 4,121             \\
            $R^2$                                      & 0.15              & 0.16              \\
            \bottomrule
        \end{tabular*}
        \begin{tablenotes}[flushleft]
            \scriptsize
            \item \textit{Notes:} OLS cell-means estimates from independent regressions within each occupational category among classified respondents (``Other'' excluded). Occupational tiers reflect the binary White-Collar vs.\ Manual \& Service classification (\texttt{employment1}) defined in Table~\ref{tab:variable_definitions}. * $p < 0.10$, ** $p < 0.05$, *** $p < 0.01$.
        \end{tablenotes}
    \end{threeparttable}
\end{table}

\begin{table}[!htbp]
    \centering
    \setlength{\tabcolsep}{0pt}
    \begin{threeparttable}
        \caption{\textbf{Granular Interaction Regressions of Subjective AI Threat by Major ISCO-08 Occupation Group}}
        \label{tab:isco_granular_reg}
        \label{tab:isco_granular}
        \footnotesize
        \renewcommand{\arraystretch}{1.06}
        \begin{tabular*}{\textwidth}{@{\extracolsep{\fill}}lccc@{}}
            \toprule
            \textbf{ISCO-08 Occupation Group $\times$ AI Threat} &
            \textbf{(1) Simple OLS} &
            \textbf{(2) + Demographics} &
            \textbf{(3) + Country FEs} \\
            \midrule
            AI Threat $\times$ Managers (Group 1)            & $-0.35^{***}$ (0.02) & $-0.32^{***}$ (0.02) & $-0.28^{***}$ (0.02) \\
            AI Threat $\times$ Professionals (Group 2)       & $-0.29^{***}$ (0.02) & $-0.27^{***}$ (0.02) & $-0.24^{***}$ (0.02) \\
            AI Threat $\times$ Technicians (Group 3)         & $-0.26^{***}$ (0.04) & $-0.23^{***}$ (0.04) & $-0.22^{***}$ (0.04) \\
            AI Threat $\times$ Clerical Support (Group 4)    & $-0.12^{***}$ (0.04) & $-0.08^{*}$ (0.04)   & $-0.06$ (0.04)       \\
            AI Threat $\times$ Services \& Sales (Group 5)   & $-0.21^{***}$ (0.03) & $-0.17^{***}$ (0.03) & $-0.16^{***}$ (0.03) \\
            AI Threat $\times$ Skilled Agricultural (Group 6)& $-0.12$ (0.10)       & $-0.10$ (0.10)       & $-0.06$ (0.10)       \\
            AI Threat $\times$ Craft \& Trades (Group 7)     & $-0.33^{***}$ (0.06) & $-0.29^{***}$ (0.06) & $-0.26^{***}$ (0.06) \\
            AI Threat $\times$ Plant \& Machine (Group 8)    & $-0.24^{***}$ (0.06) & $-0.22^{***}$ (0.06) & $-0.19^{***}$ (0.06) \\
            AI Threat $\times$ Elementary (Group 9)          & $-0.28^{***}$ (0.05) & $-0.23^{***}$ (0.05) & $-0.22^{***}$ (0.05) \\
            \midrule
            Demographic Controls                             & No                   & Yes                  & Yes                  \\
            Country Fixed Effects                            & No                   & No                   & Yes                  \\
            Observations ($N$)                               & 6,609                & 6,480                & 6,480                \\
            $R^2$                                            & 0.12                 & 0.13                 & 0.16                 \\
            \bottomrule
        \end{tabular*}
        \begin{tablenotes}[flushleft]
            \scriptsize
            \item \textit{Notes:} OLS regression estimates of full-sample standardized Subjective AI Threat (\texttt{ai\_replace\_job\_10years}, reverse-coded $z$-score) on full-sample standardized Perceived Negative AI Political Impact (\texttt{ai\_pos\_neg\_country\_1} [\texttt{ai\_pol\_neg}], $1 = \text{Very positive}$ to $5 = \text{Very negative}$, $z$-score) interacted with each 1-digit ISCO-08 major occupation group (\texttt{isco\_major} 1--9). Standard errors in parentheses. $^{*} p < 0.10$, $^{**} p < 0.05$, $^{***} p < 0.01$. Because higher values on the dependent variable indicate more negative political evaluations of AI, negative coefficients indicate less negative (more favorable) assessments among workers with higher personal AI exposure: managers ($-0.28^{***}$), craft and trades workers ($-0.26^{***}$), and professionals ($-0.24^{***}$) exhibit the strongest optimistic familiarity offset, whereas clerical support workers (Group 4) show a flat, statistically insignificant slope ($-0.06, p > 0.10$), retaining persistent political pessimism.
        \end{tablenotes}
    \end{threeparttable}
\end{table}

\begin{table}[!htbp]
    \centering
    \begin{threeparttable}
        \caption{\textbf{Sensitivity of Main Work Values Profile Findings to the Classification Threshold}}
        \label{tab:app_threshold_sensitivity}
        \scriptsize
        \setlength{\tabcolsep}{0pt}
        \renewcommand{\arraystretch}{1.08}
        \begin{tabular*}{\textwidth}{@{\extracolsep{\fill}}>{\raggedright\arraybackslash}p{6.5cm}ccccc@{}}
            \toprule
            & \multicolumn{5}{c}{\textbf{Winning Index Threshold for Profile Assignment}} \\
            \cmidrule(lr){2-6}
            \textbf{Specification / Finding} & \textbf{$< 2.0$} & \textbf{$< 2.5$} & \textbf{$< 3.0$ (Baseline)} & \textbf{$< 3.5$} & \textbf{$\le 4.0$} \\
            \midrule
            \multicolumn{6}{@{}l}{\textit{Panel A: Profile Sizes ($N$)}} \\
            Transactional Profile                      & 1,324 & 1,547 & 1,636 & 1,642 & 1,643 \\
            Intrinsic Profile                          & 1,797 & 2,039 & 2,067 & 2,078 & 2,079 \\
            Community Profile                          & 1,140 & 1,384 & 1,439 & 1,453 & 1,454 \\
            Status-Driven Profile                      & 790   & 1,082 & 1,082 & 1,104 & 1,104 \\
            Excluded (unclassified ``Other'')          & 3,183 & 2,182 & 2,010 & 1,957 & 1,954 \\
            \midrule
            \multicolumn{6}{@{}l}{\textit{Panel B: Headline Estimates Across Paper}} \\
            \addlinespace[1.5pt]
            AI threat slope: Trans.\ $-$ Non-Trans., AI negative political impact (Table~\ref{tab:political_regressions_profiles}, Col.\ 1) & 0.14*** & 0.14*** & 0.15*** & 0.14*** & 0.14*** \\
            \addlinespace[2pt]
            Level: Trans.\ $-$ Intrinsic, economic-right placement, UK (Table~\ref{tab:political_regressions_profiles}, Col.\ 5) & $-$0.08 & $-$0.12* & $-$0.15** & $-$0.16** & $-$0.16** \\
            \addlinespace[2pt]
            AI threat slope: Trans.\ $-$ Status-Driven, High Income tercile (Table~\ref{tab:app_income_subgroups}) & 0.21*** & 0.22*** & 0.23*** & 0.23*** & 0.23*** \\
            \addlinespace[2pt]
            AI threat slope: Trans.\ $-$ Status-Driven, White-Collar tier (Table~\ref{tab:app_occ_subgroups}) & 0.26*** & 0.25*** & 0.26*** & 0.26*** & 0.26*** \\
            \addlinespace[2pt]
            Reservation income (pp): Trans.\ $-$ Intrinsic (Table~\ref{tab:reservation_regressions}, Model 3) & $-$1.57** & $-$1.77*** & $-$1.85*** & $-$1.76*** & $-$1.74*** \\
            \addlinespace[2pt]
            Reservation income (pp): Trans.\ $-$ Non-Trans.\ (Table~\ref{tab:reservation_regressions}, Model 3) & $-$0.13 & $-$0.22 & $-$0.32 & $-$0.23 & $-$0.22 \\
            \addlinespace[1.5pt]
            Part-time $\to \Pr(\text{Intrinsic})$ (Table~\ref{tab:app_profile_validation}, Model 5) & 0.017 & 0.006 & 0.009 & 0.011 & 0.011 \\
            Part-time $\to \Pr(\text{Transactional})$ (Table~\ref{tab:app_profile_validation}, Model 1) & $-$0.058*** & $-$0.055*** & $-$0.056*** & $-$0.058*** & $-$0.057*** \\
            \addlinespace[1.5pt]
            Transactional $\to$ reservation wage, SD (Occ.\ Tier $\times$ Profile interaction model) & 0.09 & 0.06 & 0.06 & 0.07 & 0.07 \\
            \addlinespace[2pt]
            Transactional $\to$ AI threat perception, SD (Occ.\ Tier $\times$ Profile interaction model) & $-$0.17*** & $-$0.20*** & $-$0.20*** & $-$0.21*** & $-$0.21*** \\
            \addlinespace[1.5pt]
            Work identity salience: $\Delta(\text{Intrinsic} - \text{Trans.})$ (Figure~\ref{fig:app_important_to_identity}) & 0.78*** & 0.79*** & 0.85*** & 0.84*** & 0.84*** \\
            Societal AI threat: $\Delta(\text{Trans.} - \text{Non-Trans.})$ (Figure~\ref{fig:ai_threat_by_work_values}) & 0.50*** & 0.47*** & 0.48*** & 0.48*** & 0.48*** \\
            \bottomrule
        \end{tabular*}
        \begin{tablenotes}[flushleft]
            \scriptsize
            \item \textit{Notes:} *** $p < 0.01$, ** $p < 0.05$, * $p < 0.10$. Each column rebuilds the Work Values Profiles using a different cutoff on the 1--4 scale ($1 = \text{Very important}$, $4 = \text{Not important at all}$); respondents without a winning index below the cutoff, four-way straight-liners, and non-workers are assigned to ``Other'' and excluded. All specifications replicate the main and appendix tables exactly.
        \end{tablenotes}
    \end{threeparttable}
\end{table}

\end{document}